\documentclass[twocolumn,pdftex]{aastex63}
\usepackage[T1]{fontenc}
\usepackage[utf8]{inputenc}
\usepackage{multirow}
\usepackage{xspace}
\usepackage{grffile}
\usepackage{xstring}

\graphicspath{{./}{figures/}}

\newcommand{\kms}{km\,s$^{-1}$\xspace}

\def\lya{\mbox {Ly$\alpha$}}
\def\lyb{\mbox {Ly$\beta$}}

\def\cm3{~cm$^{-3}$}

\shorttitle{Highly Ionized N~V and O~VI Outflows in the QUEST Quasars}
\shortauthors{Veilleux et al.}

\begin{document}

\title{Galactic Winds across the Gas-Rich Merger Sequence. \\ I. Highly
  Ionized N~V and O~VI Outflows in the QUEST Quasars \footnote{Based on
    observations made with the NASA/ESA {\em Hubble Space Telescope},
    obtained from the data archive at the Space Telescope Science
    Institute. STScI is operated by the Association of Universities
    for Research in Astronomy, Inc. under NASA contract NAS 5-26555.}}

\correspondingauthor{Sylvain Veilleux}
\email{veilleux@astro.umd.edu}

\author[0000-0002-3158-6820]{Sylvain Veilleux} 
\affiliation{Department of Astronomy, University of Maryland, College
  Park, MD 20742, USA} 
\affiliation{Joint Space-Science Institute, University of Maryland,
  College Park, MD 20742, USA}

\author[0000-0002-1608-7564]{David S. N. Rupke}
\affiliation{Department of Physics, Rhodes College, Memphis, TN 38112,
  USA}

\author{Weizhe Liu} 
\affiliation{Department of Astronomy, University of Maryland, College
  Park, MD 20742, USA} 

\author{Anthony To}
\affiliation{Department of Physics, Rhodes College, Memphis, TN 38112, USA}
\affiliation{Department of Physics, University of Hawaii, Honolulu, HI
  96822, USA}

\author{Margaret Trippe} 
\affiliation{Department of Astronomy, University of Maryland, College
  Park, MD 20742, USA}
\affiliation{Johns Hopkins University Applied Physics Laboratory,
  Laurel, MD 20723 USA}
\affiliation{Lincoln Laboratory, Massachusetts Institute of Technology, Lexington, MA 02421-6426 USA}

\author{Todd M. Tripp}
\affiliation{Department of Astronomy, University of Massachussetts,
  Amherst, MA 01003, USA}

\author{Fred Hamann}
\affiliation{Department of Physics and Astronomy, University of
  California, Riverside, CA 92507, USA}

\author{Reinhard Genzel}
\affiliation{Max-Planck-Institut f\"ur Extraterrestrische Physik, Giessenbachstrasse 1, 85748 Garching, Germany}

\author{Dieter Lutz}
\affiliation{Max-Planck-Institut f\"ur Extraterrestrische Physik, Giessenbachstrasse 1, 85748 Garching, Germany}

\author{Roberto Maiolino}
\affiliation{Cavendish Laboratory, University of Cambridge, 19 J.J. Thomson Avenue, Cambridge CB3 0HE, United Kingdom}
\affiliation{Kavli Institute for Cosmology, University of Cambridge, Madingley Road, Cambridge CB3 0HA, United Kingdom}

\author{Hagai Netzer}
\affiliation{School of Physics and Astronomy, Tel-Aviv University, Tel Aviv 69978, Israel}

\author{Kenneth R. Sembach}
\affiliation{Space Telescope Science Institute, Baltimore, MD 21218, USA}

\author{Eckhard Sturm}
\affiliation{Max-Planck-Institut f\"ur Extraterrestrische Physik, Giessenbachstrasse 1, 85748 Garching, Germany}

\author{Linda Tacconi}
\affiliation{Max-Planck-Institut f\"ur Extraterrestrische Physik, Giessenbachstrasse 1, 85748 Garching, Germany}

\author{Stacy H.\ Teng} 
\affiliation{Department of Astronomy, University of Maryland, College
  Park, MD 20742, USA}
\affiliation{Institute for Defense Analyses, Alexandria, MD 22311, USA}



\begin{abstract}
This program is part of QUEST (Quasar/ULIRG Evolutionary Study) and
seeks to examine the gaseous environments of $z \la 0.3$ quasars and
ULIRGs as a function of host galaxy properties and age across the
merger sequence from ULIRGs to quasars. This first paper in the series
focuses on 33 quasars from the QUEST sample and on the kinematics of
the highly ionized gas phase traced by the N~V
$\lambda\lambda$1238,1243 and O~VI $\lambda\lambda$1032,1038
absorption lines in high-quality {\em Hubble Space Telescope ({\em
    HST}) } Cosmic Origins Spectrograph (COS) data.
N~V and O~VI outflows are present in about 60\% of the QUEST quasars
and span a broad range of properties, both in terms of equivalent
widths (from 20 m\AA\ to 25 \AA) and kinematics (outflow velocities
from a few $\times$ 100 \kms\ up to $\sim$10,000 \kms). The rate of
incidence and equivalent widths of the highly ionized outflows are
higher among X-ray weak or absorbed sources. The weighted outflow
velocity dispersions are highest among the X-ray weakest sources. No
significant trends are found between the weighted outflow velocities
and the properties of the quasars and host galaxies although this may
be due to the limited dynamic range of properties of the
current sample.
These results will be re-examined in an upcoming paper where the
sample is expanded to include the QUEST ULIRGs. Finally, a lower limit
of $\sim$0.1\% on the ratio of time-averaged kinetic power to
bolometric luminosity is estimated in the 2$-$4 objects with
blueshifted P~V $\lambda\lambda$1117,1128 absorption features.
\end{abstract}



\section{Introduction} 
\label{sec:introduction}

Large multiwavelength surveys of the local and distant universe have
shown that major mergers of gas-rich galaxies\footnote{In this paper,
  we define major mergers as those involving galaxies with $\la$ 4:1
  stellar mass ratios.} may trigger spectacular bursts of star
formation, accompanied with quasar-like episodes of rapid growth of
the supermassive black holes (SMBHs), and result in merger remnants
that follow tight SMBH-host scaling relations and resemble today's
quiescent early-type galaxies
\citep[e.g.][]{sanders1988,hickox2018}. Modern simulations of galaxy
formation and evolution
\citep[e.g.][]{nelson2019,oppenheimer2020,nelson2021} largely
reproduce these observations. However, the root cause of the fast
``quenching'' of the star formation activity in the merger remnants
depends on the detailed, sub-grid scale, implementation of how the
mass, momentum, and energy from stellar winds, supernova explosions,
and SMBH-related processes are injected into, and interact with, the
interstellar medium (ISM) and circumgalactic medium (CGM) of the host
galaxies. Over the past several years, nearby gas-rich galaxy mergers
have emerged as excellent laboratories to study in detail these
stellar and quasar feedback processes \citep[see][for a recent
  review]{veilleux2020}. These objects are the focus of the present
study.

Locally, major gas-rich galaxy mergers often coincide with obscured
ultraluminous infrared galaxies (ULIRGs). As these systems evolve, the
obscuring gas and dust, funneled to the center by the dissipative
collapse and tidal forces during the merger, are either transformed
into stars or expelled out of the nucleus by powerful winds driven by
the central quasar and starburst, giving rise to dusty quasars and
finally to completely exposed quasars. Galactic-scale winds are
ubiquitous in local ULIRGs and dusty quasars \citep[e.g.][and
  references
  therein]{sturm2011,veilleux2013a,cicone2014,rupke2017,veilleux2017,fluetsch2019,
  fluetsch2020,lutz2020,veilleux2020}.  The outflows detected in
ULIRGs extend over a large range of distances from the central energy
source, seamlessly blending with the circumgalactic medium at $>$10
kpc \citep[][and references therein]{veilleux2020}.

In these objects, the outflow masses and energetics are often
dominated by the outer ($\ga$kpc) cool dusty molecular or neutral
atomic gas phase, but the driving mechanism is best probed by
examining the inner ($\la$ sub-kpc) ionized phase. ULIRG F11119$+$3257
is the first and still the best case among local ULIRGs where a fast
($>$0.1 c), highly ionized (Fe XXV/XXVI at $\sim$7 keV),
accretion-disk scale ($<$1 pc) quasar wind appears to be driving a
massive ($>$ 100 $M_\odot$ yr$^{-1}$), large-scale (1$-$10+ kpc)
molecular and neutral-gas outflow
\citep{tombesi2015,tombesi2017,veilleux2017}.  Unfortunately, the
search for hot winds in a statistically significant sample of ULIRGs
is not feasible at present since most ULIRGs are too faint at $\sim$7
keV for current X-ray observatories.

This is where the excellent far-ultraviolet (FUV) spectroscopic
sensitivity of the {\em Hubble Space Telescope} ({\em HST}) becomes
handy. The FUV band is rich in spectroscopic diagnostics of the
neutral, low-ionization, and high-ionization gas phases
\citep{haislmaier2021}; thus {\em HST} can probe all three phases at
once. So far, only about a dozen ULIRGs and IR-bright quasars have
been studied with {\em HST}, but the results have been promising.
Prominent, blueshifted \lya\ emission out to
$-$1000 km s$^{-1}$ has been detected in half of these
ULIRGs. Blueshifted absorption features from high-ionization species
like N~V and O~VI (77 and 114 eV are needed to produce N$^{4+}$ and
O$^{5+}$ ions, respectively) and/or low-ionization species like Si~II,
Si~III, Fe~II, N~II, and Ar~I have provided additional unambiguous
signatures of outflows in a few of these objects.
\citet{martin2015} have argued that the FUV-detected outflows
represent clumps of gas condensing out of a fast, hot wind generated
by the central starburst \citep{thompson2016}. This picture is also
consistent with the blast-wave model for quasar feedback. In this
model, a fast, hot wind shocks the surrounding ISM, which then
eventually cools to reform the molecular gas after having acquired a
significant fraction of the initial kinetic energy of the hot wind
\citep[e.g.][]{weymann1985,zubovas2012,zubovas2014,faucher-giguere2012,nims2015,richings2018a,richings2018b,richings2021,girichidis2021}. An
alternative explanation is radiative acceleration
\citep[e.g.][]{ishibashi2018,ishibashi2021}, which may dominate
  the dynamics of outflows on a wide variety of scales
  \citep[e.g.][]{stern2016,revalski2018,somalwar2020}.

So far, the published data set on ULIRGs and IR-bright quasars is too
small to draw strong conclusions about the properties of the
FUV-detected winds. There is tantalizing evidence that UV-detected
AGN/starburst-driven winds
are present in most ULIRGs,
but the sample is very incomplete, particularly among ULIRGs with AGN
and matched quasars. A more diverse sample of ULIRGs and quasars is
needed to study the gaseous environments of nearby quasars and ULIRGs
as a function of host properties and age across the merger sequence
from ULIRGs to quasars.  This issue is addressed in the present study.

In this first paper, we focus our efforts on studying the highly
ionized gas, traced by N~V $\lambda\lambda$1238, 1243 and O~VI
$\lambda\lambda$1032, 1038,
in a sample of 33 local quasars, while Paper II (Liu et al. 2021, in
prep.)  will present the results on our sample of ULIRGs with AGN and
compare them with those on the quasars. As stated in
\citet{hamann2019}, the quasars in the present sample are valuable for
outflow studies in and of themselves because: 1) they fill a
largely-unexplored niche between luminous quasars with strong broad
absorption lines (BALs) with outflow velocities of up to 0.1 $-$ 0.2
$c$ and low-luminosity Seyfert 1 galaxies with exclusively narrower
outflow lines, 2) their low redshift minimizes contamination by the
\lya\ forest, and 3) the outflow lines are relatively narrow so
blending is less severe. Indeed, as we discuss below, the detected
outflows often are ``mini-BALs'' instead of BALs because their
velocity widths lie below or near the threshold of 2000 \kms\ used for
BALs \citep{weymann1981,weymann1991,hamann2004,gibson2009a}.

Our quasar sample is discussed in Section \ref{sec:sample}. The
extensive set of ancillary data on these quasars is summarized in
Section \ref{sec:ancillary}.
The {\em HST} spectra used for this study are described in Section
\ref{sec:data} and the methods applied to analyze these data are
detailed in Section \ref{sec:analysis}. The results from this analysis
are presented in Section \ref{sec:results}, and discussed in more
detail
in Section \ref{sec:discussion}. Section \ref{sec:summary} provides a
summary of the main results from this paper.


\section{Quasar Sample}
\label{sec:sample}

The quasars in our sample are selected using four criteria: (1) They
must be part of the QUEST (Quasar/ULIRG Evolutionary Study) sample of
local ($z \la 0.3$) ULIRGs and quasars. The QUEST sample has already
been described in detail in \citet{veilleux2009a,veilleux2009b} and
references therein. All 33 objects in the present sample are
Palomar-Green (PG) quasars from the Bright Quasar Sample
\citep{schmidt1983}, except Mrk~231, the nearest quasar known, whose
UV spectrum has already been analyzed by
\citet{veilleux2013b,veilleux2016} and will not be discussed here any
further. As part of the QUEST sample, the quasars are carefully
matched in terms of redshifts, bolometric luminosities, and host
galaxy masses with the QUEST ULIRGs of Paper II.  (2) Their bolometric
luminosity must be quasar-like, $\ga 10^{45}$ ergs s$^{-1}$, and
dominated by the quasar rather than the starburst based on the {\em
  Spitzer} data (see criterion \#3 below) or, equivalently, have
25-to-60 $\mu$m flux ratios $f_{25}/f_{60} \ga 0.15$
\citep{veilleux2009a}. This criterion also automatically selects
UV-detected late-stage mergers or non-mergers \citep{veilleux2009b}.
(3) A strong preference is also given to the QUEST quasars with {\em
  Spitzer} mid-infrared spectra to provide valuable information on the
AGN contribution to the bolometric luminosities of these objects. (4)
High-quality COS spectra covering systemic N~V and/or O~VI must exist
for each object in the sample. Only COS data are considered to ease
comparisons between spectra and avoid possible systematic errors
associated with comparing data sets from different instruments. As
described in Section \ref{sec:data}, both our own and archival data
are used for this study.

\begin{figure}[!htb]   
\epsscale{1.1}
\plotone{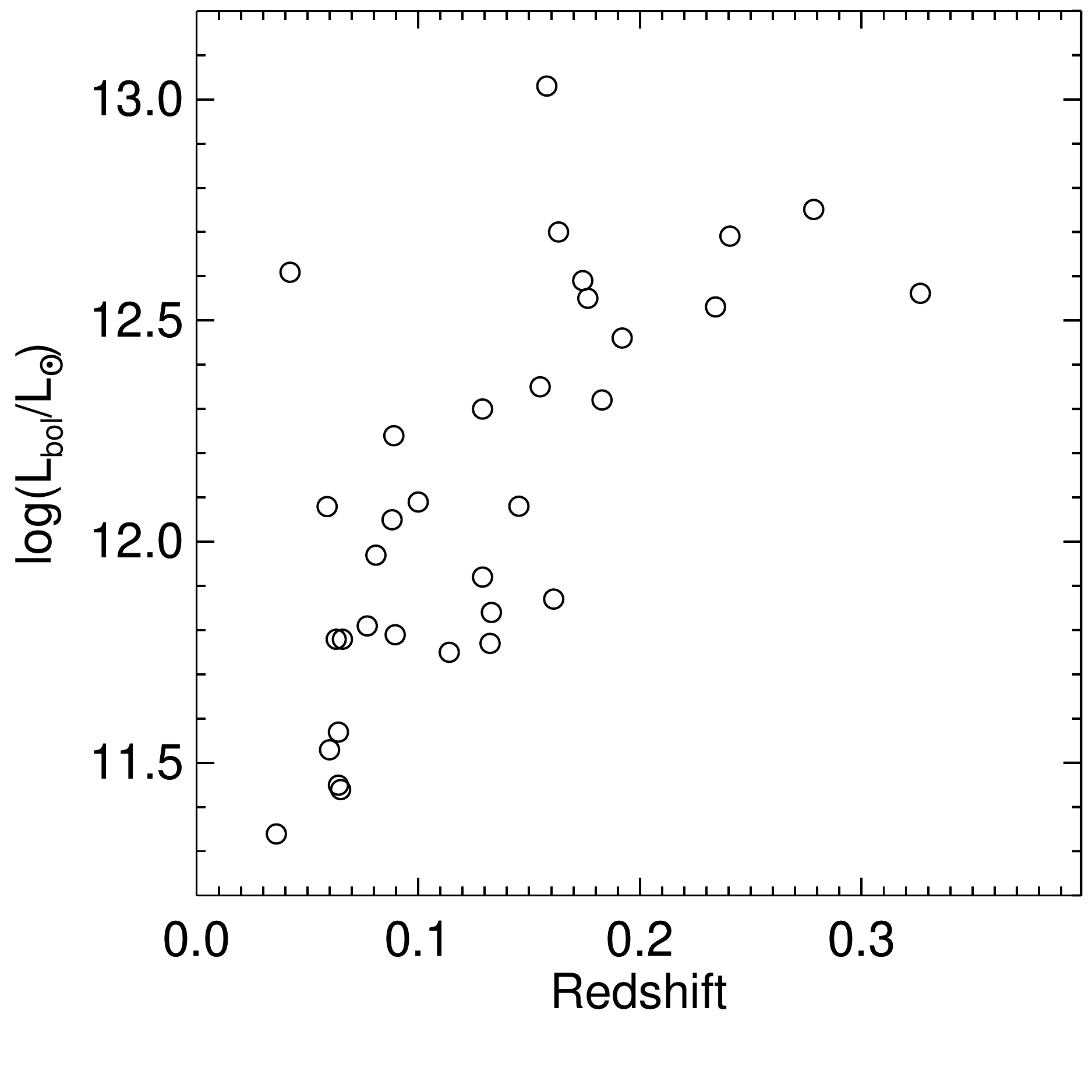}
\caption{Bolometric (AGN + starburst) luminosities of the QUEST
  quasars in the sample as a function of their redshifts.
}
\label{fig:lbol_vs_z}
\end{figure}

These criteria result in a sample of 33 objects. Table
\ref{tab:sample} lists the key properties of the quasars in our
sample, many of which are derived from our extensive set of ancillary
data on these objects, discussed in Section \ref{sec:ancillary}.  As
shown in Figure \ref{fig:lbol_vs_z}, these quasars cover the
low-redshift and low bolometric luminosity ends of the PG quasar
sample. They are well matched in redshift with the QUEST ULIRGs which
will be studied in Paper II (Liu et al. 2021, in prep), and are
representative of the entire PG~quasars sample in terms of infrared
excess (defined here as the infrared-to-bolometric luminosity ratio,
$L_{\rm IR}/L_{\rm BOL}$) and FIR brightness \citep[$L_{\rm 60~\mu
    m}/L_{\rm 15~\mu m}$ from][]{netzer2007}.


\section{Ancillary Data}
\label{sec:ancillary}

An extensive set of spectroscopic and photometric data exist on all of
the objects in the sample. Sloan Digital Sky Survey (SDSS) optical
spectra are available for all of them. High-quality optical spectra
also exist in \citet{boroson1992}, and \citet{krug2013} presents spectra
centered on Na I~D $\lambda\lambda$5890, 5896. As mentioned in Section
\ref{sec:sample}, most of these quasars have also been studied
spectroscopically in the mid-infrared with {\em Spitzer}
\citep{schweitzer2006,schweitzer2008,netzer2007,veilleux2009a}.
In addition, nearly all of the quasars in this sample are part of
X-QUEST, an archival {\em XMM-Newton} and {\em Chandra} X-ray
spectroscopic survey of the QUEST sample (Teng \& Veilleux 2010;
Columns 13-17 in Table \ref{tab:sample}). VLT and Keck near-infrared
spectroscopic data exist for a number of these objects
\citep{dasyra2007}.

Optical and near-infrared images of these objects have been obtained
from the ground \citep{surace2001,veilleux2002,guyon2006} and with
{\em HST} \citep{veilleux2006,kim2008,hamilton2008,veilleux2009b},
providing photometric and morphological measurements on both the
quasars and host galaxies (e.g., morphological type, quasar-to-host
luminosity ratio, strength of tidal features).  Far-infrared
photometry obtained with the {\em Herschel} PACS instrument exists for
all of these objects \citep{lani2017,shangguan2018}, while
far-infrared spectra centered on the OH 119 $\mu$m feature exist for
five of them \citep{veilleux2013a}. Finally, Green Bank Telescope
(GBT) H~I 21-cm line emission and absorption spectra are available for
16 of these quasars \citep{teng2013}.

The connection between UV and X-ray properties is critical, so we
  searched the literature for additional X-ray measurements (ignoring
  older ones from \emph{ROSAT/ASCA}). These are listed and referenced
  in Table \ref{tab:sample}. For \emph{Chandra} observations from
  \citet{teng2010} where no absorbing column was detected, more recent
  constraints from \citet{ricci2017} (based on \emph{Swift/BAT}
  detections) are available in some cases. In these cases, we
  substitute the newer measurement of absorbing column.

\begin{longrotatetable}
\movetabledown=0.25in
\begin{deluxetable*}{ccccc ccccc cccc ccc}
\tabletypesize{\tiny}
\tablecolumns{17}
\tablecaption{Properties of the QUEST Quasars in the Sample\label{tab:sample}}
\tabletypesize{\scriptsize}
\setlength{\tabcolsep}{2pt}

\tablehead{\colhead{Name} & \colhead{Other} & \colhead{$z$} &
  \colhead{log~$\nu$L$_{\nu}$(UV)} & \colhead{log~$R$} &
  \colhead{Radio} & \colhead{$\alpha_{\rm OX}$} &
  \colhead{log($\frac{L_{\rm bol}}{L_\odot}$)} &
  \colhead{$\alpha_{\rm AGN}$} &
  \colhead{log($\frac{L_{\rm IR}}{L_\odot}$)} &
  \colhead{log($\frac{M_{\rm BH}}{M_\odot}$)} &
  \colhead{log~$\eta_{\rm Edd}$} &
  \colhead{log($L_{\rm SX}$)} &
  \colhead{log($L_{\rm HX}$)} &
  \colhead{$\Gamma_X$} &
  \colhead{$N_{\rm H}$} & \colhead{Ref.}\\
  & Name & & [erg~s$^{-1}$] & & Class & & & & & & & [erg~s$^{-1}$] & [erg~s$^{-1}$] & & [10$^{22}$ cm$^{-2}$] & }
\colnumbers
\startdata
PG~0007$+$106 & III~Zw~2    & 0.0893 & 44.55 & $+$2.29 & Flat  &    $-$1.43 & 12.24 & 1\tablenotemark{a}        & 11.63 & 8.07$_{-0.46}^{+0.45}$ & $-$0.34$_{-0.45}^{+0.46}$ & 43.94                      & 44.18                      & 1.73$_{-0.04}^{+0.04}$ & 0.11$_{-0.01}^{0.02}$     & a, 4, 5 \\
PG~0026$+$129 & ...         & 0.1454 & 45.32 & $+$0.03 & Quiet &    $-$1.50 & 12.08 & 0.986$_{-0.027}^{+0.014}$ & 11.71 & 8.49$_{-0.45}^{+0.44}$ & $-$0.93$_{-0.44}^{+0.45}$ &                            & 44.40                      & 2.00$_{-0.11}^{+0.13}$ & $<$0.01                   & g, 5 \\
PG~0050$+$124 & I~Zw~1      & 0.0589 & 44.24 & $-$0.48 & Quiet &    $-$1.56 & 12.08 & 0.925$_{-0.094}^{+0.075}$ & 12.04 & 7.33$_{-0.62}^{+0.62}$ &    0.20$_{-0.62}^{+0.62}$ & 44.04$_{-0.09}^{+0.05}$    & 43.88$_{-0.09}^{+0.05}$    & 2.25$_{-0.03}^{+0.05}$ & 0.09$_{-0.02}^{+0.08}$    & f, 6 \\
~             &             &        &       &         &       &            &       &                           &       &                        &                           & 43.78$_{-0.12}^{+0.02}$    & 43.64$_{-0.04}^{+0.01}$    & 2.09$_{-0.03}^{+0.03}$ & 0.04$_{-0.02}^{+0.03}$    & 6 \\
~             &             &        &       &         &       &            &       &                           &       &                        &                           &                            & 43.633$_{-0.005}^{+0.005}$ & 2.37$_{-0.04}^{+0.08}$ & 0.045                     & 3 \\
PG~0157$+$001 & Mrk~1014    & 0.1633 & 45.34 & $+$0.33 & Quiet &    $-$1.60 & 12.70 & 0.727$_{-0.271}^{+0.236}$ & 12.67 & 8.06$_{-0.62}^{+0.61}$ & $-$0.01$_{-0.65}^{+0.63}$ & 43.92$_{-0.04}^{+0.04}$    & 43.83$_{-0.25}^{+0.11}$    & 2.54$_{-0.09}^{+0.09}$ & $<$0.009                  & e, 3, 6 \\
~             &             &        &       &         &       &            &       &                           &       &                        &                           & 44.00$_{-0.08}^{+0.04}$    & 43.86$_{-0.04}^{+0.04}$    & 2.1$_{-0.1}^{+0.1}$    & $<$0.009                  & 3, 6 \\
PG~0804$+$761 & ...         & 0.100  & 45.54 & $-$0.22 & Quiet &    $-$1.52 & 12.09 & 0.996$_{-0.004}^{+0.004}$ & 11.98 & 8.73$_{-0.43}^{+0.43}$ & $-$1.16$_{-0.43}^{+0.43}$ & 44.54                      & 44.45                      & 2.27$_{-0.20}^{+0.09}$ & 0.044$_{-0.010}^{+0.007}$ & a \\
PG~0838$+$770 & VII~Zw~244  & 0.1324 & 44.83 & $-$0.96 & Quiet &    $-$1.54 & 11.77 & 0.945$_{-0.018}^{+0.027}$ & 11.66 & 8.05$_{-0.62}^{+0.61}$ & $-$0.82$_{-0.61}^{+0.62}$ & 44.152$_{-0.014}^{+0.009}$ & 43.54$_{-0.05}^{+0.04}$    & 1.49$_{-0.08}^{+0.08}$ & $<$0.1\tablenotemark{b}   & d, 4, 5 \\
PG~0844$+$349 & ...         & 0.064  & 44.59 & $-$1.52 & Quiet &    $-$1.54 & 11.45 & 0.971$_{-0.049}^{+0.029}$ & 11.18 & 7.86$_{-0.49}^{+0.46}$ & $-$0.94$_{-0.46}^{+0.49}$ & 44.152$_{-0.014}^{+0.009}$ & 43.80$_{-0.03}^{+0.03}$    & 2.66$_{-0.06}^{+0.05}$ & 6.13$_{-1.39}^{+3.03}$    & a, 6 \\
PG~0923$+$201 & ...         & 0.192  & 45.42 & $-$0.85 & Quiet &    $-$1.57 & 12.46 & 0.990$_{-0.000}^{+0.000}$ & 12.05 & 7.90$_{-0.62}^{+0.61}$ & 0.04$_{-0.61}^{+0.62}$    &                            &                            &                        &                           & i \\
PG~0953$+$414 & ...         & 0.2341 & 45.95 & $-$0.36 & Quiet &    $-$1.50 & 12.53 & 0.982$_{-0.018}^{+0.018}$ & 12.20 & 8.33$_{-0.44}^{+0.44}$ & $-$0.33$_{-0.44}^{+0.44}$ & 45.037$_{-0.009}^{+0.006}$ & 44.81$_{-0.03}^{+0.02}$    & 2.44$_{-0.03}^{+0.03}$ & 18.52$_{-5.6}^{+9.84}$    & h, 6 \\
PG~1001$+$054 & ...         & 0.1611 & 44.93 & $-$0.30 & Quiet &    $-$2.13 & 11.87 & 0.836$_{-0.100}^{+0.123}$ & 11.66 & 7.63$_{-0.62}^{+0.61}$ & $-$0.35$_{-0.62}^{+0.62}$ & 43.00$_{-0.30}^{+0.18}$    & 43.08$_{-0.60}^{+0.15}$    & 2.01$_{-0.48}^{+0.67}$ & 8.09$_{-3.57}^{+5.47}$    & e, 6 \\
PG~1004$+$130 & 4C~$+$13.41 & 0.2406 & 45.30 & $+$2.36 & Steep & $<$$-$2.01 & 12.69 & 0.963$_{-0.036}^{+0.037}$ & 12.22 & 9.16$_{-0.62}^{+0.61}$ & $-$1.01$_{-0.61}^{+0.62}$ & 43.48$_{-0.05}^{+0.05}$    & 43.76$_{-0.13}^{+0.08}$    & 1.67$_{-0.11}^{+0.20}$ & 2.99$_{-1.37}^{+2.67}$    & e, 6 \\
~             &             &        &       &         &       &            &       &                           &       &                        &                           & 43.51$_{-0.15}^{+0.04}$    & 43.89$_{-0.22}^{+0.07}$    & 1.52$_{-0.26}^{+0.17}$ & 1.44$_{-0.69}^{+0.64}$    & 6 \\
PG~1116$+$215 & ...         & 0.1765 & 45.79 & $-$0.14 & Quiet &    $-$1.57 & 12.55 & 0.991$_{-0.008}^{+0.009}$ & 12.28 & 8.42$_{-0.62}^{+0.61}$ & $-$0.39$_{-0.61}^{+0.62}$ & 44.927$_{-0.006}^{+0.021}$ & 44.65$_{-0.04}^{+0.03}$    & 2.53$_{-0.03}^{+0.04}$ & 27.21$_{-11.26}^{+16.01}$ & h, 6 \\
~             &             &        &       &         &       &            &       &                           &       &                        &                           & 44.922$_{-0.004}^{+0.004}$ & 44.65$_{-0.04}^{+0.03}$    & 2.49$_{-0.01}^{+0.01}$ & 31.61$_{-4.13}^{+5.14}$   & 6 \\
~             &             &        &       &         &       &            &       &                           &       &                        &                           & 44.93$_{-0.01}^{+0.01}$    & 44.67$_{-0.03}^{+0.03}$    & 2.51$_{-0.04}^{+0.04}$ & 20.21$_{-5.10}^{+5.94}$   & 6 \\
PG~1126$-$041 & Mrk~1298    & 0.060  & 44.29 & $-$0.77 & Quiet &    $-$2.13 & 11.53 & 0.962$_{-0.075}^{+0.038}$ & 11.52 & 7.64$_{-0.62}^{+0.61}$ & $-$0.65$_{-0.61}^{+0.62}$ & 43.04$_{-0.05}^{+0.05}$    & 43.11$_{-0.12}^{+0.05}$    & 1.95$_{-0.10}^{+0.10}$ & 4.66$_{-0.39}^{+0.42}$    & a, 6 \\
PG~1211$+$143 & ...         & 0.0809 & 44.96 & $+$1.39 & Steep &    $-$1.57 & 11.97 & 1.000$_{-0.000}^{+0.000}$ & 11.74 & 7.85$_{-0.62}^{+0.61}$ & $-$0.40$_{-0.61}^{+0.62}$ & 44.328$_{-0.007}^{+0.005}$ & 43.94$_{-0.01}^{+0.01}$    & 2.83$_{-0.02}^{+0.02}$ & 12.98$_{-0.90}^{+0.94}$   & h, 6 \\
~             &             &        &       &         &       &            &       &                           &       &                        &                           & 44.201$_{-0.005}^{+0.005}$ & 43.89$_{-0.01}^{+0.01}$    & 2.63$_{-0.02}^{+0.02}$ & 12.40$_{-1.49}^{+1.64}$   & 6 \\
PG~1226$+$023 & 3C~273      & 0.158  & 46.50 & $+$3.06 & Flat  &    $-$1.47 & 13.03 & 0.949$_{-0.128}^{+0.051}$ & 12.80 & 8.41$_{-0.24}^{+0.15}$ & 0.08$_{-0.16}^{+0.24}$    & 45.491$_{-0.002}^{+0.002}$ & 45.742$_{-0.008}^{+0.008}$ & 2.07$_{-0.01}^{+0.01}$ & $<$0.01                   & a, 5, 6 \\
~             &             &        &       &         &       &            &       &                           &       &                        &                           & 45.461$_{-0.004}^{+0.004}$ & 45.722$_{-0.006}^{+0.005}$ & 1.81$_{-0.01}^{+0.01}$ & $<$0.01                   & 5, 6 \\
~             &             &        &       &         &       &            &       &                           &       &                        &                           & 45.591$_{-0.002}^{+0.002}$ & 45.820$_{-0.005}^{+0.004}$ & 2.28$_{-0.01}^{+0.01}$ & $<$0.01                   & 5, 6 \\
~             &             &        &       &         &       &            &       &                           &       &                        &                           & 45.663$_{-0.001}^{+0.002}$ & 45.825$_{-0.006}^{+0.006}$ & 2.08$_{-0.01}^{+0.01}$ & $<$0.01                   & 5, 6 \\
~             &             &        &       &         &       &            &       &                           &       &                        &                           & 45.461$_{-0.004}^{+0.003}$ & 45.67$_{-0.02}^{+0.01}$    & 2.13$_{-0.02}^{+0.02}$ & $<$0.01                   & 5, 6 \\
~             &             &        &       &         &       &            &       &                           &       &                        &                           & 45.544$_{-0.003}^{+0.003}$ & 45.941$_{-0.007}^{+0.010}$ & 1.96$_{-0.02}^{+0.01}$ & $<$0.01                   & 5, 6 \\
PG~1229$+$204 & Mrk~771     & 0.064  & 44.42 & $-$0.96 & Quiet &    $-$1.49 & 11.57 & 0.985$_{-0.030}^{+0.015}$ & 11.27 & 7.76$_{-0.48}^{+0.46}$ & $-$0.71$_{-0.46}^{+0.48}$ & 43.785$_{-0.008}^{+0.007}$ & 43.61$_{-0.02}^{+0.02}$    & 2.38$_{-0.03}^{+0.03}$ & 13.52$_{-3.36}^{+5.77}$   & g, 6 \\
Mrk~231       & ...         & 0.04217& 42.70 &         &       &    $-$1.92 & 12.61 & 0.709$_{-0.067}^{+0.066}$ & 12.54 & 8.58$_{-0.50}^{+0.50}$ & $-$0.63$_{-0.50}^{+0.50}$ & 42.13$_{-0.04}^{+0.01}$    & 42.58$_{-0.11}^{+0.01}$    & 1.40$_{-0.1}^{+0.03}$  & 9.5$_{-1.9}^{+2.3}$       & b, 7, 8 \\
~             &             &        &       &         &       &            &       &                           &       &                        &                           &                            &                            &                        & 19.4$_{-4.4}^{+5.7}$      & 7, 8 \\
PG~1302$-$102 &PKS~1302-102 & 0.2784 & 45.83 & $+$2.27 & Flat  &    $-$1.58 & 12.75 & 0.982$_{-0.037}^{+0.018}$ & 12.49 & 8.77$_{-0.62}^{+0.61}$ & $-$0.55$_{-0.61}^{+0.62}$ &                            & 44.81                      & 1.66$_{-0.11}^{+0.10}$ & $<$0.06                   & h, 1 \\
PG~1307$+$085 & ...         & 0.155  & 45.35 & $-$1.00 & Quiet &    $-$1.52 & 12.35 & 0.952$_{-0.066}^{+0.048}$ & 11.76 & 8.54$_{-0.46}^{+0.44}$ & $-$0.72$_{-0.44}^{+0.46}$ & 44.02$_{-0.02}^{+0.02}$    & 44.16$_{-0.09}^{+0.05}$    & 1.89$_{-0.10}^{+0.11}$ & 5.64$_{-1.48}^{+2.62}$    & a, 6 \\
PG~1309$+$355 & ...         & 0.1829 & 45.05 & $+$1.26 & Flat  &    $-$1.71 & 12.32 & 0.870$_{-0.127}^{+0.130}$ & 12.05 & 8.24$_{-0.62}^{+0.61}$ & $-$0.49$_{-0.62}^{+0.62}$ & 43.87$_{-0.01}^{+0.02}$    & 43.88$_{-0.05}^{+0.04}$    & 2.19$_{-0.06}^{+0.07}$ & 6.02$_{-1.84}^{+3.68}$    & i, 6 \\
PG~1351$+$640 & ...         & 0.0882 & 45.22 & $+$0.64 & Quiet &    $-$1.78 & 12.05 & 0.779$_{-0.221}^{+0.143}$ & 11.87 & 8.72$_{-0.62}^{+0.61}$ & $-$1.30$_{-0.63}^{+0.62}$ & 43.398$_{-0.023}^{+0.007}$ & 43.23$_{-0.04}^{+0.03}$    & 2.42$_{-0.04}^{+0.04}$ & 14.61$_{-3.81}^{+5.72}$   & h, 6 \\
PG~1411$+$442 & ...         & 0.0896 & 44.34 & $-$0.89 & Quiet &    $-$2.03 & 11.79 & 1.000$_{-0.000}^{+0.000}$ & 11.66 & 8.54$_{-0.46}^{+0.45}$ & $-$1.27$_{-0.45}^{+0.46}$ & 43.60$_{-0.06}^{+0.05}$    & 43.41$_{-0.18}^{+0.06}$    & 2.41$_{-0.15}^{+0.18}$ & 26.29$_{-4.08}^{+3.76}$   & h, 6 \\
PG~1435$-$067 & ...         & 0.129  & 45.12 & $-$1.15 & Quiet &    $-$1.63 & 11.92 & 0.976$_{-0.040}^{+0.024}$ & 11.47 & 8.26$_{-0.62}^{+0.61}$ & $-$0.86$_{-0.61}^{+0.62}$ & 44.11$_{-0.04}^{+0.04}$    & 43.94$_{-0.08}^{+0.07}$    & 2.36$_{-0.10}^{+0.11}$ & $<$0.1\tablenotemark{b}         & a, 6 \\
PG~1440$+$356 & Mrk~478     & 0.077  & 45.18 & $-$0.43 & Quiet &    $-$1.38 & 11.81 & 0.836$_{-0.081}^{+0.071}$ & 11.76 & 7.36$_{-0.62}^{+0.61}$ & $-$0.15$_{-0.61}^{+0.62}$ & 44.33$_{-0.01}^{+0.01}$    & 43.74$_{-0.06}^{+0.05}$    & 3.02$_{-0.04}^{+0.04}$ & 8.92$_{-3.51}^{+8.66}$    & a, 6 \\
~             &             &        &       &         &       &            &       &                           &       &                        &                           & 44.375$_{-0.004}^{+0.005}$ & 43.90$_{-0.01}^{+0.01}$    & 2.86$_{-0.01}^{+0.01}$ & 14.24$_{-2.34}^{+2.88}$   & 6 \\
~             &             &        &       &         &       &            &       &                           &       &                        &                           & 44.299$_{-0.006}^{+0.006}$ & 43.74$_{-0.03}^{+0.03}$    & 2.98$_{-0.02}^{+0.02}$ & 8.225$_{-1.49}^{+2.03}$   & 6 \\
~             &             &        &       &         &       &            &       &                           &       &                        &                           & 44.127$_{-0.008}^{+0.007}$ & 43.64$_{-0.03}^{+0.02}$    & 2.86$_{-0.02}^{+0.02}$ & 12.67$_{-2.70}^{+3.45}$   & 6 \\
PG~1448$+$273 & ...         & 0.065  & 43.78 & $-$0.60 & Quiet &    $-$1.59 & 11.44 & 0.997$_{-0.007}^{+0.003}$ & 11.19 & 6.86$_{-0.62}^{+0.61}$ & 0.06$_{-0.61}^{+0.62}$    & 43.949$_{-0.007}^{+0.006}$ & 43.49$_{-0.02}^{+0.02}$    & 2.80$^{+0.02}_{-0.01}$ & 16.72$^{+6.24}_{-4.39}$   & a, 6 \\
PG~1501$+$106 & Mrk~841     & 0.036  & 44.03 & $-$0.44 & Quiet &    $-$1.64 & 11.34 & 1.000$_{-0.000}^{+0.000}$ & 11.13 & 8.42$_{-0.62}^{+0.61}$ & $-$1.59$_{-0.61}^{+0.62}$ & 44.049$_{-0.002}^{+0.005}$ & 43.833$_{-0.008}^{+0.008}$ & 2.46$^{+0.02}_{-0.02}$ & 23.12$^{+6.40}_{-4.56}$   & a, 6 \\
~             &             &        &       &         &       &            &       &                           &       &                        &                           & 44.090$_{-0.005}^{+0.002}$ & 43.845$_{-0.007}^{+0.008}$ & 2.50$^{+0.02}_{-0.02}$ & 18.66$^{+3.60}_{-2.77}$   & 6 \\
~             &             &        &       &         &       &            &       &                           &       &                        &                           & 44.068$_{-0.004}^{+0.002}$ & 43.851$_{-0.006}^{+0.007}$ & 2.45$^{+0.02}_{-0.02}$ & 15.88$^{+2.42}_{-2.07}$   & 6 \\
~             &             &        &       &         &       &            &       &                           &       &                        &                           & 43.740$_{-0.004}^{+0.004}$ & 43.672$_{-0.006}^{+0.006}$ & 2.26$^{+0.01}_{-0.01}$ & 13.01$^{+0.57}_{-0.50}$   & 6 \\
~             &             &        &       &         &       &            &       &                           &       &                        &                           & 43.623$_{-0.006}^{+0.004}$ & 43.65$_{-0.01}^{+0.01}$    & 2.11$^{+0.02}_{-0.02}$ & 11.48$^{+1.69}_{-1.56}$   & 6 \\
PG~1613$+$658 & Mrk~876     & 0.129  & 45.43 & $+$0.00 & Quiet &    $-$1.21 & 12.30 & 0.820$_{-0.092}^{+0.106}$ & 12.25 & 8.34$_{-0.51}^{+0.46}$ & $-$0.64$_{-0.46}^{+0.51}$ & 44.28$_{-0.03}^{+0.06}$    & 44.38$_{-0.09}^{+0.07}$    & 1.95$_{-0.10}^{+0.10}$ & 28.24$_{-20.67}^{+107.78}$& a, 6 \\
~             &             &        &       &         &       &            &       &                           &       &                        &                           & 44.44$_{-0.02}^{+0.03}$    & 44.43$_{-0.06}^{+0.05}$    & 2.12$_{-0.08}^{+0.08}$ & 10.45$_{-4.77}^{+10.78}$  & 6 \\
PG~1617$+$175 & Mrk~877     & 0.114  & 44.93 & $-$0.14 & Quiet &    $-$1.64 & 11.75 & 0.903$_{-0.081}^{+0.097}$ & 11.55 & 8.67$_{-0.45}^{+0.44}$ & $-$1.48$_{-0.44}^{+0.45}$ &                            &                            &                        &                           & a \\
PG~1626$+$554 & ...         & 0.133  & 45.17 & $-$0.96 & Quiet &    $-$1.37 & 11.84 & 0.976$_{-0.019}^{+0.024}$ & 10.90 & 8.39$_{-0.62}^{+0.61}$ & $-$1.08$_{-0.61}^{+0.62}$ & 44.17$_{-0.08}^{+0.08}$    & 44.21$_{-0.07}^{+0.07}$    & 2.04$_{-0.14}^{+0.15}$ & $<$0.01                   & a, 5, 6 \\
PG~2130$+$099 & II~Zw~136   & 0.063  & 44.46 & $-$0.49 & Quiet &    $-$1.47 & 11.78 & 0.995$_{-0.010}^{+0.005}$ & 11.63 & 7.43$_{-0.43}^{+0.43}$ & $-$0.17$_{-0.43}^{+0.43}$ & 43.708$_{-0.009}^{+0.011}$ & 43.62$_{-0.02}^{+0.02}$    & 2.29$_{-0.05}^{+0.05}$ & 5.91$_{-0.62}^{+0.73}$    & j, 6 \\
PG~2214$+$139 & Mrk~304     & 0.0658 & 44.45 & $-$1.30 & Quiet &    $-$2.02 & 11.78 & 0.998$_{-0.004}^{+0.002}$ & 11.46 & 8.44$_{-0.62}^{+0.62}$ & $-$1.18$_{-0.62}^{+0.62}$ & 43.46$_{-0.07}^{+0.04}$    & 43.64$_{-0.17}^{+0.05}$    & 1.80$_{-0.16}^{+0.16}$ & 4.48$_{-0.68}^{+0.68}$    & e, 6 \\
PG~2233$+$134 & ...         & 0.3265 & 45.94 & $-$0.55 & Quiet &    $-$1.66 & 12.56 & 1\tablenotemark{a}        & 12.33 & 7.93$_{-0.62}^{+0.61}$ & 0.12$_{-0.61}^{+0.62}$    &                            & 44.52                      & 2.41$_{-0.18}^{+0.18}$ & $<$0.01                   & e, 2 \\
PG~2349$-$014 & 4C$-$01.61  & 0.1742 & 45.51 &         &       &            & 12.59 & 0.904$_{-0.054}^{+0.045}$ & 11.90 & 9.14$_{-0.50}^{+0.50}$ & $-$1.11$_{-0.50}^{+0.50}$ &                            & 44.57                      & 1.78$_{-0.35}^{+0.20}$ & $<$0.01                   & e, 5 \\
\enddata

\tablecomments{Column (1): Object name. Column (2): Other name. Column
  (3): Redshifts, with reference listed in Column (17). Where
  available, redshifts are based on the [\ion{O}{3}] narrow line.  For
  3 quasars, we use \ion{H}{1}
  \citep{devaucouleurs1991,carilli1998,springob2005} instead; for two,
  full-spectrum fits to SDSS spectra \citep{schneider2010}; and for a
  single case, CO data \citep{evans2006}. Column (4): Logarithm of the
  monochromatic luminosity at rest-frame 1125 \AA\ derived using the
  Galactic extinctions from \citet{schlafly2011} and the reddening
  curve with $R_V =
  3.1$ of \citet{fitzpatrick1999}. Column (5): logarithm of
  $R$, the ratio of radio-to-optical luminosity from
  \citet{boroson1992}. Column (6): Radio class
  $-$ Quiet, Steep, or Flat depending on
  log~$R$ and radio spectral index from \citet{boroson1992}. Column
  (7): X-ray to optical spectral index $\alpha_{\rm
    OX}$ = 0.372 log~($f_{\rm 2~keV}/f_{\rm
    3000~A}$) from \citet{brandt2000}, where $f_{\rm
    2~keV}$ and $f_{\rm
    3000~A}$ are the rest-frame flux densities at 2 keV and 3000 \AA,
  respectively. (For Mrk~231, we report the value from
  \citet{teng2014}.) Column (8): Bolometric luminosity in solar units
  calculated from 7 $\times$ $L$(5100 \AA) + $L_{\rm
    IR}$ \citep{netzer2007}, where
  $L$(5100 \AA) is the continuum luminosity $\lambda
  L_\lambda$ at 5100 \AA\ rest wavelength and $L_{\rm
    IR}$ is the 1 $-$ 1000
  $\mu$m infrared luminosity listed in column (10). We adopt
  a cosmology of $H_0$ $=$ 69.3 km~s$^{-1}$;
  $\Omega_m$ = 0.287;
  $\Omega_{\lambda}$ = 0.713 (WMAP9). Column (9): Fraction of the bolometric
  luminosity produced by the AGN, i.e.\ $\alpha_{\rm AGN} = L_{\rm
    AGN}/L_{\rm
    BOL}$, based on the {\em Spitzer} results \citep{veilleux2009a}. The error bars are computed from the lowest and highest values among the six methods from this paper. 
  Column (10): Logarithm of the 1 $-$ 1000
  $\mu$m infrared luminosity in solar units from \citet{zhuang2018},
  except Mrk~231 \citep{u2013},
  PG~1626$+$554 \citep{lyu2017}, and
  PG~2349$-$014 \citep{veilleux2009b}. Column (11): Logarithm of the
  black hole mass in solar units from reverberation mapping (RM)
  measurements from The AGN Black Hole Mass Database \citep{bentz2015}
  or, if unavailable, single-epoch measurements from
  \citet{vestergaard2006}, normalized down from
  $f$ = 5.5 \citep{Onken2004} to
  $f$ = 4.3 to match RM scaling. According to \citet{vestergaard2006},
  single-epoch measurements should have an extra 0.43 error added in
  quadrature. For
  PG~2349$-$014, we used the more uncertain photometric measurement of
  \citet{veilleux2009b}. For 3C~273, we recorded the GRAVITY
  measurement \citep{gravity2018}. Column (12): Logarithm of the ratio
  of the bolometric luminosity to the Eddington luminosity. Column
  (13): Luminosity in the soft X-rays (0.5
  $-$ 2 keV). Column (14): Luminosity of the hard X-rays (2
  $-$ 10 keV). Column (15): Photon index of the best-fit absorbed
  power-law distribution to the X-ray emission (d$N$/d$E$ $\propto
  E^{-\Gamma_X}$ where
  $E$ is the X-ray photon energy). Column (16): Column density in
  units of 10$^{22}$
  cm$^{-2}$. For most data from \citet{teng2010}, this is from the
  best-fit absorbed power-law. Where this best fit is unabsorbed and
  \emph{Swift/BAT} data are available, we substitute
  $\mathrm{N_H}$ from \citet{ricci2017}. For X-ray related quantities,
  different rows = different observations, dates. Column (17):
  References for the redshift and X-ray measurements.}

\tablerefs{Redshift: (a) \citealt{boroson1992}; (b)
  \citealt{carilli1998}; (c) \citealt{devaucouleurs1991}; (d)
  \citealt{evans2006}; (e) \citealt{hewett2010}; (f) \citealt{ho2009};
  (g) \citealt{hu2020}; (h) \citealt{marziani1996}; (i)
  \citealt{schneider2010}; (j) \citealt{springob2005}; X-ray:
  (1) \citealt{inoue2007}; (2) \citealt{jin2012}; (3)
  \citealt{laha2018}; (4) \citealt{piconcelli2005}; (5)
  \citealt{ricci2017}; (6) \citealt{teng2010}; (7) \citealt{teng2014};
  (8) \citealt{veilleux2014}; (9) \citealt{waddell2020} }

\tablenotetext{a}{No measurement; we assume a value of 1.}
\tablenotetext{b}{Best-fit value is 0; we assume an upper limit of
  10$^{21}$~cm$^{-2}$.}

\end{deluxetable*}
\end{longrotatetable}

\section{HST Data}
\label{sec:data}

We obtained high-quality spectra for 19 quasars using the Cosmic
Origins Spectrograph (COS) with grating G130M under {\em HST} PID
12569 in Cycle 19 (PI Veilleux). We acquired multi-epoch COS/G130M
data on PG~1411$+$442 under programs 13451, 14460, and 14885 (PI
Hamann). We searched for archival COS/G130M spectra of other quasars
in the QUEST sample, as well as for multi-epoch data on the subsample
we observed. We found archival data on 14 additional QUEST quasars
and recent multi-epoch exposures for three quasars in the subsample
first observed in Cycle 19. We list the characteristics of these
observations in Table \ref{tab:observations}.

\begin{deluxetable*}{ccccccc}
\tablecolumns{7}
\tabletypesize{\tiny} \tablecaption{Summary of the Observations\label{tab:observations}}
\tablehead{\colhead{Name} & \colhead{Range [\AA]} & \colhead{CENWAVE [\AA]} & \colhead{t$_{exp}$[s]} & \colhead{Date} & \colhead{PID} &
  \colhead{PI} } \colnumbers \startdata
PG~0007$+$106  & 1151--1470 & 1309/1327           &  1868 & 2011-12-14 & 12569 & S.\ Veilleux\\
PG~0026$+$129  & 1133--1451 & 1291/1309           &  1868 & 2011-10-25 & 12569 & S.\ Veilleux\\
PG~0050$+$124  & 1151--1470 & 1309/1327           &  1868 & 2012-11-01 & 12569 & S.\ Veilleux\\
               & 1133--1465 & 1291/1309/1327      &  7621 & 2015-01-20 & 13811 & E.\ Costantini\\
PG~0157$+$001  & 1151--1469 & 1309/1327           &  1828 & 2012-01-25 & 12569 & S.\ Veilleux\\
PG~0804$+$761  & 1136--1458 & 1291/1300/1309/1318 &  5510 & 2010-06-12 & 11686 & N.\ Arav\\
PG~0838$+$770  & 1136--1458 & 1291/1300/1309/1318 &  8865 & 2009-09-24 & 11520 & J.\ Green\\
PG~0844$+$349  & 1151--1470 & 1309/1327           &  1900 & 2012-03-06 & 12569 & S.\ Veilleux\\
PG~0923$+$201  & 1133--1451 & 1291/1309           &  1860 & 2012-03-14 & 12569 & S.\ Veilleux\\
PG~0953$+$414  & 1136--1458 & 1291/1300/1309/1318 &  4785 & 2011-10-18 & 12038 & J.\ Green\\
PG~1001$+$054  & 1066--1367 & 1222                &  2068 & 2014-04-04 & 13423 & R.\ Cooke\\
               & 1140--1455 & 1291/1300/1309/1318 &  3165 & 2014-06-19 & 13347 & J.\ Bregman\\
               & 1131--1429 & 1291                &  2902 & 2019-03-26 & 15227 & J.\ Burchett\\
PG~1004$+$130  & 1133--1451 & 1291/1309           &  4107 & 2011-12-21 & 12569 & S.\ Veilleux\\
PG~1116$+$215  & 1136--1458 & 1291/1300/1309/1318 &  4677 & 2011-10-25 & 12038 & J.\ Green\\
PG~1126$-$041  & 1152--1470 & 1309/1327           &  1856 & 2012-04-15 & 12569 & S.\ Veilleux\\
               &  901--1200 & 1055                &  1874 & 2014-06-01 & 13429 & M.\ Giustini\\
               & 1171--1467 & 1327                &  1580 & 2014-06-01 & 13429 & M.\ Giustini\\
               &  900--1200 & 1055                &  1874 & 2014-06-12 & 13429 & M.\ Giustini\\
               & 1171--1467 & 1327                &  1580 & 2014-06-12 & 13429 & M.\ Giustini\\
               &  900--1200 & 1055                &  1874 & 2014-06-28 & 13429 & M.\ Giustini\\
               & 1171--1467 & 1327                &  1580 & 2014-06-28 & 13429 & M.\ Giustini\\
               &  901--1200 & 1055                &  1837 & 2015-06-14 & 13836 & M.\ Giustini\\
               & 1171--1468 & 1327                &  1540 & 2015-06-14 & 13429 & M.\ Giustini\\
PG~1211$+$143  & 1171--1472 & 1327                &  2320 & 2015-04-14 & 13947 & J.\ Lee\\
PG~1226$+$023  & 1135--1470 & 1291/1300/1309/1318/1327 &  4002 & 2012-04-22 & 12038 & J.\ Green\\
PG~1229$+$204  & 1152--1469 & 1309/1327           &  1868 & 2012-04-26 & 12569 & S. Veilleux\\
Mrk~231        & 1152--1472 & 1309/1327           & 12536 & 2011-10-15 & 12569 & S. Veilleux\\
PG~1302$-$102  & 1136--1458 & 1291/1300/1309/1318 &  5979 & 2011-08-16 & 12038 & J.\ Green\\
PG~1307$+$085  & 1152--1470 & 1309/1327           &  1836 & 2012-06-16 & 12569 & S.\ Veilleux\\
PG~1309$+$355  & 1133--1451 & 1291/1309           &  1896 & 2011-12-06 & 12569 & S.\ Veilleux\\
PG~1351$+$640  & 1152--1470 & 1309/1327           &  2108 & 2011-10-21 & 12569 & S.\ Veilleux\\
PG~1411$+$442  & 1152--1470 & 1309                &  1936 & 2011-10-23 & 12569 & S.\ Veilleux\\
               &  941--1241 & 1096                &  4954 & 2015-02-12 & 13451 & F. Hamann\\
               & 1152--1453 & 1309                &  1917 & 2015-02-12 & 13451 & F. Hamann\\
               &  941--1241 & 1096                &  2407 & 2016-04-16 & 14460 & F. Hamann\\
               & 1152--1453 & 1309                &  1954 & 2016-04-16 & 14460 & F. Hamann\\
               &  941--1241 & 1096                &  1783 & 2017-06-10 & 14885 & F. Hamann\\
               & 1152--1453 & 1309                &  1847 & 2017-06-10 & 14885 & F. Hamann\\
PG~1435$-$067  & 1133--1451 & 1291/1309           &  1864 & 2012-02-29  & 12569 & S.\ Veilleux\\
PG~1440$+$356  & 1152--1470 & 1309/1327           &  1924 & 2012-01-26  & 12569 & S.\ Veilleux\\
PG~1448$+$273  & 1136--1448 & 1291/1309           &  2946 & 2011-06-18  & 12248 & J.\ Tumlinson\\
PG~1501$+$106  & 1132--1434 & 1291                &  3121 & 2014-07-06  & 13448 & A.\ Fox\\
PG~1613$+$658  & 1145--1467 & 1300/1309/1318/1327 &  9499 & 2010-04-08  & 11524 & J.\ Green\\
               & 1133--1429 & 1291                &  3080 & 2010-04-09  & 11686 & N.\ Arav\\
PG~1617$+$175  & 1133--1451 & 1291/1309           &  1844 & 2012-06-16  & 12569 & S. Veilleux\\
PG~1626$+$554  & 1136--1458 & 1291/1300/1309/1318 &  3318 & 2011-06-15  & 12029 & J.\ Green\\
PG~2130$+$099  & 1135--1458 & 1291/1300/1309/1318 &  5513 & 2010-10-28  & 11524 & J.\ Green\\
PG~2214$+$139  & 1152--1463 & 1309/1327           &  1401 & 2011-11-08  & 12569 & S.\ Veilleux\\
               & 1138--1434 & 1291                &  2082 & 2012-09-21  & 12604 & A.\ Fox\\
PG~2233$+$134  & 1171--1472 & 1327                &  2104 & 2014-06-18  & 13423 & R.\ Cooke\\
PG~2349$-$014  & 1152--1470 & 1309/1327           &  1844 & 2011-10-20  & 12569 & S.\ Veilleux\\
\enddata

\tablecomments{Column (1): Name of object; Column (2): Wavelength
  range, in \AA; Column (3): CENWAVE setting(s); Column (4): Exposure
  time, in seconds; Column (5): Start date; Column (6): Proposal ID;
  Column (7): Program principal investigator.}
\end{deluxetable*}

Since most of the Cycle 19 COS data have not yet been the subject of a
paper \citep[the exceptions are Mrk~231 and
  PG~1411$+$442;][]{veilleux2013b,veilleux2016,hamann2019}, we briefly
summarize here how they were obtained. A total of 24 orbits were
allocated for these 19 targets with most targets requiring one
orbit. The exceptions are PG~1004$+$130 (2 orbits) and Mrk 231 (5
orbits). All but Mrk~231 are point sources with accurate positions;
they were acquired directly using ACQ/PEAKXD and ACQ/PEAKD. For
Mrk~231, a NUV image was obtained with ACQ/IMAGE. All observations
were observed in time-tag mode to allow us to exclude poor quality
data and improve thermal correction and background removal. We split
the exposures into four segments of similar durations at two FP\_POS
settings (\#2 and \#4) and two wavelength settings (CENWAVE) separated
by $\sim$20 \AA. This observing strategy reduces the fixed pattern
noise and fills up the chip gap without excessive overheads.

The observations include at least 1150--1450 \AA\ in the observer's
frame. This range includes redshifted O~VI $\lambda\lambda$ 1032, 1038,
N~V $\lambda\lambda$1238, 1243, \lya\ $\lambda$1216 and/or \lyb\
$\lambda$1025 in emission and/or absorption. In at least two cases
(PG~1126$-$041, PG~1411$+$442, and perhaps also PG~1001$+$054 and
PG~1004$+$130; see Sec.\ \ref{sec:discussion_pg1126}), the weaker P~V
$\lambda\lambda$1117, 1128 absorption lines are also detected. The
specific lines covered depend on the quasar redshift.  The
short-wavelength cutoff of the COS prevents us from searching for O~VI
systems in quasars with $z$ $\la$ 0.11, while N~V systems are
redshifted out of the COS data in quasars with $z$ $\ga$ 0.18. It is
therefore possible to study both O~VI and N~V only over a limited
range of quasar redshifts. Nevertheless, we achieve our science goals
by covering at least one H~I Lyman series line and one
high-ionization doublet (O~VI and/or N~V). In at least two cases
(PG~1411$+$442 and PG~1004$+$130), weaker and/or lower-ionization
lines, such as C~II $\lambda$1335, C~III $\lambda$977, N~III
$\lambda$990, O~I $\lambda$1304, Si~II $\lambda$1260, Si~III
$\lambda$1206, and Si~IV $\lambda\lambda$1394, 1403, are also present
in the spectra. These lines may be used to help constrain the
location, ionization, total column densities (${\mathrm N_H}$) and
metal abundances in the absorbing gas
\citep[e.g.][]{hamann2019}. PG~1004$+$130, one of the highest-redshift
sources in our sample, also shows S~VI $\lambda\lambda$933, 945.

All of our sample have data with CENWAVE of 1291, 1300, 1309, 1318,
and/or 1327, and almost all of the data (with the exception of the
final round of data on PG~1001$+$054) were obtained in COS lifetime
positions (LP) 1--3. For these CENWAVE settings, the spectral
resolution of COS increases with wavelength and has degraded somewhat
with changes in LP, but is still $>$10$^4$ at all wavelengths. This
corresponds to resolution better than 30~km~s$^{-1}$ FWHM at all
wavelengths, ranging up to a peak of $\sim$15~km~s$^{-1}$ at LP1 and
1450~\AA. Three quasars have additional data from CENWAVE 1055, 1096,
or 1222 observations. For these CENWAVE values, the spectral
resolution peaks in the FUVB (blue) segment at $>$10$^4$, but is lower
in FUVA, with average values of $\sim$3000, 5000, and $10^4$,
respectively.

We downloaded all exposures from the Hubble Legacy Archive and
determined that they were processed by CALCOS v3.3.10. For each
quasar, we coadded all exposures with CENWAVE 1222--1327 into a single
spectrum using v3.3 of {\em coadd\_x1d.pro} \citep{danforth2010},
setting BIN=3. The resulting median S/N per binned pixel over
1290--1310~\AA\ is 11.5, with a standard deviation of 10.9 and a range
of 2--52. (The low end of the range arises in a strong N~V BAL in
PG~1126$-$041.) We separately coadded the two quasar datasets with
CENWAVE 1055 and 1096.


\section{Data Analysis}
\label{sec:analysis}

\vskip 0.1in

We conducted a uniform analysis of the high-ionization absorbers in
our sample. In this section, we describe the methods we used to
identify and characterize these absorption features.

We used v0.5 of the publicly available, IDL-based IFSFIT package
\citep{rupke2014,rupke2015} to model the absorption lines. The rest of
the software used to model the data (continuum fitting, plotting,
regressions) is contained in or called from our public COSQUEST
  repository on GitHub \citep{david_rupke_2021_5659382}.

\subsection{Model Fitting}
\label{sec:fitting}

\begin{figure}[!htb]   
\epsscale{1.2}
\plotone{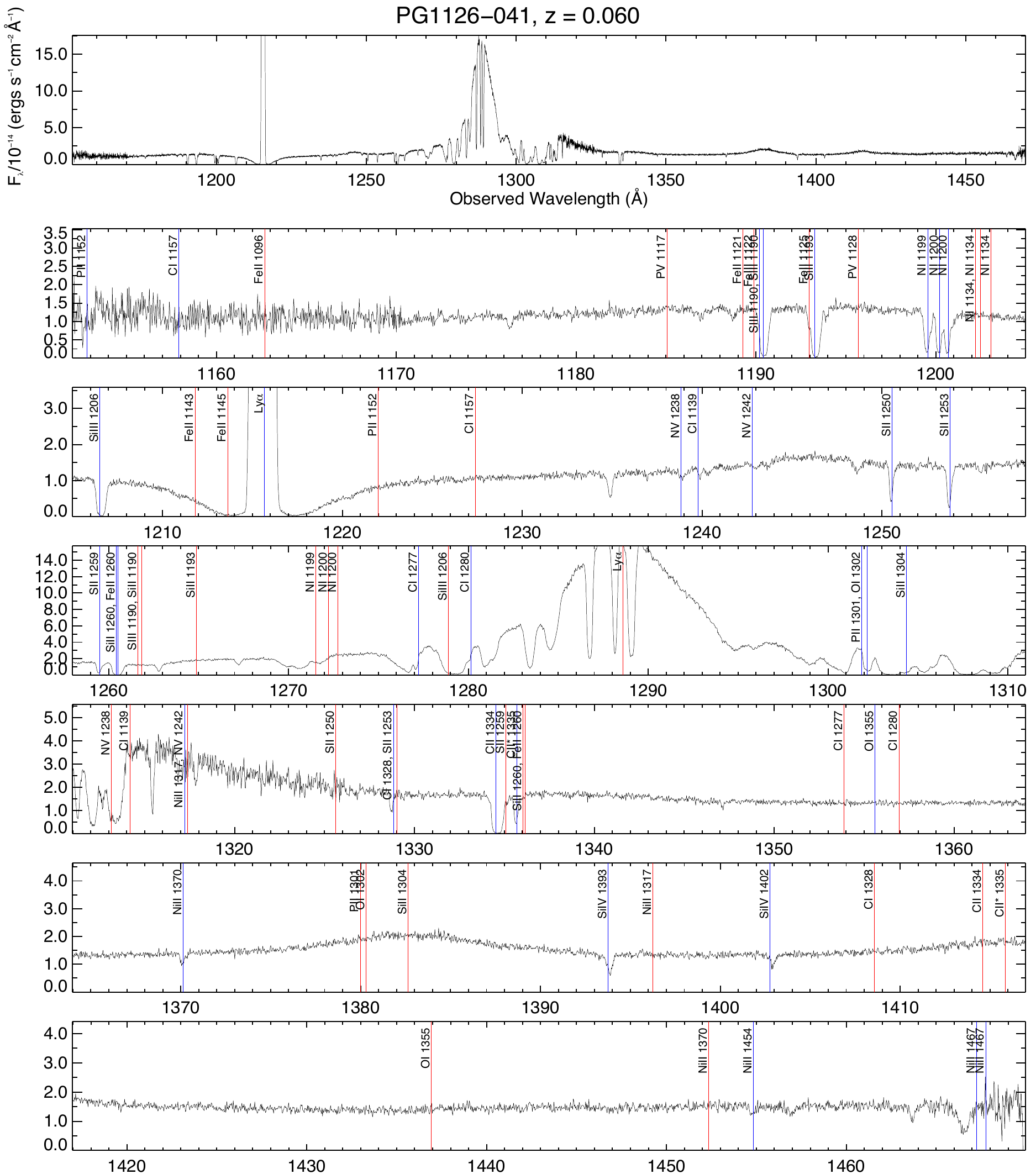}
\caption{An example of a FUV spectrum used in the study. Shown here is
  the COS spectrum of PG~1126$-$041, where the data are displayed in
  black and the expected positions of the features in the Milky Way
  and quasar rest-frames are indicated in blue and red,
  respectively.}
\label{fig:example_pg1126_spec}
\end{figure}

The starting point of the analysis is to identify the various emission
and absorption lines produced by the quasars and their
environments. Since our program is focused on QSO and ULIRG outflows,
we only identify and measure absorption lines within $\sim$10,000
\kms\ of the QSO redshifts. We refer to these lines as ``associated''
absorbers. Identifications of the foreground ``intervening'' absorbers
can be found in \citet{tripp2008}, \citet{savage2014}, and
\citet{danforth2016}. We first compare each quasar spectrum against a
list of common UV absorbers in quasar spectra \citep[][Figure
\ref{fig:example_pg1126_spec}]{prochaska2001}.
We list the quasar redshifts in Table \ref{tab:sample}. Most of these
redshifts are derived from narrow optical emission lines
  ([\ion{O}{3}], H$\beta$) and may underestimate the true recession
velocities since some fraction of the line emission may arise from the
outflowing material itself \citep[e.g.][]{rupke2017}. See
\citet{teng2013} for a comparison of these measurements with the H~I
21-cm emission and absorption line profiles.

Next, we fit the continuum and broad line emission (\lya, N~V,
and O~VI) in three separate spectral windows around \lya, N~V,
and O~VI$+$\lyb. In the two quasars in which we fit P~V, this
continuum region is also fit separately. Within each of these windows,
we use a piecewise function of 1--4 segments in the majority of
cases. In relatively featureless spectral regions, these segments are
low-order polynomials. In more complex spectral regions we employ
cubic B-splines. The B-splines are themselves piecewise polynomials,
and we separate the spline knots by a typical interval of 3~\AA. We
invoke BSPLINE\_ITERFIT from the SDSS IDLUTILS library to fit the
B-splines.

In seven cases, the fits with piecewise functions are poorly
constrained. For PG~1001$+$054, PG~1004$+$130, PG~1411$+$442, and
PG~1617$+$175, PG2130$+$099, and PG2214$+$139 this is due to broad,
deep absorption features over which it is difficult to fit polynomials
or splines. For a seventh quasar---PG~1351$+$640---the poor
constraints are due to several narrow absorbers near the peak of
\lya. In two of these cases, we instead use a Lorentzian profile to
fit \lya. For the five others we use the BOSS template from
\citet{harris2016}, scale it multiplicatively by a low-order power
law, and add a linear pedestal.

After fitting the continuum, we normalize the data in each spectral
window by dividing by this fit.

\begin{figure}[!htb]   
\epsscale{0.75}
\plotone{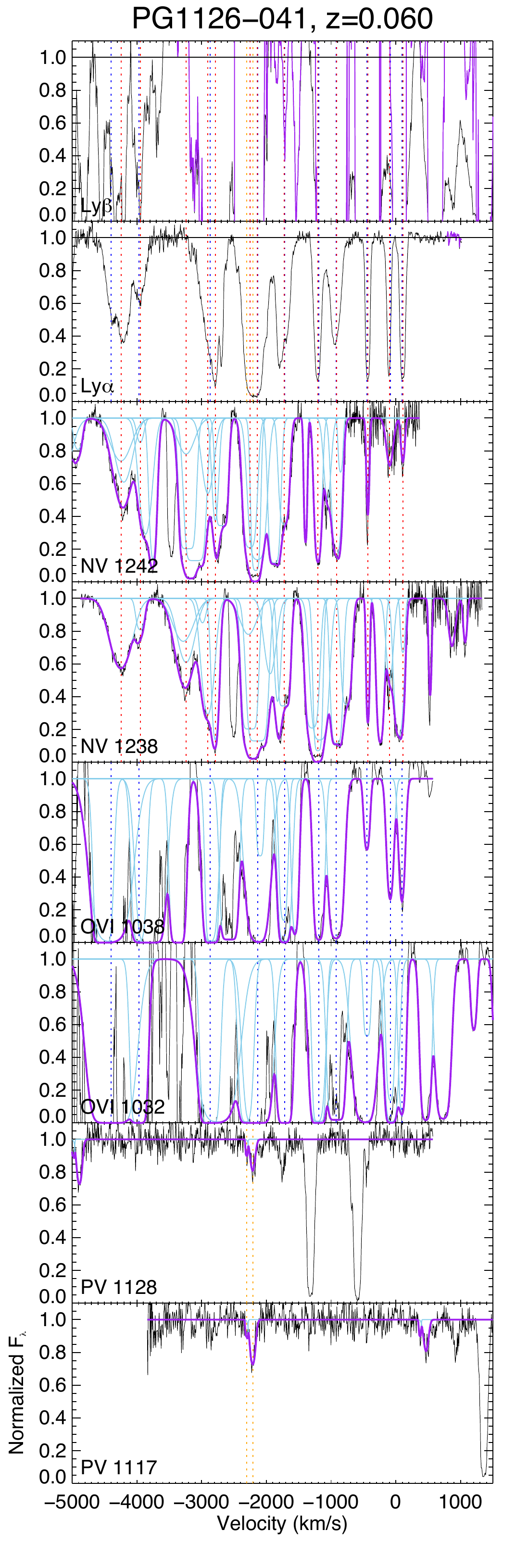}
\caption{An example of interline comparison that is used to identify
  absorbing systems associated with the quasars. The results shown
  here are for PG~1126$-$041, produced by dividing the spectrum shown
  in Fig.\ \ref{fig:example_pg1126_spec} by a smooth
  polynomical/spline/template fit to the continuum near the key
  absorption lines of our study and plotting the results in velocity
  space in the quasar rest frame. The data are in black, the
  components used to fit the absorption profiles are shown in blue,
  and the overall fit is shown in purple. The velocity centroids of
  the main absorbing systems are indicated by blue (O~VI) or red
  (Ly$\alpha$, Ly$\beta$, N~V, P~V) vertical dotted lines. The two
  strong lines in the panel labeled P~V $\lambda$1128 are Galactic ISM
  features without counterparts in P~V $\lambda$1117 or any of the
  other lines. }
\label{fig:example_pg1126_vel}
\end{figure}

We characterize the doublet absorption features (N~V
$\lambda\lambda$1238, 1243; O~VI $\lambda\lambda$1032, 1038; and P~V
$\lambda\lambda$1117, 1128) in the quasar spectra using simple model
fits. {\em Our primary objectives are to estimate the overall
  equivalent widths and kinematics of the outflowing gas associated
  with these features} (mass, momentum, and energy estimates are
beyond the scope of the present paper, except for a few special cases
discussed in Section \ref{sec:discussion_pg1126}). We are not aiming
to derive precise column densities from the (often saturated)
absorption line profiles, so the use of the precise COS line-spread
function (LSF) is not required here (we return to this point at
  the end of this section). If the lines within these doublets were
unblended, fits to the intensity profiles of the individual lines
would thus be sufficient. However, the doublet lines are often
strongly blended because of (1) strong blueshifts due to high outflow
velocities and (2) broad line profiles due to multiple clouds along
the line of sight and/or large linewidths. We thus adopt the doublet
fitting procedure of \citet{rupke2005}, which is optimized for blended
doublets. In this method, the total absorption profiles of a feature
are fit as the product of multiple doublet components. Each component
is a Gaussian in optical depth $\tau$ {\em vs} wavelength with a
constant covering factor $C_f$. Within each doublet the two lines have
a constant $\tau$ ratio. This allows us to simultaneously fit $\tau$
and $C_f$, which are otherwise degenerate in the fit of a single
line. The free parameters in the fit to each doublet component are
thus $C_f$, peak $\tau$, velocity width, and central wavelength. The
determination of the number of components needed in the fit is
subjective and non-linear $-$ it depends on the line complexity and
data quality. The main goal here is to get a good fit to the
absorption features to derive the equivalent widths and kinematics of
the outflowing gas associated with these features. We do not attach a
physical meaning to the individual components in the fit.

The general expression for the normalized intensity of a doublet
component is
\begin{eqnarray}
  I(\lambda) = 1 - C_f + C_f e^{-\tau_\mathrm{low}(\lambda)-\tau_\mathrm{high}(\lambda)},
\label{eq:I_lambda}
\end{eqnarray}
where $C_f$ is the line-of-sight covering factor (or the fraction of
the background source producing the continuum that is covered by the
absorbing gas; though scattering into the line of sight can also play
a role) and $\tau_\mathrm{low}$ and $\tau_\mathrm{high}$ are the
intrinsic optical depths of the lower- and higher-wavelength lines in
the doublet \citep{rupke2005}. The background light source is assumed
to be spatially uniform. The covering factor is the same for both
lines of the doublet. The peak (and total) optical depths of the
resonant doublet lines in O~VI, N~V, and P~V are related by a constant
factor $\tau_{low}/\tau_{high} = 2.00$ because of the 4-fold
degeneracy in the upper state of the higher energy transition compared
to the 2-fold degeneracy in the lower state. (The higher degeneracy is
due in turn to its higher total angular momentum quantum number
$j$). For more than one doublet component, we use the product of the
intensities of the individual components, which is the
partially-overlapping case of \citet{rupke2005}.

Because the doublet profile shape--i.e., relative depths of the two
lines and trough shape--does not change significantly above optical
depths $\tau_{high}$ of a few, we set a limit of $\tau_{high} \le
5$. Out of 59 O~VI components, 19 have $\tau_{high} = 5$, or 32\%. For
N~V, 13 of 62 components have $\tau_{high} = 5$, or 21\%.

The results from these fits are also used to calculate the total
velocity-integrated equivalent widths of the absorbers in the object's
rest frame,
\begin{eqnarray}
  W_{\rm eq} = \int [1 - f(v)] dv, 
\label{eq:weq}
\end{eqnarray}
the weighted average outflow velocity, 
\begin{eqnarray}
  v_{\rm wtavg} = \frac{\int v [1 - f(v)]dv }{W_{\rm eq}}, 
\label{eq:vwtavg}
\end{eqnarray}
and the weighted outflow velocity dispersion,
\begin{eqnarray}
  \sigma_{\rm rms} = \left(\frac{\int (v - v_{\rm wtavg})^2 [1 -
  f(v)] dv}{W_{\rm eq}}\right)^{\frac{1}{2}},
\label{eq:sigmawtavg}
\end{eqnarray}
a measure of the second moment in velocity space of the absorbers in
each quasar. These quantities are similar to those defined by
\citet{trump2006}, but without the constraints on depth, width, or
velocity.
These constraints have little effect on the results for our sample,
but we find it useful to include possibly inflowing absorbers. 
  Note that $W_{\rm eq}$, $v_{\rm wtavg}$, and $\sigma_{\rm rms}$ are
  not corrected for partial covering. To test the impact of this
  assumption on our results, we have recomputed them after changing
  the absorption lines so that they have $C_f = 1$ instead of the
  measured $C_f$, and then redid the regression analysis discussed in
  Sec.\ \ref{sec:regressions}. Only very small changes of order 1\% in
  the $p$-values are observed if we correct for partial covering.

Figure \ref{fig:example_pg1126_vel} shows the fits to the spectrum
presented in Figure \ref{fig:example_pg1126_spec}. The fits to all of
the features detected in the FUV spectra of the 33 quasars in our
sample are presented in Appendix \ref{appendix:detailed_results}, and
the results derived from these fits are tabulated in Table
\ref{tab:fits}.

We computed errors in best-fit parameters and derived model
  quantities by refitting the model spectrum 1000 times. In each case
  we added Gaussian-distributed random errors to each pixel in the
  model with $\sigma$ equal to the measurement error. These formal
  errors are small due to the high S/N in our data. Errors due to
  continuum placement are likely to dominate the true error budget.

  We also estimated upper limits to the doublet equivalent width in
  cases where we did not detect N~V and/or O~VI. To do so, we assumed
  an optically-thick ($\tau_{1243}$ or $\tau_{1038}=5$), $v=0$,
  $\sigma = 50$~\kms\ absorption line. We set the covering factor
  equal to half the root-mean-square deviation in the continuum within
  $\pm0.5$~\AA\ of the expected rest-frame location of each line in
  the doublet. (The factor-of-2 accounts for fitting 2 lines instead
  of 1.) We set the limit equal to the resulting model equivalent
  width.

The optical depths and covering factors derived from our fitting
scheme are approximations. Though it is a physically-motivated way to
decompose strongly-blended doublets, the method implicitly assumes
that the velocity dependences of $C_f$ and $\tau$ can be described as
the sum of discrete independent Gaussians. In reality, they are
probably more complex functions of velocity \citep[e.g.][]{arav2005,
  arav2008}. In several cases--the N~V absorbers in PG~1001$+$054,
PG~1411$+$442, PG~1617$+$175, and PG2214$+$139, and the O~VI absorbers
in PG~1001$+$054 and PG~1004$+$130--the fits include very broad
components that cannot be distinguished from complexes of narrower
lines given the data quality. In two O~VI absorbers (PG~0923$+$201 and
PG~1309$+$355), there are no data on the blue line because it is
contaminated by geocoronal \lya, so any constraints on $\tau$ and
$C_f$ come solely from line shape. Finally, in four O~VI fits
(PG~1001$+$054, PG~1004$+$130, PG~1126$-$041, and PG~1617$+$175), the
\lyb\ and O~VI absorption lines blend together and cannot be easily
separated in the fit. In three of these cases (all but PG~1001$+$054),
we simply fit the visible absorption as due solely to O~VI at
wavelengths in which there is at least some O~VI absorption
contributing to the spectrum. For the fourth case, we are able to
roughly separate the lines by fitting only down to a specific
wavelength. A detailed object-by-object discussion is given in
Appendix \ref{appendix:detailed_results}.

Despite these caveats, the fitting procedure is sufficient to meet our
primary objectives of estimating the overall equivalent widths and
kinematics of these features. The 3-$\sigma$ detection limit on the
doublet equivalent widths is typically $\sim$ 20 m\AA\ in our data
although it varies from one spectrum to the other.

We have conducted detailed tests of the impact of the COS LSF on
  our measurements to verify that the use of the precise COS LSF is
  not required here.  In one series of simulations, we created a
  series of fake, saturated Voigt line profiles with a median S/N of 5
  per pixel and line widths ranging from $\sigma$ = 10 km~s$^{-1}$ to
  50 km~s$^{-1}$. We convolved the profiles with the LSF downloaded
  from the COS website. We find that the LSF causes a difference of up
  to only $\sim$10\% on the line width and covering fraction
  measurements for lines with $\sigma \ge$ 20 km~s$^{-1}$ (the
  corresponding Doppler $b$ parameter of the Voigt profile is
  $\sqrt{2}~\sigma$ $\simeq$ 28 km~s$^{-1}$), which is smaller than
  the values measured for nearly all of the absorbers detected in our
  objects (Table \ref{tab:fits}). We have also run a COS LSF analysis
  on a ULIRG with narrow N~V absorption features, taken from the
  sample of Paper II. Using the method described here, we get a
  Doppler $b$ parameter of 78 km~s$^{-1}$ and covering fraction of
  0.84, while the COS LSF gives 83 km~s$^{-1}$ and 0.82, respectively,
  confirming that the results for the relatively broad absorbers
  reported in the present paper are reliable.

\begin{deluxetable*}{lccccc}
\footnotesize
\tabletypesize{\scriptsize}
\tablecolumns{6}
\tablecaption{Results from the Multi-Component Fits to the
  Absorbers\label{tab:fits}}
\tablehead{\colhead{Name} & \colhead{Line} & \colhead{$W_{\rm eq}$} &
  \colhead{$v_{\rm wtavg}$} & \colhead{$\sigma_{\rm rms}$} &
  \colhead{\# comp.} \\
& & \AA\ & \kms\ & \kms\ &  }
\colnumbers
\startdata
  PG0007+106  &  N~V  &$<$0.16  &  &  & \\
  PG0026+129  &  N~V  &$<$0.09  &  &  & \\
              & O~VI  &$<$0.12  &  &  & \\
  PG0050+124  &  N~V  &  0.88$_{-0.017}^{+0.019}$  &-1106.3$_{-18.7 }^{+19.0 }$  &  595.0$_{-16.4 }^{+18.0 }$  &  6 \\
  PG0157+001  &  N~V  &$<$0.13  &  &  & \\
              & O~VI  &$<$0.09  &  &  & \\
  PG0804+761  &  N~V  &  0.02$_{-0.003}^{+0.003}$  &  591.0$_{-1.7  }^{+1.8  }$  &   12.3$_{-1.8  }^{+1.9  }$  &  1 \\
              & O~VI  &  0.18$_{-0.007}^{+0.009}$  &  571.0$_{-1.5  }^{+1.6  }$  &   27.8$_{-1.1  }^{+1.1  }$  &  1 \\
  PG0838+770  &  N~V  &$<$0.12  &  &  & \\
              & O~VI  &$<$0.06  &  &  & \\
  PG0844+349  &  N~V  &  0.73$_{-0.011}^{+0.011}$  &  151.4$_{-0.8  }^{+0.8  }$  &   31.8$_{-0.5  }^{+0.5  }$  &  2 \\
  PG0923+201  & O~VI  &  4.48$_{-0.072}^{+0.072}$  &-3048.3$_{-11.9 }^{+11.4 }$  &  335.6$_{-8.7  }^{+9.3  }$  &  1 \\
  PG0953+414  & O~VI  &  0.20$_{-0.006}^{+0.006}$  & -825.4$_{-13.9 }^{+14.7 }$  &  418.3$_{-7.1  }^{+7.2  }$  &  2 \\
  PG1001+054  &  N~V  &  6.07$_{-0.059}^{+0.058}$  &-5969.9$_{-37.4 }^{+24.0 }$  & 1326.5$_{-11.7 }^{+18.0 }$  &  4 \\
              & O~VI  & 11.61$_{-0.072}^{+0.080}$  &-5743.9$_{-83.1 }^{+33.2 }$  & 1092.4$_{-27.6 }^{+32.7 }$  &  6 \\
  PG1004+130  & O~VI  & 24.29$_{-0.277}^{+0.245}$  &-5335.7$_{-75.4 }^{+80.0 }$  & 2970.2$_{-47.4 }^{+72.3 }$  & 12 \\
  PG1116+215  & O~VI  &  0.27$_{-0.008}^{+0.008}$  &-2271.6$_{-46.3 }^{+43.2 }$  &  908.7$_{-25.6 }^{+22.1 }$  &  2 \\
  PG1126-041  &  N~V  & 10.66$_{-0.034}^{+0.037}$  &-2085.5$_{-9.8  }^{+10.0 }$  & 1076.1$_{-7.2  }^{+6.9  }$  & 13 \\
              & O~VI  & 16.80$_{-1.616}^{+1.891}$  &-2559.2$_{-251.5}^{+298.8}$  & 1482.2$_{-169.4}^{+132.6}$  & 10 \\
              &  P~V  &  0.21$_{-0.017}^{+0.017}$  &-2234.2$_{-6.1  }^{+5.9  }$  &   48.2$_{-3.8  }^{+4.3  }$  &  2 \\
  PG1211+143  &  N~V  &$<$0.05  &  &  & \\
  PG1226+023  &  N~V  &$<$0.04  &  &  & \\
              & O~VI  &$<$0.04  &  &  & \\
  PG1229+204  &  N~V  &$<$0.09  &  &  & \\
     Mrk 231  &  N~V  &$<$0.18  &  &  & \\
  PG1302-102  & O~VI  &$<$0.05  &  &  & \\
  PG1307+085  &  N~V  &$<$0.10  &  &  & \\
              & O~VI  &  0.15$_{-0.016}^{+0.017}$  &-3406.2$_{-14.1 }^{+11.9 }$  &   66.8$_{-14.1 }^{+11.8 }$  &  2 \\
  PG1309+355  & O~VI  &  8.57$_{-0.031}^{+0.038}$  & -893.9$_{-4.7  }^{+4.9  }$  &  364.5$_{-3.1  }^{+3.2  }$  &  7 \\
  PG1351+640  &  N~V  &  4.36$_{-0.034}^{+0.036}$  &-1264.4$_{-11.4 }^{+10.0 }$  &  428.4$_{-3.8  }^{+4.8  }$  &  9 \\
  PG1411+442  &  N~V  & 10.30$_{-0.018}^{+0.019}$  &-1594.8$_{-2.4  }^{+2.6  }$  &  562.7$_{-2.3  }^{+2.5  }$  &  4 \\
              &  P~V  &  0.83$_{-0.028}^{+0.028}$  &-1754.6$_{-5.5  }^{+5.9  }$  &  131.0$_{-3.6  }^{+3.8  }$  &  2 \\
  PG1435-067  &  N~V  &$<$0.15  &  &  & \\
              & O~VI  &$<$0.08  &  &  & \\
  PG1440+356  &  N~V  &  0.89$_{-0.021}^{+0.023}$  &-1478.4$_{-26.7 }^{+28.1 }$  &  775.3$_{-25.8 }^{+27.0 }$  &  3 \\
  PG1448+273  &  N~V  &  3.22$_{-0.033}^{+0.038}$  & -229.8$_{-2.1  }^{+2.2  }$  &  164.1$_{-1.0  }^{+1.0  }$  &  4 \\
  PG1613+658  &  N~V  &  0.14$_{-0.004}^{+0.005}$  &-3714.6$_{-5.5  }^{+6.1  }$  &  121.8$_{-3.6  }^{+3.2  }$  &  2 \\
              & O~VI  &  0.68$_{-0.008}^{+0.008}$  &-3691.1$_{-1.8  }^{+2.0  }$  &  126.6$_{-0.8  }^{+0.8  }$  &  2 \\
  PG1617+175  &  N~V  &  3.00$_{-0.055}^{+0.059}$  &-3094.7$_{-23.8 }^{+19.9 }$  &  526.8$_{-22.5 }^{+34.3 }$  &  5 \\
              & O~VI  &  6.08$_{-0.173}^{+0.173}$  &-3323.7$_{-133.0}^{+113.3}$  &  920.2$_{-71.6 }^{+83.2 }$  &  8 \\
              &  P~V  &  0.06$_{-0.033}^{+0.040}$  &-3355.0$_{-38.0 }^{+55.1 }$  &   42.1$_{-25.8 }^{+46.1 }$  &  1 \\
  PG1626+554  &  N~V  &$<$0.08  &  &  & \\
              & O~VI  &$<$0.06  &  &  & \\
  PG2130+099  &  N~V  &  0.76$_{-0.013}^{+0.013}$  &-1312.3$_{-9.6  }^{+9.9  }$  &  540.3$_{-9.3  }^{+9.5  }$  &  3 \\
  PG2214+139  &  N~V  &  8.01$_{-0.023}^{+0.024}$  &-1461.1$_{-17.9 }^{+17.9 }$  &  681.7$_{-20.2 }^{+21.9 }$  &  5 \\
  PG2233+134  & O~VI  &  0.17$_{-0.024}^{+0.025}$  & -211.2$_{-3.0  }^{+3.0  }$  &   17.3$_{-2.9  }^{+3.2  }$  &  1 \\
  PG2349-014  &  N~V  &$<$0.12  &  &  & \\
\enddata

\tablecomments{Column (1): Name of object. Column (2): N~V means N~V
  $\lambda$1238, 1243, O~VI means O~VI $\lambda$1032, 1038, and P~V
  means P~V $\lambda$1117, 1128. N~V or O~VI is not listed when it
  lies outside of the spectral range of the data. Column (3):
  Velocity-integrated equivalent widths (eq.\ \ref{eq:weq}). Column
  (4): Average depth-weighted outflow velocity (eq.\ \ref{eq:vwtavg}),
  which is a measure of the average velocity of the outflow systems in
  each object. Column (5): Average depth-weighted outflow velocity
  dispersion (eq.\ \ref{eq:sigmawtavg}), which is a measure of the
  range in velocity of the outflow systems in each quasar. Column (6):
  Number of absorption components.
}
\end{deluxetable*}

\subsection{Regressions}
\label{sec:regressions}

To search for connections between outflow and quasar/host
  properties, we computed linear regressions between the properties in
  Tables \ref{tab:sample} and \ref{tab:fits}. In most cases, we apply
  the Bayesian model in LINMIX\_ERR \citep{kelly2007}. We use the
  Metropolis-Hastings sampler and a single Gaussian to represent the
  distribution of quasar/host parameters (except for $W_{\rm eq}$
  vs. AGN fraction, for which we used NGAUSS$=$3). LINMIX\_ERR permits
  censored y-values, which is the case for $W_{\rm eq}$.

  When we compute the regressions for the independent variable
  $\mathrm{N_H}$, however, the x-axis values are also censored. In
  this case we turn to the method of \citet{isobe1986} for computing
  the Kendall tau correlation coefficient with censored data in both
  axes. We use the implementation of pymccorrelation
  \citep{privon2020}, which in turn perturbs the data in Monte Carlo
  fashion to compute the errors in the correlation coefficient
  \citep{curran2014}.

  For both regression methods, we computed the significance of a
  correlation as the fraction of cross-correlation values $r < 0$ ($r
  > 0$) for a positive (negative) best-fit $r$. For LINMIX\_ERR, the
  $r$ values are draws from the posterior distribution, while for
  pymccorrelation they are results of the Monte Carlo perturbations.
  
  We do not consider the N~V and O~VI points independent for the
  purposes of the regressions. Therefore, where both doublets are
  present in the data for a given quasar, we compute the average
  measurement (either detection or limit) from the two lines. If only
  one line is detected, we use that measurement rather than averaging
  a detection and a limit. Where multiple X-ray measurements exist for
  a quasar, we take the average. Errors in $L_{\rm BOL}$,
  $L_{\rm IR}/L_{\rm BOL}$, $L_{\rm FIR}/L_{\rm BOL}$, and
  $\alpha_{\rm OX}$ are unknown, so for the purposes of regression we
  fix the errors to 0.1~dex. For $\nu L_\nu$(UV), we ignore the
  negligible statistical measurement errors.


\section{Results}
\label{sec:results}

\vskip 0.1in

The results from our spectral analysis of the {\em HST} spectra are
summarized in Table \ref{tab:fits}. In this section, we investigate
whether the presence or nature of quasar-driven outflows and starburst
winds
correlate with the properties of the quasars and host galaxies.  The
quantities that we consider in our correlation matrix are listed in
Table \ref{tab:sample} and defined in the notes to that table. 
  The results from the statistical and regression analyses are
  summarized in Tables \ref{tab:incidence} and \ref{tab:regressions}.

Note that we do not make a distinction between quasar-driven outflows
and starburst-driven winds in this section, and we only consider
absorption lines within 10,000 \kms\ of the QSO redshift (inclusion of
lines at greater displacements leads to unacceptable contamination by
intervening absorbers). A more comprehensive assessment of the
detected outflows is conducted in Section \ref{sec:discussion}, after
we have considered the line profiles more fully, including signs of
saturation and/or partial covering of the continuum source
(Sec.\ \ref{sec:tau_cf}) and the overall kinematics of the outflowing
gas (Sec.\ \ref{sec:kinematics}). Similarly, the comparison of our
results with those from previous studies is postponed until Section
\ref{sec:discussion}, once the results from our spectral analysis have
been fully presented.

\subsection{Rate of Incidence of Outflows}
\label{sec:incidence}

\begin{figure}[!htb]   
\epsscale{1.1}
\plotone{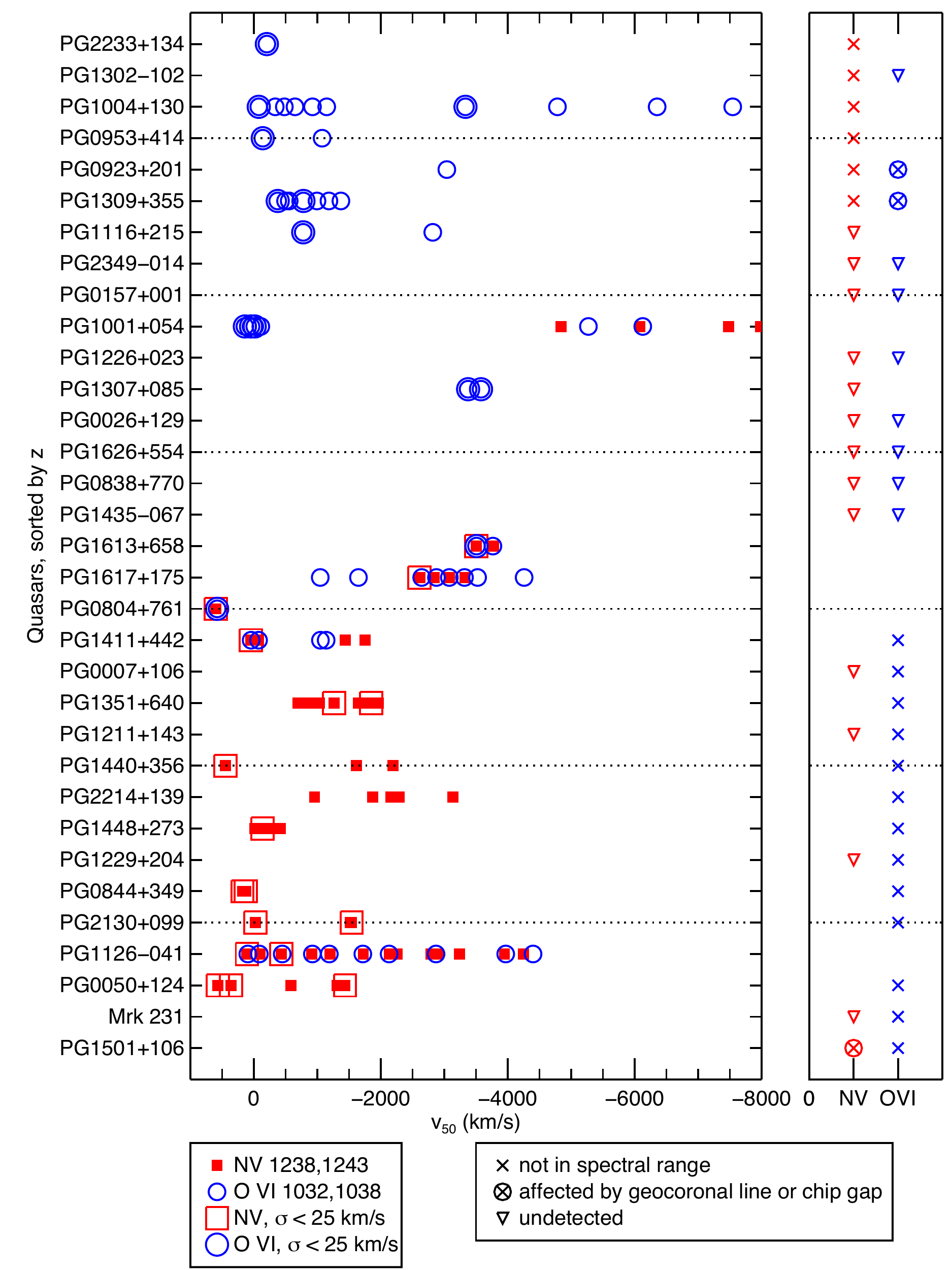}
\caption{Median velocities of the N~V and O~VI absorbing systems
  detected in the QUEST quasars of our sample. Note that the faster
  outflows with more negative velocities lie on the right in this
  figure. The objects are sorted from top to bottom in order of
  decreasing redshift. Red symbols mark N~V $\lambda\lambda$1238, 1243
  and blue symbols mark O~VI $\lambda\lambda$1032, 1038. Open symbols
  indicate systems with velocity dispersion $\sigma <$ 25 \kms. The
  two columns on the right indicate whether N~V (red) or O~VI (blue)
  is within the spectral range of the data (``x'' indicates that it is
  not), affected by geocoronal line or chip gap (encircled ``x''), or
  simply undetected (downward-pointing triangle). A lack of symbol
  marks a detection.
}
\label{fig:quasar_zsort_vs_v50}
\end{figure}

\begin{figure}[!htb]   
\epsscale{1.1}
\plotone{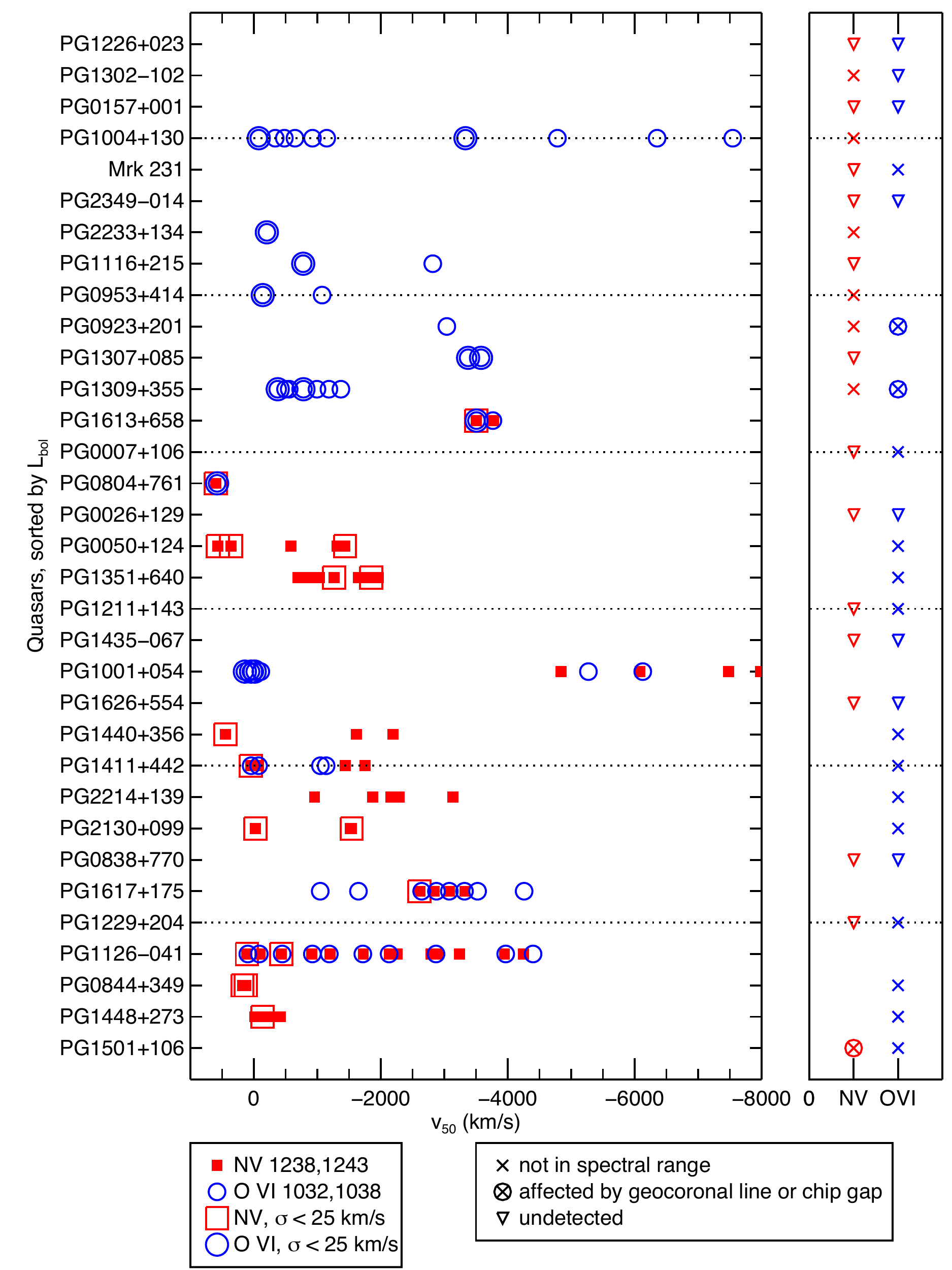}
\caption{Same as Fig.\ \ref{fig:quasar_zsort_vs_v50}, but the objects
  are sorted from top to bottom in order of decreasing bolometric
  luminosity. 
}
\label{fig:quasar_lbolsort_vs_v50}
\end{figure}

Figures \ref{fig:quasar_zsort_vs_v50} and
\ref{fig:quasar_lbolsort_vs_v50} show the median velocities in the
quasar rest frame of all of the detected N~V and O~VI absorption-line
systems, sorted from top to bottom by decreasing redshift and
bolometric luminosity, respectively. The first of these figures clearly
illustrates the fact mentioned in Section \ref{sec:data} that our
ability to detect the N~V and O~VI features is limited by the spectral
coverage of the COS data to $z \la 0.18$ and $z \ga 0.11$,
respectively (systems outside of the spectral range are indicated by
an ``x'' in this figure and Fig.\ \ref{fig:quasar_lbolsort_vs_v50}).

A cursory examination of Figures \ref{fig:quasar_zsort_vs_v50} and
\ref{fig:quasar_lbolsort_vs_v50} shows that blueshifted N~V or O~VI
absorption systems suggestive of outflows (with equivalent widths
above our 3-$\sigma$ detection limit of $\sim$ 20 m\AA) are detected
in about 60\% of the quasars in our sample, and there is no obvious
trend in the rate of incidence with redshift or bolometric luminosity.

The results of a more quantitative analysis based on $\beta$
distributions (Cameron 2011) are listed in Table \ref{tab:incidence}.
The overall rate of incidence of N~V or O~VI absorbers is 61\% with a
1-$\sigma$ range of (52\% $-$ 68\%), once taking into account the
spectral coverage of the data. This rate is virtually the same for N~V
and O~VI. Among quasars with log~$L_{\rm BOL}/L_\odot$ $>$ 12.0,
this rate is 61\% with a 1-$\sigma$ range of (49\% $-$ 71\%), while it
is 60\% (47\% $-$ 71\%) among the systems of lower luminosities.
These rates are thus not significantly different from each other, and
are similar to the rate of incidence of O~VI outflows in local Seyfert
1 galaxies \citep{kriss2004a,kriss2004b} as well as C~IV
\citep{crenshaw1999} or X-ray \citep{reynolds1997,george1998}
absorption.

\begin{deluxetable}{lccc}[!htb]
\tabletypesize{\scriptsize}
\tablecolumns{4}
\tablecaption{Rate of Incidence of Outflows\label{tab:incidence}}
\tablehead{
\colhead{Line} & \colhead{Detection} & \colhead{Total} & \colhead{Fraction (1-$\sigma$ range)} \\
\colhead{(1)} & \colhead{(2)} & \colhead{(3)}& \colhead{(4)}}
\startdata
\cutinhead{All Quasars}
  N V  & 13  & 27  & 0.48 (0.39 $-$ 0.58)\\
 O VI  & 12  & 20  & 0.60 (0.49 $-$ 0.70)\\
 Both  &  5  & 14  & 0.36 (0.25 $-$ 0.50)\\
  Any  & 20  & 33  & 0.61 (0.52 $-$ 0.68)\\
\cutinhead{log $L_{\rm BOL}/L_\odot$ $\geq$ 12.0}
  N V  &  4  & 12  & 0.33 (0.23 $-$ 0.48)\\
 O VI  &  9  & 14  & 0.64 (0.50 $-$ 0.75)\\
 Both  &  2  &  8  & 0.25 (0.16 $-$ 0.44)\\
  Any  & 11  & 18  & 0.61 (0.49 $-$ 0.71)\\
\cutinhead{log $L_{\rm BOL}/L_\odot$ $<$ 12.0}
  N V  &  9  & 15  & 0.60 (0.47 $-$ 0.71)\\
 O VI  &  3  &  6  & 0.50 (0.32 $-$ 0.68)\\
 Both  &  3  &  6  & 0.50 (0.32 $-$ 0.68)\\
  Any  &  9  & 15  & 0.60 (0.47 $-$ 0.71)\\
\cutinhead{$N_{\rm H}$ $>$ $10^{22}$ cm$^{-2}$}
  N V  & 10  & 16  & 0.62 (0.50 $-$ 0.73)\\
 O VI  &  8  &  8  & 1.00 (0.81 $-$ 0.98)\\
 Both  &  3  &  5  & 0.60 (0.38 $-$ 0.76)\\
  Any  & 15  & 19  & 0.79 (0.67 $-$ 0.85)\\
\cutinhead{$N_{\rm H}$ $\leq$ $10^{22}$ cm$^{-2}$}
  N V  &  2  & 10  & 0.20 (0.13 $-$ 0.37)\\
 O VI  &  2  & 10  & 0.20 (0.13 $-$ 0.37)\\
 Both  &  1  &  8  & 0.12 (0.08 $-$ 0.32)\\
  Any  &  3  & 12  & 0.25 (0.17 $-$ 0.41)\\
\cutinhead{$\alpha_{ox}$ $\geq$ $-$1.6}
  N V  &  7  & 17  & 0.41 (0.31 $-$ 0.53)\\
 O VI  &  6  & 12  & 0.50 (0.37 $-$ 0.63)\\
 Both  &  2  &  9  & 0.22 (0.14 $-$ 0.41)\\
  Any  & 11  & 20  & 0.55 (0.44 $-$ 0.65)\\
\cutinhead{$\alpha_{ox}$ $<$ $-$1.6}
  N V  &  6  &  9  & 0.67 (0.49 $-$ 0.78)\\
 O VI  &  6  &  7  & 0.86 (0.64 $-$ 0.91)\\
 Both  &  3  &  4  & 0.75 (0.48 $-$ 0.85)\\
  Any  &  9  & 12  & 0.75 (0.59 $-$ 0.83)\\
\enddata

\tablecomments{Column (1): Feature(s) used in the statistical
  analysis. ``Both'' means both N~V and O~VI doublets and ``Any''
  means either N~V or O~VI doublet or both; Column (2): Number of
  objects with detected outflows; Column (3): Number of objects in
  total with the appropriate redshift; Column (4): Fraction of objects
  with detected outflows. The two numbers in parentheses indicate the
  1-$\sigma$ range (68\% probability) of the fraction of objects with
  detected outflows, computed from the $\beta$ distribution
  \citep{cameron2011}.}
\end{deluxetable}

\begin{deluxetable}{ccccc}
\footnotesize
\tabletypesize{\tiny}
\tablecolumns{5}
\tablecaption{Linear Regression Results\label{tab:regressions}}
\tablehead{\colhead{$y$} & \colhead{$x$} & \colhead{$N$} &
  \colhead{$p$} & \colhead{$r$} }
\colnumbers
\startdata
        $W_{\rm eq}$  &\underline{                                 log($L_{\rm BOL}/L_\sun$)}  & 32  &  0.046  & -0.34$_{-0.17}^{+0.19}$ \\
        $W_{\rm eq}$  &                              log[$\lambda L_{1125}/{\rm erg~s}^{-1}$]  & 32  &  0.140  & -0.22$_{-0.18}^{+0.21}$ \\
        $W_{\rm eq}$  &                                                          AGN fraction  & 32  &  0.094  &  0.78$_{-0.61}^{+0.19}$ \\
        $W_{\rm eq}$  &                                             log($L_{\rm AGN}/L_\sun$)  & 32  &  0.054  & -0.33$_{-0.16}^{+0.19}$ \\
        $W_{\rm eq}$  &                                              log($M_{\rm BH}/M_\sun$)  & 32  &  0.289  & -0.33$_{-0.43}^{+0.65}$ \\
        $W_{\rm eq}$  &                                                       Eddington Ratio  & 32  &  0.168  & -0.42$_{-0.38}^{+0.44}$ \\
        $W_{\rm eq}$  &\underline{                                         $\alpha_{\rm OX}$}  & 31  &  0.002  & -0.62$_{-0.13}^{+0.17}$ \\
        $W_{\rm eq}$  &                                         log($L_{\rm IR}/L_{\rm BOL}$)  & 32  &  0.067  &  0.37$_{-0.25}^{+0.20}$ \\
        $W_{\rm eq}$  &                                        log($L_{\rm FIR}/L_{\rm BOL}$)  & 30  &  0.094  & -0.29$_{-0.19}^{+0.22}$ \\
        $W_{\rm eq}$  &\underline{                           log[$N({\rm H})/{\rm cm}^{-2}$]}  & 30  &$<$0.001 &  0.19$_{-0.03}^{+0.03}$ \\
        $W_{\rm eq}$  &                                                              $\Gamma$  & 30  &  0.143  &  0.25$_{-0.23}^{+0.20}$ \\
        $W_{\rm eq}$  &\underline{  log[$F(0.5-2~{\rm keV})/{\rm erg~s}^{-1}~{\rm cm}^{-2}$]}  & 26  &  0.005  & -0.54$_{-0.14}^{+0.18}$ \\
        $W_{\rm eq}$  &\underline{   log[$F(2-10~{\rm keV})/{\rm erg~s}^{-1}~{\rm cm}^{-2}$]}  & 29  &  0.002  & -0.55$_{-0.14}^{+0.18}$ \\
        $W_{\rm eq}$  &                            log[$L(0.5-2~{\rm keV})/{\rm erg~s}^{-1}$]  & 26  &  0.072  & -0.33$_{-0.19}^{+0.22}$ \\
        $W_{\rm eq}$  &\underline{                 log[$L(2-10~{\rm keV})/{\rm erg~s}^{-1}$]}  & 30  &  0.009  & -0.51$_{-0.15}^{+0.20}$ \\
        $W_{\rm eq}$  &                           log[$L(0.5-10~{\rm keV})/{\rm erg~s}^{-1}$]  & 26  &  0.051  & -0.38$_{-0.18}^{+0.23}$ \\
        $W_{\rm eq}$  &                         log[$L(0.5-2~{\rm keV})/L(0.5-10~{\rm keV})$]  & 26  &  0.192  &  0.21$_{-0.24}^{+0.22}$ \\
        $W_{\rm eq}$  &                                log[$L(0.5-10~{\rm keV})/L_{\rm BOL}$]  & 26  &  0.211  & -0.20$_{-0.21}^{+0.25}$ \\
     $v_{\rm wtavg}$  &                                             log($L_{\rm BOL}/L_\sun$)  & 20  &  0.138  & -0.27$_{-0.22}^{+0.25}$ \\
     $v_{\rm wtavg}$  &                              log[$\lambda L_{1125}/{\rm erg~s}^{-1}$]  & 20  &  0.283  & -0.15$_{-0.22}^{+0.25}$ \\
     $v_{\rm wtavg}$  &                                                     $\alpha_{\rm OX}$  & 20  &  0.103  &  0.31$_{-0.24}^{+0.21}$ \\
     $v_{\rm wtavg}$  &                                         log($L_{\rm IR}/L_{\rm BOL}$)  & 20  &  0.115  &  0.37$_{-0.30}^{+0.25}$ \\
     $v_{\rm wtavg}$  &                                        log($L_{\rm FIR}/L_{\rm BOL}$)  & 20  &  0.496  &  0.00$_{-0.26}^{+0.25}$ \\
     $v_{\rm wtavg}$  &                                                          AGN fraction  & 20  &  0.340  &  0.35$_{-0.80}^{+0.49}$ \\
     $v_{\rm wtavg}$  &                                             log($L_{\rm AGN}/L_\sun$)  & 20  &  0.150  & -0.24$_{-0.21}^{+0.23}$ \\
     $v_{\rm wtavg}$  &                                              log($M_{\rm BH}/M_\sun$)  & 20  &  0.286  & -0.31$_{-0.42}^{+0.56}$ \\
     $v_{\rm wtavg}$  &                                                       Eddington Ratio  & 20  &  0.414  &  0.13$_{-0.59}^{+0.53}$ \\
      $v_{\rm wtavg}$ &\underline{                           log[$N({\rm H})/{\rm cm}^{-2}$]}  & 18  &   0.034 & -0.15$_{-0.08}^{+0.08}$ \\
     $v_{\rm wtavg}$  &\underline{                                                  $\Gamma$}  & 18  &  0.011  &  0.67$_{-0.22}^{+0.14}$ \\
     $v_{\rm wtavg}$  &\underline{  log[$F(0.5-2~{\rm keV})/{\rm erg~s}^{-1}~{\rm cm}^{-2}$]}  & 17  &  0.008  &  0.61$_{-0.21}^{+0.15}$ \\
     $v_{\rm wtavg}$  &\underline{   log[$F(2-10~{\rm keV})/{\rm erg~s}^{-1}~{\rm cm}^{-2}$]}  & 17  &  0.016  &  0.57$_{-0.23}^{+0.16}$ \\
     $v_{\rm wtavg}$  &                            log[$L(0.5-2~{\rm keV})/{\rm erg~s}^{-1}$]  & 17  &  0.087  &  0.36$_{-0.26}^{+0.22}$ \\
     $v_{\rm wtavg}$  &                             log[$L(2-10~{\rm keV})/{\rm erg~s}^{-1}$]  & 18  &  0.264  &  0.17$_{-0.28}^{+0.25}$ \\
     $v_{\rm wtavg}$  &                           log[$L(0.5-10~{\rm keV})/{\rm erg~s}^{-1}$]  & 17  &  0.211  &  0.23$_{-0.28}^{+0.25}$ \\
     $v_{\rm wtavg}$  &\underline{             log[$L(0.5-2~{\rm keV})/L(0.5-10~{\rm keV})$]}  & 17  &  0.006  &  0.69$_{-0.20}^{+0.13}$ \\
     $v_{\rm wtavg}$  &\underline{                    log[$L(0.5-10~{\rm keV})/L_{\rm BOL}$]}  & 17  &  0.009  &  0.64$_{-0.20}^{+0.14}$ \\
  $\sigma_{\rm rms}$  &                                             log($L_{\rm BOL}/L_\sun$)  & 20  &  0.238  &  0.18$_{-0.25}^{+0.23}$ \\
  $\sigma_{\rm rms}$  &                              log[$\lambda L_{1125}/{\rm erg~s}^{-1}$]  & 20  &  0.425  & -0.04$_{-0.24}^{+0.24}$ \\
  $\sigma_{\rm rms}$  &\underline{                                         $\alpha_{\rm OX}$}  & 20  &  0.004  & -0.55$_{-0.15}^{+0.20}$ \\
  $\sigma_{\rm rms}$  &                                         log($L_{\rm IR}/L_{\rm BOL}$)  & 20  &  0.305  & -0.16$_{-0.30}^{+0.32}$ \\
  $\sigma_{\rm rms}$  &                                        log($L_{\rm FIR}/L_{\rm BOL}$)  & 20  &  0.433  & -0.04$_{-0.25}^{+0.25}$ \\
  $\sigma_{\rm rms}$  &                                                          AGN fraction  & 20  &  0.389  & -0.13$_{-0.58}^{+0.55}$ \\
  $\sigma_{\rm rms}$  &                                             log($L_{\rm AGN}/L_\sun$)  & 20  &  0.247  &  0.17$_{-0.25}^{+0.23}$ \\
  $\sigma_{\rm rms}$  &                                              log($M_{\rm BH}/M_\sun$)  & 20  &  0.301  &  0.29$_{-0.59}^{+0.43}$ \\
  $\sigma_{\rm rms}$  &                                                       Eddington Ratio  & 20  &  0.374  & -0.18$_{-0.49}^{+0.58}$ \\
  $\sigma_{\rm rms}$  &                                       log[$N({\rm H})/{\rm cm}^{-2}$]  & 18  &   0.086 &  0.10$_{-0.07}^{+0.08}$ \\
  $\sigma_{\rm rms}$  &\underline{                                                  $\Gamma$}  & 18  &  0.048  & -0.46$_{-0.20}^{+0.26}$ \\
  $\sigma_{\rm rms}$  &\underline{  log[$F(0.5-2~{\rm keV})/{\rm erg~s}^{-1}~{\rm cm}^{-2}$]}  & 17  &  0.016  & -0.56$_{-0.16}^{+0.22}$ \\
  $\sigma_{\rm rms}$  &\underline{   log[$F(2-10~{\rm keV})/{\rm erg~s}^{-1}~{\rm cm}^{-2}$]}  & 17  &  0.015  & -0.55$_{-0.16}^{+0.22}$ \\
  $\sigma_{\rm rms}$  &                            log[$L(0.5-2~{\rm keV})/{\rm erg~s}^{-1}$]  & 17  &  0.066  & -0.40$_{-0.21}^{+0.26}$ \\
  $\sigma_{\rm rms}$  &                             log[$L(2-10~{\rm keV})/{\rm erg~s}^{-1}$]  & 18  &  0.137  & -0.28$_{-0.23}^{+0.26}$ \\
  $\sigma_{\rm rms}$  &                           log[$L(0.5-10~{\rm keV})/{\rm erg~s}^{-1}$]  & 17  &  0.120  & -0.32$_{-0.23}^{+0.27}$ \\
  $\sigma_{\rm rms}$  &\underline{             log[$L(0.5-2~{\rm keV})/L(0.5-10~{\rm keV})$]}  & 17  &  0.022  & -0.54$_{-0.18}^{+0.24}$ \\
  $\sigma_{\rm rms}$  &\underline{                    log[$L(0.5-10~{\rm keV})/L_{\rm BOL}$]}  & 17  &  0.008  & -0.61$_{-0.15}^{+0.21}$ \\
\enddata

\tablecomments{Column (1): Dependent variable (absorption line
  property). Column (2): Independent variable (quasar/host
  property). Column (3): Number of points. Column (4): $p$-value of
  null hypothesis (no correlation). Column (5): Correlation
  coefficient and 1$\sigma$ errors. Underlined entries under col.\ (2)
  indicate significant correlations with $p$-values below 0.05.}
\end{deluxetable}

We have also searched for trends between the rate of incidence of
outflows and several other quantities. The detection rate of
  outflows (79\%) among quasars that have strongly absorbed X-ray
  continua ($N_{\rm H}$ $>$ $10^{22}$ cm$^{-2}$) is significantly
  higher than those that do not (25\%) (Table
  \ref{tab:incidence}). These rates differ at the 2-$\sigma$ level
  (95.4\%), where the ranges of the incidence rate are 56 $-$ 92\% for
  quasars with absorbed X-ray continua and 9 $-$ 54\% for the
  others. Using the \texttt{scipy.stats} implementation of the Fisher
  exact test, the null hypothesis that galaxies with strongly- and
  weakly-absorbed X-ray continua UV absorbers are equally likely to
  show N~V or O~VI absorbers is rejected at the 99.2\%\ level. The
  rate of incidence of outflows among quasars with a steep X-ray to
  optical spectral index ($\alpha_{\rm OX} < -1.6$; 75\%) is also
  higher than those with a shallow index (55\%), although the Fisher
  exact test shows that this difference is not significant
  ($p = 0.45$). A similar dependence on the X-ray properties of the
quasars has been reported in several studies of higher luminosity
quasars and lower luminosity Seyfert 1 galaxies using C~IV
$\lambda\lambda$1548, 1550 as a tracer of warm ionized outflows. We
return to this result in Sections \ref{sec:ew} and
\ref{sec:kinematics}, and \ref{sec:discussion}.

\subsection{Optical Depths and Covering Factors}
\label{sec:tau_cf}

\begin{figure}[!htb]   
\epsscale{1.15}
\plottwo{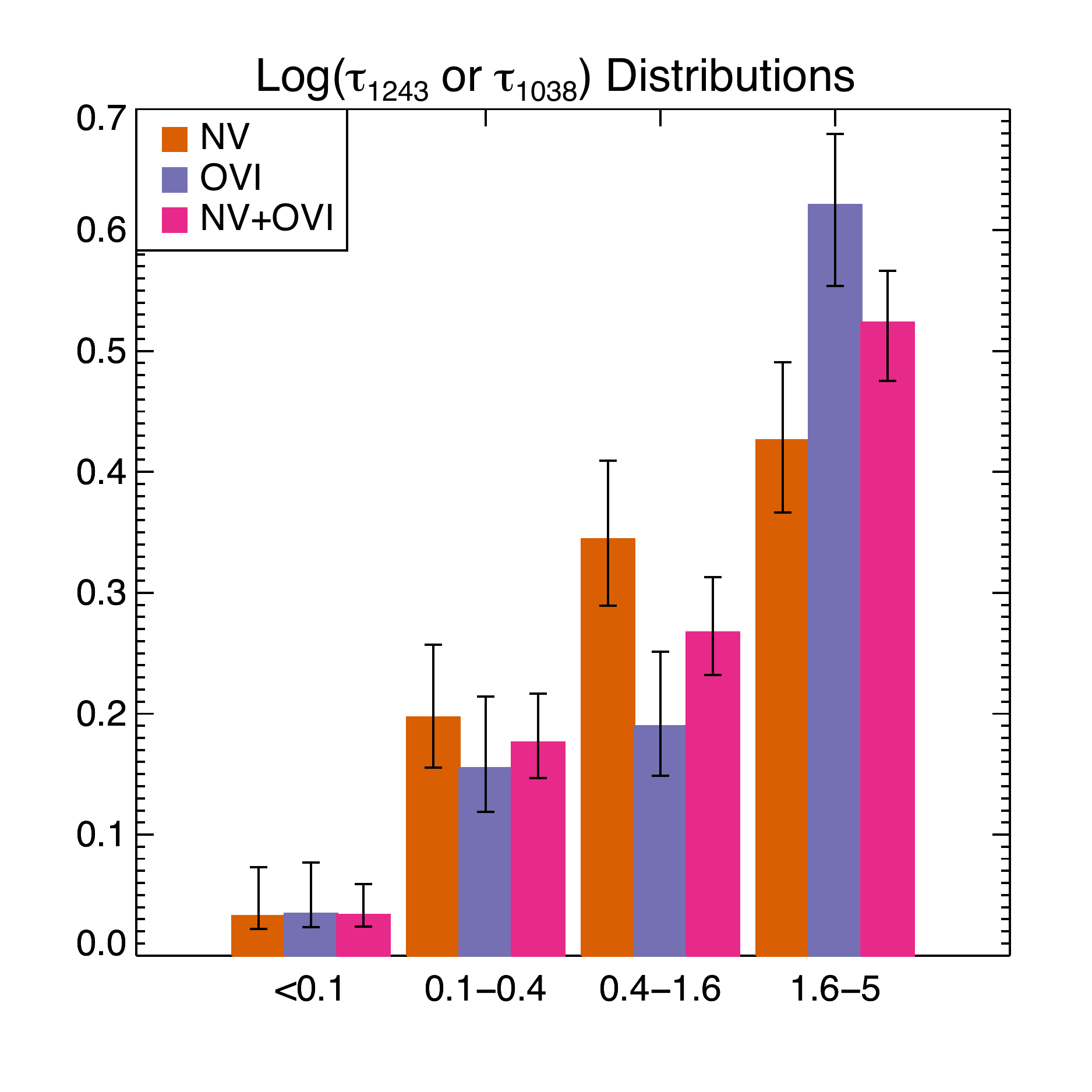}{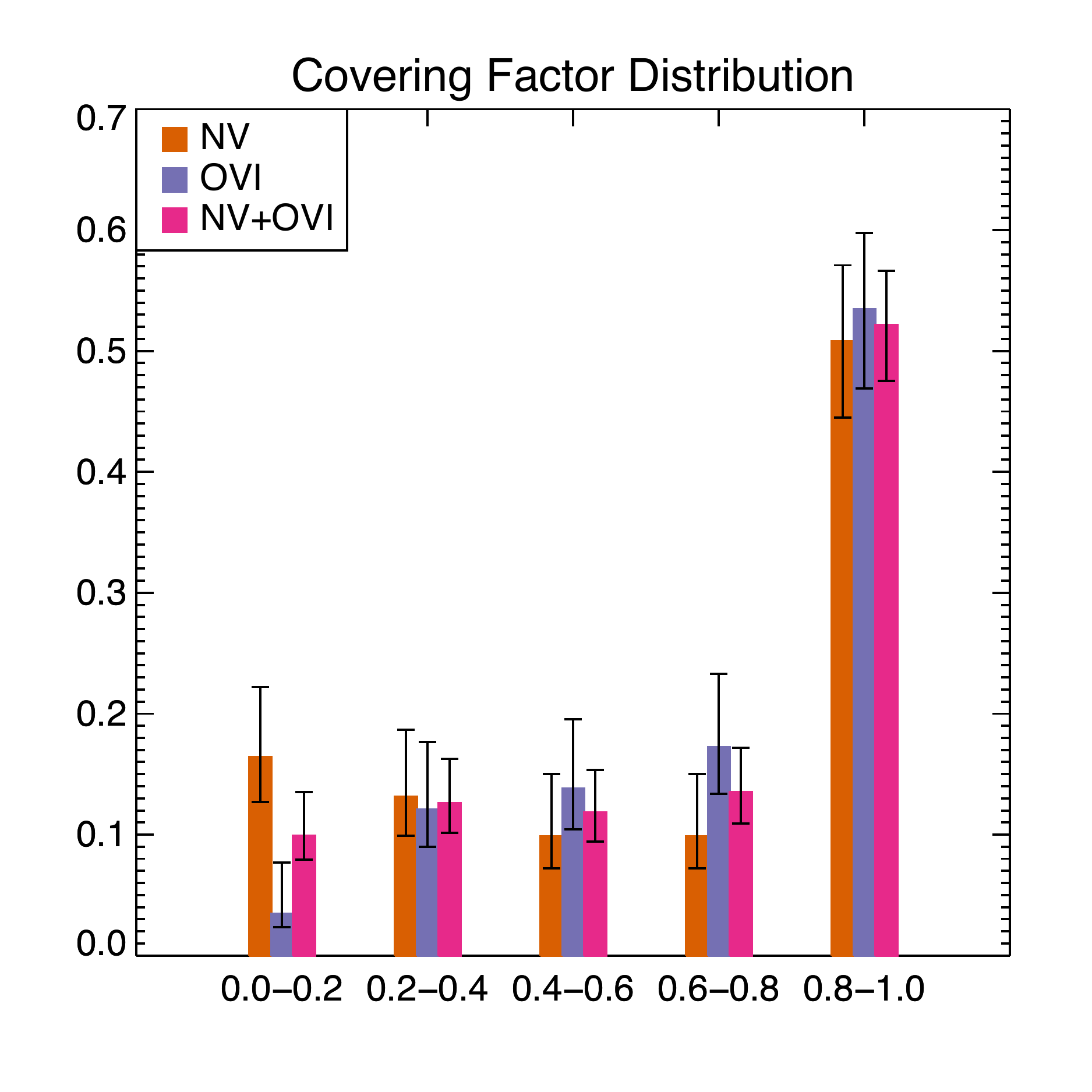}
\caption{Distributions of the optical depths (left) and covering
  factors (right) of the individual components used to fit the profiles
  of the N~V (orange), O~VI (purple), or joint N~V + O~VI (pink)
  absorption features.}
\label{fig:hist_tau_cf}
\end{figure}

The distributions of the N~V $\lambda$1243 and O~VI $\lambda$1038
optical depths and covering factors derived
from the individual components in the multi-component fits are
presented as histograms in Figure \ref{fig:hist_tau_cf}. 

Again, we repeat that the optical depths and covering factors
presented here are only approximations. Nevertheless, it is clear from
the left panel in Figure \ref{fig:hist_tau_cf} that a significant
fraction of the absorbing systems are affected by saturation effects
($\tau_{1243}$ or $\tau_{1038}$ $>$ 1), therefore making the
equivalent widths of the N~V and O~VI features unreliable indicators
of the total column densities of highly ionized gas in many of these
cases.

The right panel of Figure \ref{fig:hist_tau_cf} shows that the mode of
the distribution of covering factors is consistent with unity, but
$\sim$50\% of the N~V and O~VI absorbers only partially cover the FUV
quasar continuum emission (+ possibly the broad emission line region
$-$ BELR; Fig.\ \ref{fig:hist_tau_cf}), consistent with small clouds
located relatively near the quasars. As described at the end of
  Sec.\ \ref{sec:fitting}, emission infill of the absorption profiles
associated with the broad wings of the COS LSF is negligible and thus
does not affect this conclusion. We return to this result in Section
\ref{sec:discussion_origins}.

\subsection{Outflow Equivalent Widths}
\label{sec:ew}

The velocity-integrated equivalent widths ($W_{\rm eq}$;
eq.\ \ref{eq:weq}) of the outflow systems in each quasar are listed in
Table \ref{tab:fits}. They span a broad range from $\sim$ 25 \AA\ down
to 20 m\AA, near our 3-$\sigma$ detection limit.

\begin{figure}[!htb]   
\epsscale{1.09}
\plotone{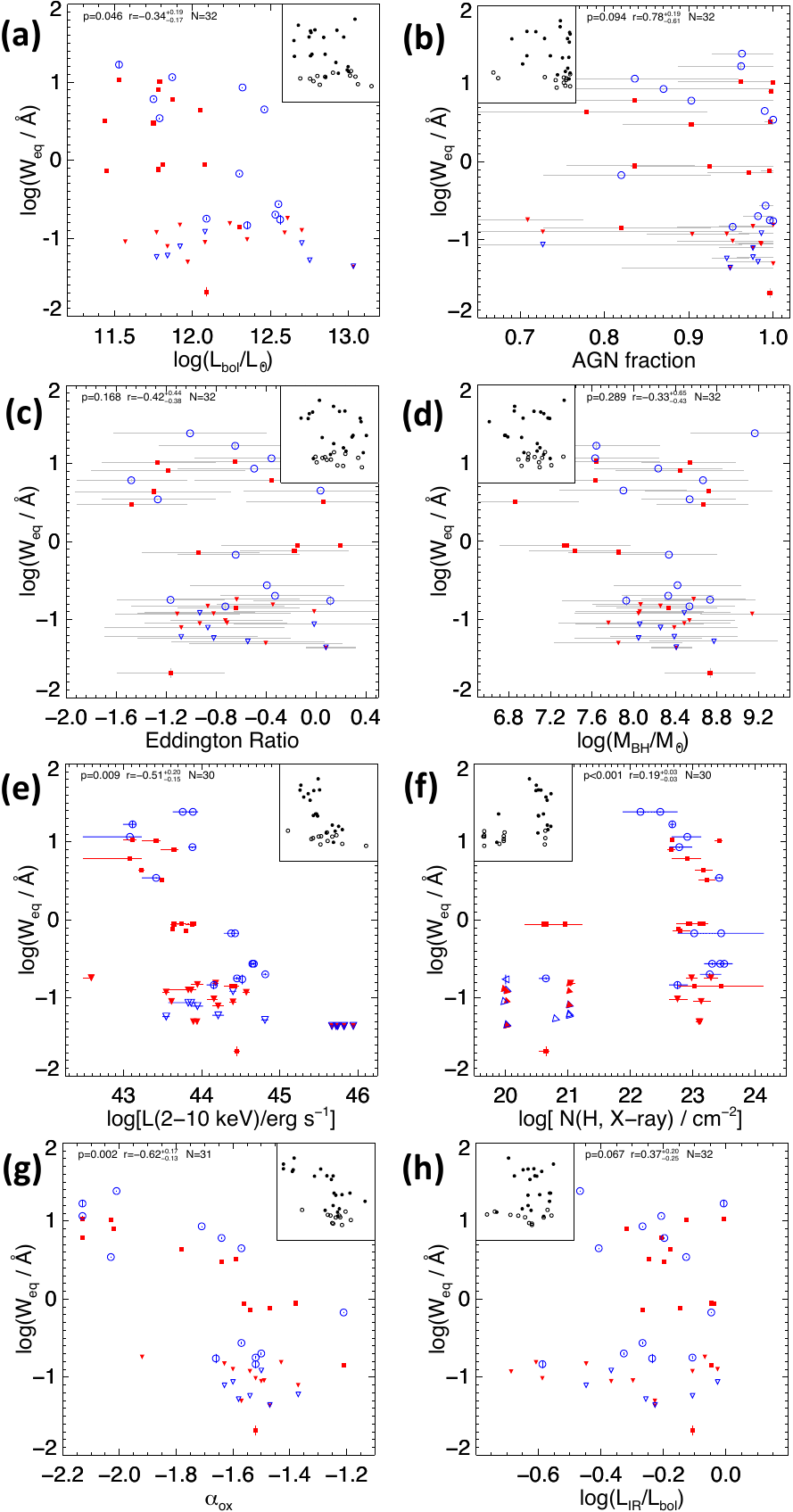}
\caption{ The velocity-integrated equivalent widths, $W_{\rm eq}$, of
  the outflow systems in the QUEST quasars are plotted as a function
  of the (a) bolometric luminosities, (b) AGN bolometric fractions,
  (c) Eddington ratios, (d) black hole masses, (e) hard X-ray (2 $-$
  10 keV) luminosities, (f) X-ray absorbing column densities, (g)
  X-ray to optical spectral indices, and (h) ratios of the infrared
  luminosities to the bolometric luminosities. Red squares mark N~V
  $\lambda\lambda$1238, 1243 and blue circles mark O~VI
  $\lambda\lambda$1032, 1038. Triangles indicate upper limits (in
    one or both quantities). The regression results ($p$-values,
    correlation coefficients $r$ with 1$\sigma$ errors, and number of
    points $N$; Section \,\ref{sec:analysis}) are shown at the top of
    each panel. The actual points used in the regression, in which N~V
    and O~VI quantities and/or X-ray measurements are averaged for a
    given quasar, are shown in each inset panel.  The solid points are
    detections, while the open points are censored values in one or
    both quantities plotted.}
\label{fig:weq_vs_quantities}
\end{figure}

The equivalent widths of the outflows were compared against the
properties of the quasars and host galaxies listed in Table
\ref{tab:sample}. Some of the results are shown in Figure
\ref{fig:weq_vs_quantities}.  By and large, we do not find any
significant trends between $W_{\rm eq}$ and any of the quasar and host
properties, except with some of the quantities that are derived from
the X-ray data (Table \ref{tab:regressions}). Taken at face
value, this result is surprising since, for instance, it means that
the equivalent width of the outflow is largely agnostic of the
properties of the central engine over a range of $\sim$ 1.5 dex in
power (FUV, bolometric, or quasar-only luminosity), $\sim$2.0 dex in
Eddington ratio, and $\sim$2.5 dex in black hole mass. The lack of
correlations with the properties of the hosts is less surprising since
the quasar sample spans a relatively narrow range of values in these
quantities so the lack of correlation with these quantities may be
attributed to the lower dynamical range.

Examples of trends between $W_{\rm eq}$ and the X-ray properties of
the quasars are shown in panels (e), (f), and (g) of Figure
\ref{fig:weq_vs_quantities}. In panel (e), the equivalent width of the
outflow decreases with increasing HX luminosity.  Panel (f) in this
figure illustrates the dependence of the rate of incidence of these
outflows on the X-ray column densities already pointed out in Section
\ref{sec:incidence}. The stronger highly ionized outflows with $W_{\rm
  eq} \ga$ 1 \AA\ are only present in quasars with X-ray column
densities above $\sim$ 10$^{22}$ cm$^{-2}$. While it is a required
condition for a strong outflow, it is not a sufficient condition since
most quasars with these X-ray absorbing column densities show either
weak outflows in the FUV ($W_{\rm eq} <$ 0.3 \AA) or none at all.
Panel (g) also shows a distinct trend for strong outflows with $W_{\rm
  eq} \ga$ 1 \AA\ among objects with $\alpha_{\rm OX} \la -1.7$. A
similar trend is observed when normalizing the X-ray luminosities to
the bolometric luminosities (not shown), but disappears when
considering only the X-ray slope (e.g.\ the SX/HX ratio or index of
the best-fit absorbed power-law distribution to the X-rays; not
shown). Similar results have been found when considering C~IV outflows
\citep[e.g.][]{brandt2000,laor2002,baskin2005,gibson2009b,gibson2009a}. We
return to this issue in Section \ref{sec:discussion} below.

\subsection{Outflow Kinematics}
\label{sec:kinematics}

\begin{figure}[!htb]   
\epsscale{1.15}
\plottwo{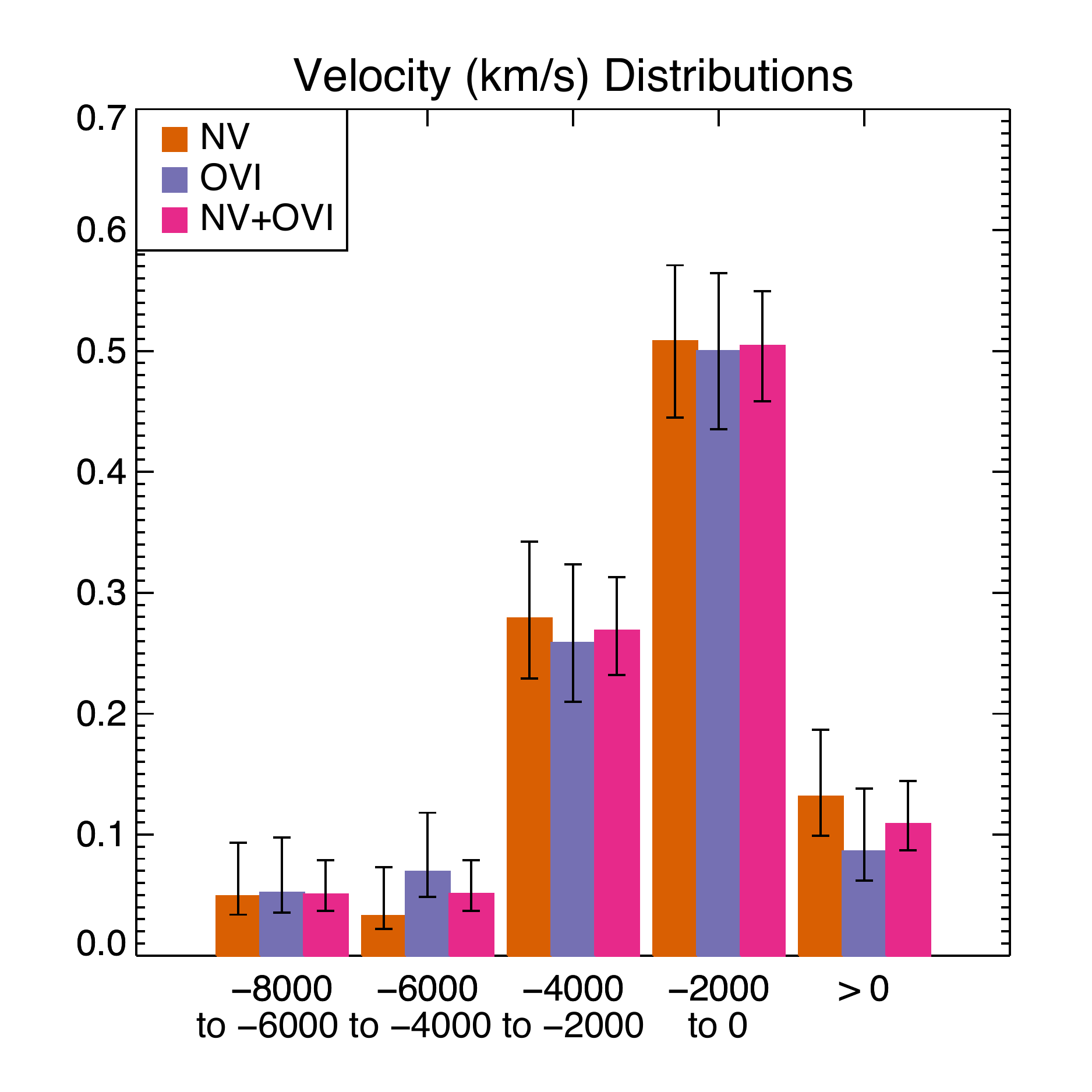}{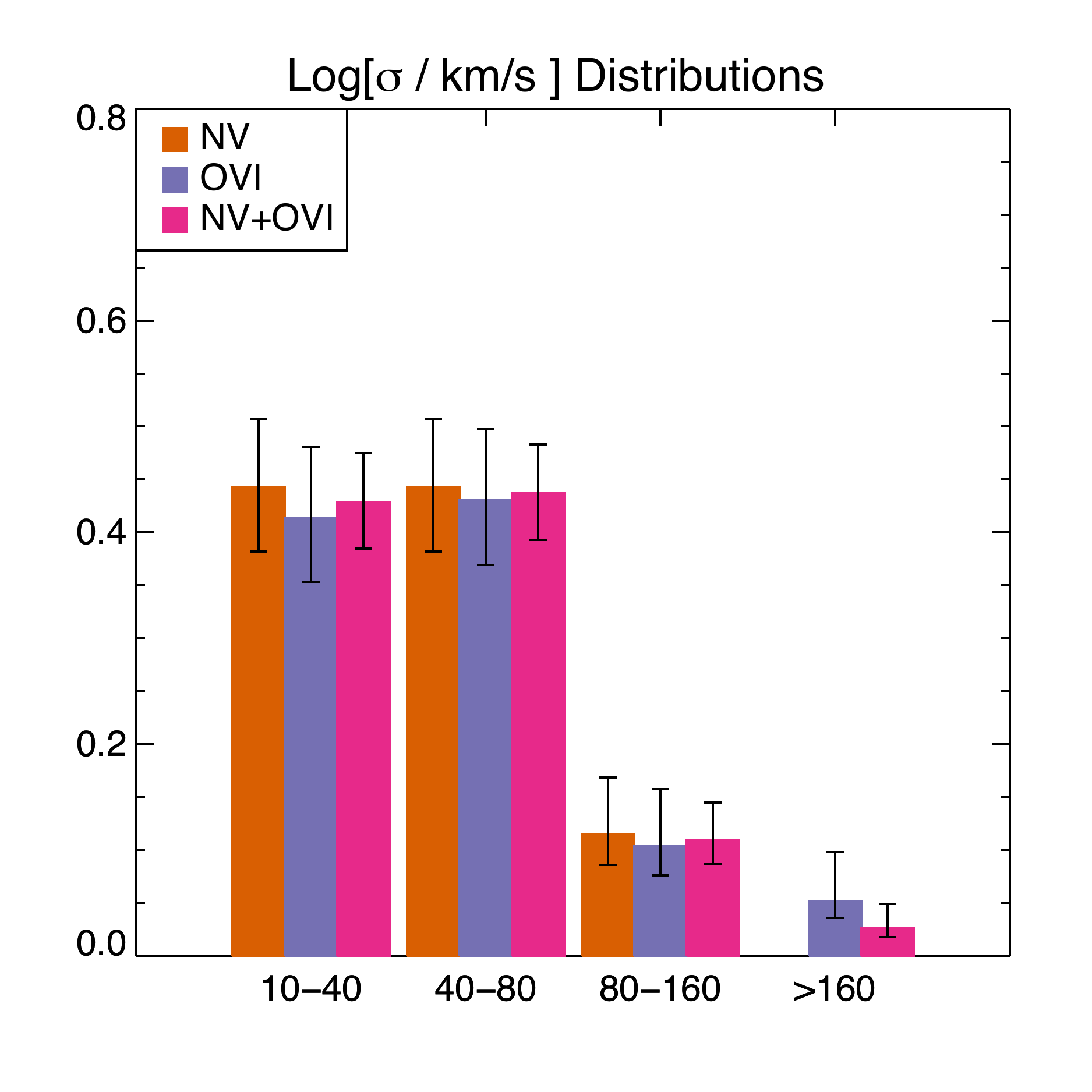}
\caption{Distributions of the median velocities (left) and velocity
  dispersions (right) of the individual components used to fit the
  profiles of the N~V (orange), O~VI (purple), and joint N~V + O~VI
  (pink) absorption features.}
\label{fig:hist_vel_sig}
\end{figure}

Figure \ref{fig:hist_vel_sig} shows the distributions of the velocity
centroids and dispersions ($\sigma$) of the various individual
components that were used to fit the N~V and O~VI absorbers in the
quasar sample. About half of all of the individual components have
blueshifted (outflow) velocities that lie between [$-$2000, 0]
\kms\ and have 1-$\sigma$ widths less than 40 \kms. In a blindly
selected sample of O~VI absorbers, \citet{tripp2008} similarly found
that the majority of associated absorbers are within 2000 \kms\ of the
QSO redshift (see their Figure 15). Likewise, they found that the O~VI
line widths are $<$ 40 \kms. Up to $\sim$10\% of the individual
components in the present survey have redshifted velocities of up to a
few $\times$ 100 \kms; some of them may be attributed to uncertain or
systematically blueshifted systemic velocities derived from the quasar
emission lines (Sec.\ \ref{sec:analysis}) rather than actual
inflows.

\begin{figure}[!htb]   
\epsscale{1.15}
\plotone{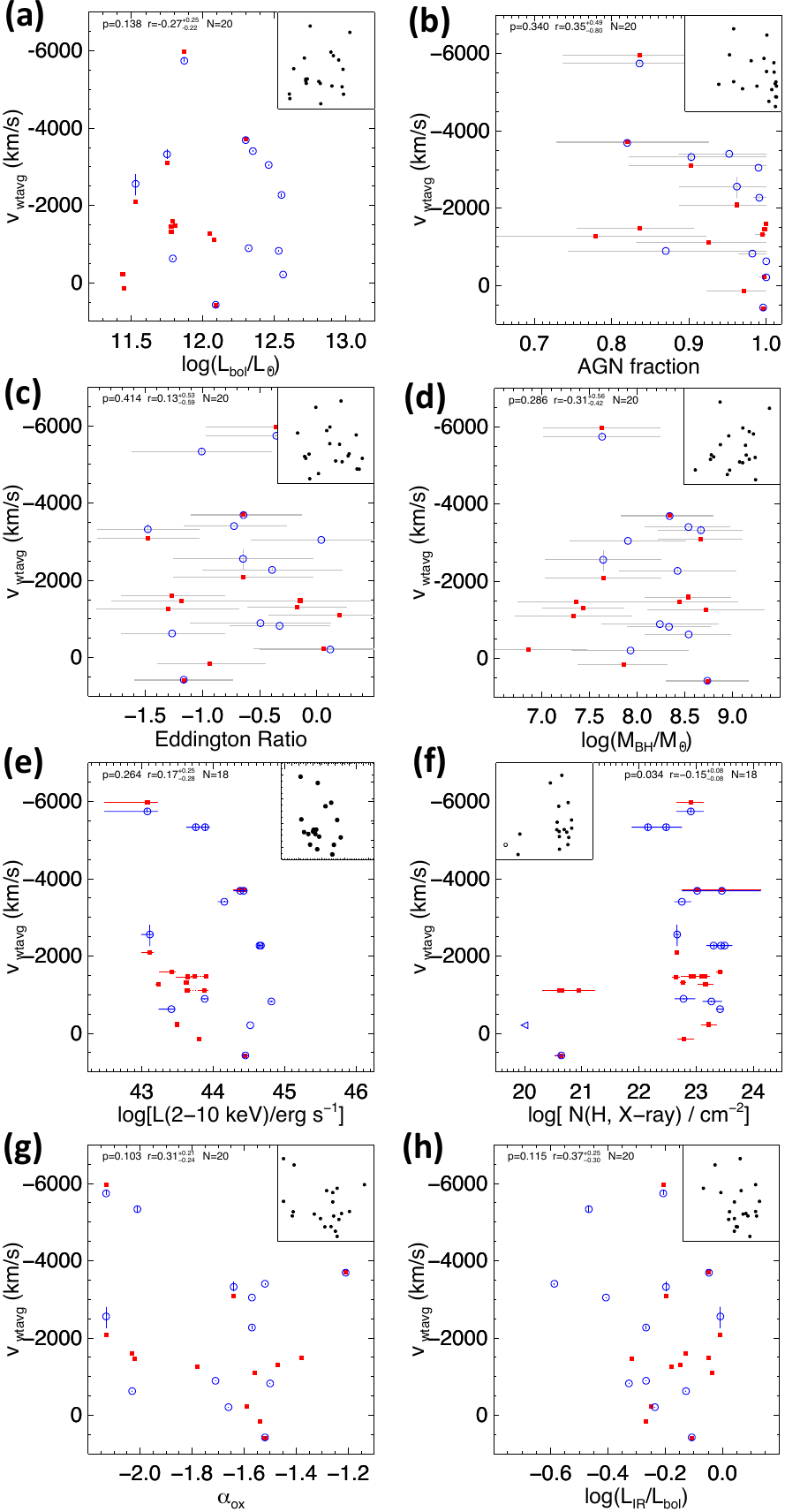}
\caption{Same as Fig.\ \ref{fig:weq_vs_quantities} but the weighted
  average velocities.
}
\label{fig:vwtavg_vs_quantities}
\end{figure}

\begin{figure}[!htb]   
\epsscale{1.15}
\plotone{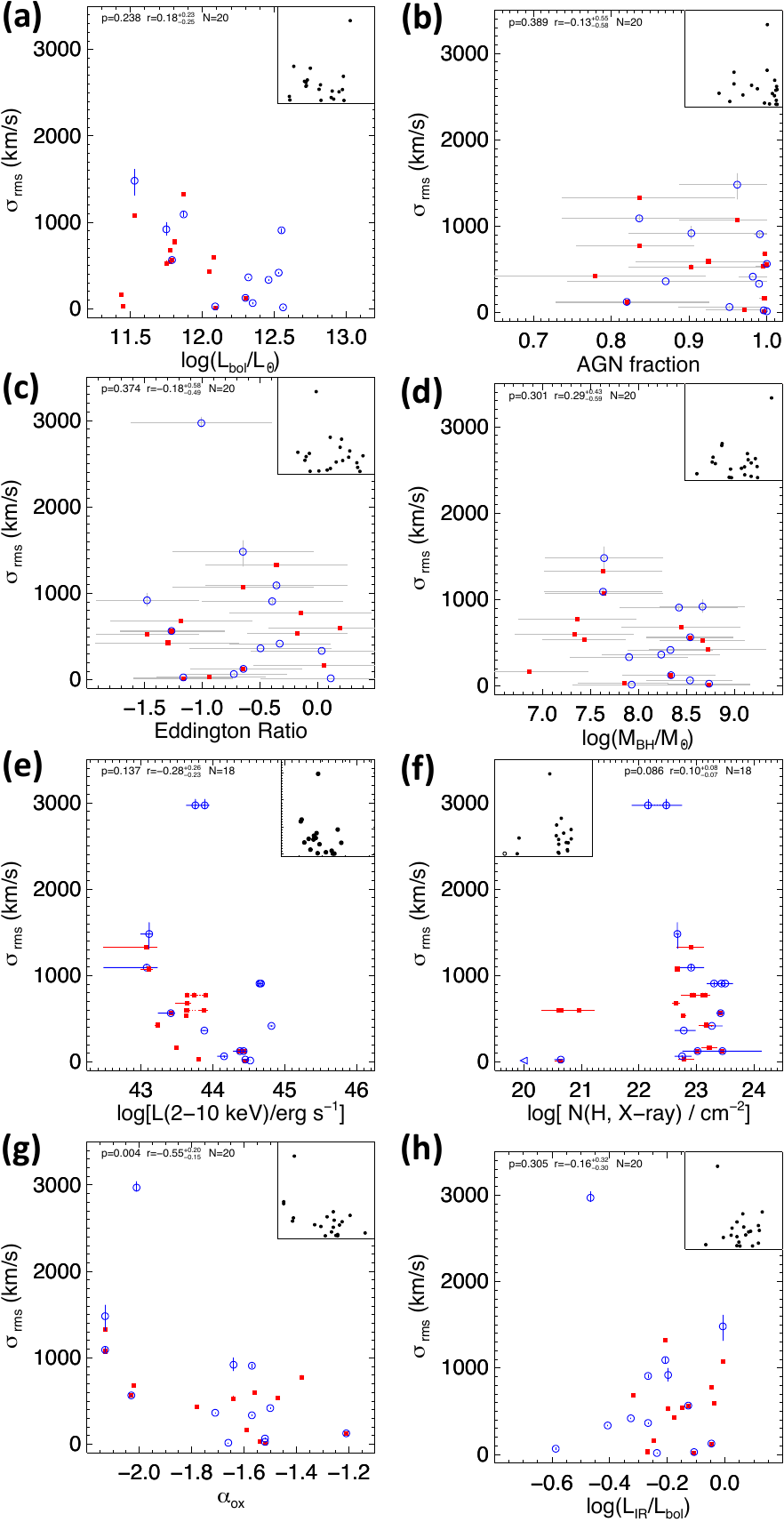}
\caption{Same as Fig.\ \ref{fig:weq_vs_quantities} but for the
  weighted velocity dispersions. }
\label{fig:vwtavg_vs_quantities}
\end{figure}

More physically meaningful kinematic quantities are the weighted
average velocities and velocity dispersions of the outflow systems in
each object (eqs.\ \ref{eq:vwtavg} and \ref{eq:sigmawtavg}). Of the 20
detected absorbers in Table \ref{tab:fits}, 8 (3) have weighted
average outflow velocities (velocity dispersion) in excess of 2000
(1000) \kms.

We find in Table \ref{tab:regressions} that there is no
distinct trend between the weighted outflow velocities and velocity
dispersions and the quasar and host properties, except for the lack of
outflows in X-ray unabsorbed quasars
(Fig.\ \ref{fig:vwtavg_vs_quantities}), pointed out in
Sec.\ \ref{sec:incidence}, and the larger weighted outflow velocity
dispersions among X-ray faint sources with $\alpha_{\rm OX} \la
-2$. The lack of a correlation between outflow velocities and the
quasar luminosities seems at odds with those from most previous C~IV
absorption-line studies
\citep[e.g.][]{perry1978,brandt2000,laor2002,ganguly2007,ganguly2008,gibson2009b,gibson2009a,zhang2014,rankine2020}
and other multi-wavelength analyses \citep[e.g., references in
  Sec.\ \ref{sec:introduction} and][]{veilleux2020}. We examine this
issue in more detail in Section \ref{sec:discussion} below.


\section{Discussion}
\label{sec:discussion}

\vskip 0.1in

\subsection{Origins of the Absorption Features}
\label{sec:discussion_origins}

The blueshifted N~V and O~VI absorption features reported in Section
\ref{sec:results} may have several origins: quasar-driven outflows,
starburst-driven winds, tidal debris from the galaxy mergers, and
intervening CGM.
Here we do not consider contamination of the quasar spectra by young
stars since none of them show the obvious spectral signatures of young
stars (e.g., narrow and shallow N~V or O~VI absorption troughs
accompanied by redshifted emission). This is only an issue among
starburst-ULIRGs \citep[e.g.][]{martin2015}.

Telltale signs that the detected lines are formed in a quasar-driven
outflow include (1) line profiles that are blueshifted, broad, and
smooth compared to the thermal line widths ($\la$ 10 $-$ 20 \kms\ for
highly ionized N$^{4+}$, P$^{4+}$, and O$^{5+}$ ions at $T \simeq
10^{4.5 - 5.5}$ K), (2) line ratios within the multiplets N~V
$\lambda$1238/$\lambda$1243, O~VI $\lambda$1032/$\lambda$1038, and P~V
$\lambda$1117/$\lambda$1128 that imply partial covering of the quasar
emission source, and (3) and large column densities in these
high-ionization ions
\citep{hamann1997a,hamann1997b,tripp2008,hamann2019}. N~V is typically
very weak or absent in intervening systems \citep{werk2016}. High
N~V/H~I and O~VI/H~I are also much higher in associated absorbers than
in intervening systems \citep[e.g.,][]{tripp2008}.

Among the 20 quasars with N~V or O~VI absorption systems suggestive of
outflows, 17 objects have absorption line profiles that meet the first
of the above criteria (the only exceptions are PG~0804+761,
PG~0844$+$349, and PG~2233+134). Many of the quasars with blueshifted
N~V or O~VI absorption lines show N~V $\lambda$1238/$\lambda$1243
and/or O~VI $\lambda$1032/$\lambda$1038 line ratios that also meet
criteria \#2 and \#3 (Fig.\ \ref{fig:hist_tau_cf}). Mrk~231 does not
formally meet these criteria (since it has no N~V absorption line and
O~VI falls outside of the spectral range of the data), but it shows
all of the characteristics of a FeLoBAL at visible and NUV wavelengths
\citep[and its \lya\ line emission is highly blueshifted;][and
  references therein]{veilleux2013b,veilleux2016}, so we include it
here among those with quasar-driven outflows. So, overall, at least 18
quasars in our sample have absorption features suggestive of
quasar-driven outflows.

In 15 of the 20 absorber detections, the velocity widths, FWHM$_{\rm
  rms} \equiv 2.355~\sigma_{\rm rms}$, are below the minimum of
2000 \kms\ generally used for BALs
\citep{weymann1981,weymann1991,hamann2004,gibson2009b,gibson2009a}, so
they fall in the category of mini-BALs (500 $<$ FWHM$_{\rm rms}$ $<$
2000 \kms) or NALs (FWHM$_{\rm rms}$ $<$ 500 \kms). Moreover, in
many cases, the profiles are highly structured rather than smooth, and
thus do not meet the ``BAL-nicity'' criterion to be true BALs.

The weak and narrow redshifted absorption features in PG~0804$+$761
and PG~0844$+$349 are good candidates for infalling tidal debris.

\subsection{Location and Structure of the Mini-BALs}
\label{sec:discussion_location_structure}

\subsubsection{Depths of the Absorption Profiles}

The depths of the mini-BALs may be used to put constraints on the
location of the outflowing absorbers. The source of the FUV continuum
in these quasars is presumed to be the accretion disk on scale of
$\sim$ few $\times$ 10$^{15}$ cm ($\la$ 0.01 pc), where we used
equation (6) in \citet{hamann-herbst2019} assuming an Eddington ratio
$\eta_{\rm Edd} = 0.1$. But it is clear from the spectra that in many
cases (e.g., PG~1001$+$054, 1004$+$130, 1126$-$041, 1309$+$355,
1351$+$640, 1411$+$442, 2214$+$139) the mini-BALs absorb not only the
FUV continuum emission but also a significant fraction of the \lya,
N~V, and O~VI line emission produced in the BELR. The gas producing
the mini-BALs must therefore be located outside of the BELR on scales
larger than
\begin{eqnarray}
r_{\rm BELR} = 0.1 \left(\frac{L_{\rm AGN}}{10^{46}~{\rm
    erg~s}^{-1}}\right)^{1/2}~{\rm pc}
\label{eq:r_BELR}
\end{eqnarray}
\citep[e.g.][]{kaspi2005,kaspi2007,bentz2013}. The radius of the outer
boundary of the BELR, $r_{\rm out}$, is likely set by dust sublimation
\citep{netzer1993,baskin2018}. For gas densities of 10$^5$ $-$
10$^{10}$ cm$^{-3}$, \citet{baskin2018} derive
\begin{eqnarray}
  r_{\rm out} &\simeq& r_{\rm subl} \simeq (3 - 6) r_{\rm BELR} \\ &=& (0.3 -
  0.6) \left(\frac{L_{\rm AGN}}{10^{46}~{\rm
      erg~s}^{-1}}\right)^{1/2}~{\rm pc},
\label{eq:r_subl}
\end{eqnarray}
where graphite grains of size $\sim$ 0.05 $\mu$m is assumed
\citep[Fig.\ 5 in][]{baskin2018}. Note that these values are smaller
than those in \citet{barvainis1987} and \citet{veilleux2020}, which
are based on silicate grains and lower gas densities (thus lower
evaporation temperatures).

More can be said about the structure of the absorbing gas from the
fact that the N~V and O~VI absorption features are optically thick
($\tau \ga 1 - 5$; Fig.\ \ref{fig:hist_tau_cf}) but are not completely
dark. The covering factors derived from the multi-component fits to
the N~V and O~VI mini-BALs range from 0.1 to 1, a direct indication
that the absorbing material is compact and spatially
inhomoneneous. The often structured velocity profiles of the N~V and
O~VI mini-BALs (N~V in PG~1411$+$442 is arguably the only exception)
also suggest a high level of kinematic sub-structures in the outflows,
different from the smooth BALs observed in high-luminosity
quasars. These properties of the mini-BALs may indicate one of two
things: (1) Our line of sight is not aligned along the direction of
the outflowing stream of gas as in the case for the BALs
\citep[e.g.][]{murray1995,elvis2000,ganguly2001}, but instead
intercepts only a small fraction of this stream and results in a
covering factor of the background emission that is highly dependent on
the inhomogeneity of the outflowing gas. (2) Another equally plausible
explanation is that (structured) mini-BALs form in more sparse outflows
or (in the unified outflow model discussed for high-z quasars) in more
sparse outflow regions, e.g., at higher latitudes above the disk.

\begin{figure*}[!htb]   
\epsscale{1.1}
\plotone{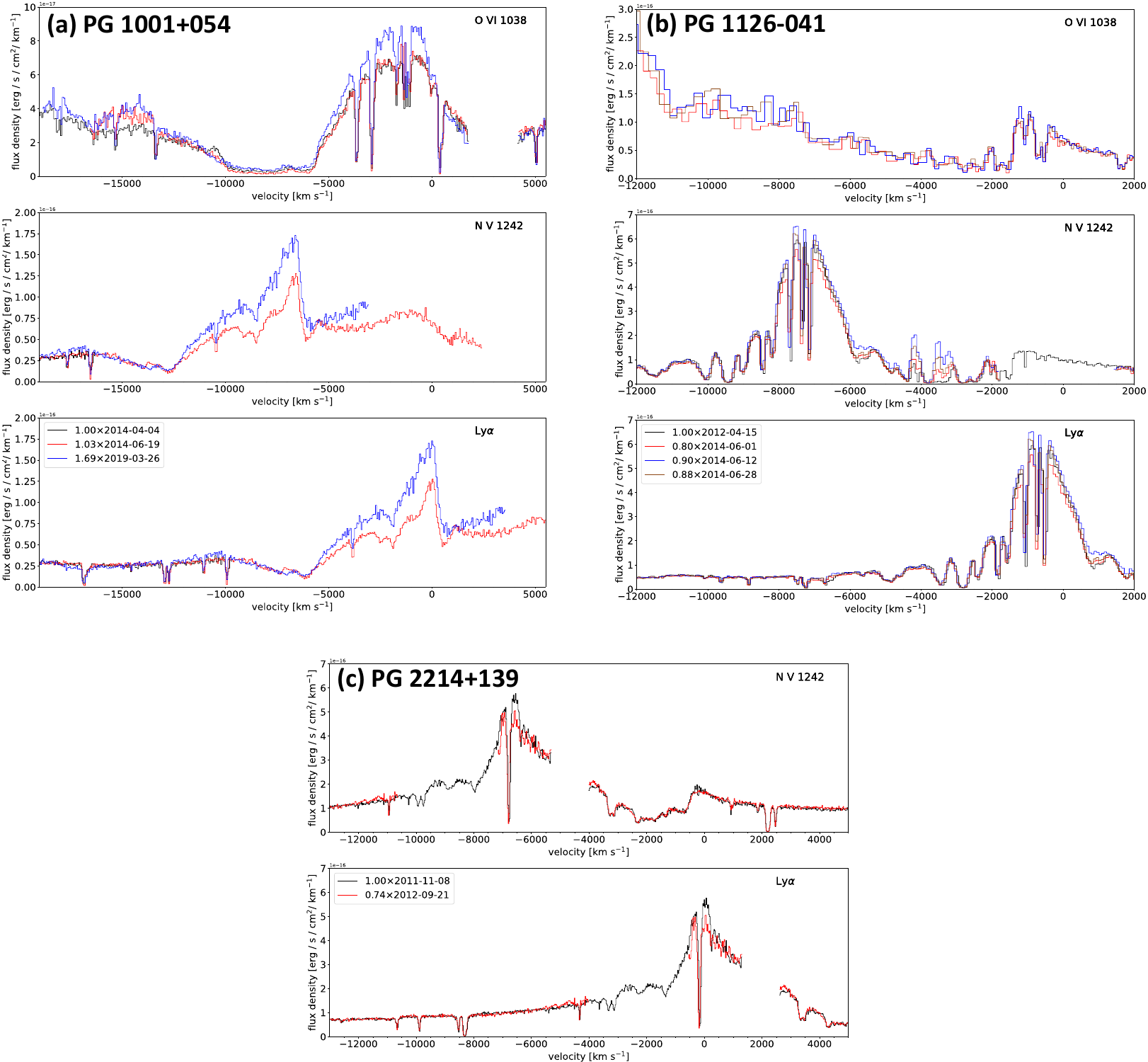}
\caption{Multi-epoch comparisons of the mini-BALs in (a)
  PG~1001$+$054, (b) PG~1126$-$041, and (c) PG~2214$+$139. All spectra
  are normalized to match the continuum level blueward of \lya\ or
  redward of N~V and O~VI. The multiplicative factor is indicated in
  the caption.}
\label{fig:mini-bal_variability}
\end{figure*}

\subsubsection{Variability of the Absorption Profiles}

Additional constraints on the location and structure of the BALs and
mini-BALS in our sample may be obtained from profile
variability. There is a vast literature on this topic
\citep[e.g.,][and references
  therein]{gibson2008,hamann2008,gibson2010,capellupo2012,filiz-ak2012,filiz-ak2013,grier2015,he2019,yi2019}.
In our sample, four of the mini-BALs (PG~1001$+$054, PG~1126$-$041,
PG~1411$+$442, and PG~2214$+$139) have been observed at two different
epochs or more, and can therefore be searched for mini-BAL profile
variations. 
The emergence of a dense [log $n_{\rm H}$(cm$^{-3}$) $\ga$ 7] new
outflow absorption-line system in PG~1411$+$442 was reported in
\citet{hamann2019} and the detailed inferrence of a distance $\la$ 0.4
pc from the quasar is not repeated here. We present the archival COS
spectra for the other three objects in Figure
\ref{fig:mini-bal_variability}, normalized to the same FUV continuum
level to emphasize absorption profile variations.

In PG~1001$+$054 (Fig.\ \ref{fig:mini-bal_variability}a), the dramatic
(72\%) decrease in the FUV continuum emission from June 2014 to March
2019 is accompanied by a strengthening of the broad Ly$\alpha$, N~V,
and O~VI emission lines in terms of equivalent widths but no obvious
change in the mini-BAL profiles.  In PG~1126$-$041
(Fig.\ \ref{fig:mini-bal_variability}b), the more modest (20\%)
decrease of the continuum emission from April 2012 to June 2014 are
not accompanied by any obvious variations in the equivalent widths of
any of the broad emission and absorption lines except for the most
blueshifted N~V absorption features below $-$3000 \kms\ which show
variations on timescales perhaps as short as 12 days.  The broad
emission and absorption lines in PG~2214$+$139
(Fig.\ \ref{fig:mini-bal_variability}c) show no variations between
November 2011 and September 2021 despite a 26\% increase in the
strength of the FUV continuum emission.

The fast 12-day variability of the high-velocity N~V mini-BAL in
PG~1126$-$041 may be interpreted in two different ways.  One
possibility is that transverse motions of the outflowing clouds across
our line of sight to the continuum source and BELR are responsible for
these changes \citep[as in PG~1411$+$442;][]{hamann2019}. A variant on
this idea is that the changes in profiles are due instead to the
dissolution and creation of the absorbing clouds/clumps in the outflow
as they transit in front of the continuum source. In the other
scenario, changes in the ionization structure of the absorbing clouds
due to changes in the incident quasar flux cause the absorbing N~V and
O~VI columns to vary and reproduce the observations. If this is the
case, the variability timescale sets a constraint on the ionization or
recombination timescale, which depends solely on the incident ionizing
continuum and gas density \citep[$\sim$ 10$^5$ yrs/$n_H$, where
  $n_H$ is the number density of the clouds in cm$^{-3}$;
  e.g.,][]{he2019}.

This last scenario predicts that changes in the FUV continuum of the
quasar will produce changes in the mini-BAL. While changes are indeed
observed in both the FUV continuum emission and high-velocity N~V
mini-BAL of PG~1126$-$041, the amplitudes of these changes are not
correlated. From 2012-04-15 to 2014-06-01, the continuum emission
strengthened while the N~V mini-BAL weakened. From 2014-06-01 to
2014-06-12, both the continuum emission and N~V mini-BAL
weakened. Finally, from 2014-06-12 to 2014-06-28, the continuum
emission remained constant to within 1\% but the N~V mini-BAL
strengthened slightly. This lack of a direct connection between
variations in the continuum and the N~V mini-BAL seems to disfavor the
scenario where the mini-BAL variations are associated with changes in
the ionization structure of the absorbing clouds, unless
log~$n_H$(cm$^{-3}$) $\la$ 5-6 in which case $r >> r_{\rm out}$ and
time delays associated with the finite recombination timescale could
be at play \citep[cf.][]{hamann2019}.

While more detailed modeling of the mini-BAL of PG~1126$-$041 is
beyond the scope of the present paper, the fact that the mini-BAL
variations are only observed in N~V and only at high velocities may be
an optical depth effect: the cloud complex that produces the
Ly$\alpha$ and O~VI mini-BALs and low-velocity N~V mini-BAL may be so
optically thick to be immune to variations in the ionizing continuum
or tangential movement of the absorbing gas across the continuum
source (we return to this topic in Sec.\ \ref{sec:discussion_pg1126}
below).

\subsection{Driving Mechanisms of the Mini-BALs}
\label{sec:discussion_driving}

As reviewed in, for instance, \citet{veilleux2020}, the absorbing
clouds making up the mini-BALs may be material (1) entrained in a hot,
fast-moving fluid, or (2) pushed outward by radiation or cosmic ray
pressure, or (3) created {\em in-situ} from the hot wind material
itself. In the first two scenarios, the equation of motion of the
outflowing absorbers of mass $M_{abs}$ that subtends a solid angle
$\Omega_{abs}$ is
\begin{eqnarray}
\frac{d}{dt}~\left[M_{abs}(r)~\dot{r}\right] =
\Omega_{abs}~r^2~(P_{th} + P_{CR} + P_{jet}) \nonumber \\ +
\left(\frac{\Omega_{abs}}{4 \pi}\right)\left(\frac{\tilde{\tau}
  L_{\rm bol}}{c}\right) - \frac{G M(r) M_{abs}(r)}{r^2},
\label{eq:eom}
\end{eqnarray}
where $M(r)$ is the galaxy mass enclosed within a radius $r$ and
$\tilde{\tau}$ is a volume- and frequency-integrated optical depth
that takes into account both single- and multiple-scattering processes
in cases of highly optically thick clouds
\citep{hopkins2014,hopkins2020}.\footnote{More explicitly,
  $\tilde{\tau} \equiv (1 - e^{-\tau_{\rm single}})(1 + \tau_{\rm
    eff,IR}).$ The value of $\tilde{\tau}$ therefore ranges from
  $\sim\tau_{\rm single} = \tau_{\rm UV/optical} << 1$ in the
  optically thin case to $\sim$ (1 + $\tau_{\rm eff, IR}$) $\gtrsim$ 1
  in the infrared optically-thick limit (the effective infrared
  optical depth, $\tau_{\rm eff, IR}$, is also sometimes called the
  ``boost factor''; \citet{veilleux2020}).} The terms on the right in
Equation \ref{eq:eom} are the forces due to the thermal, cosmic ray,
and jet ram pressures, the radiation pressure, and gravity,
respectively. Magneto-hydrodynamical effects are assumed to be
negligible at the distances of these absorbing clouds. The quasars in
our sample do not have powerful radio jets so the jet ram pressure
term can safely be neglected. Similarly, the relatively modest radio
luminosities of the mini-BAL quasars relative to their optical and
bolometric luminosities (Column 6 in Table \ref{tab:sample}) suggest
that cosmic-ray electrons do not play an important dynamical role in
accelerating the BAL clouds. Indeed the fraction of BAL quasars seems
to vary inversely with the radio loudness parameter, $R$ \citep[Column 5 
  in Table \ref{tab:sample}; e.g.,][] {becker2001,shankar2008}.
Below, we consider the remaining thermal and radiation pressure terms
separately. In reality, these pressure forces may act together to
drive the mini-BAL outflows (see Sec.\ \ref{sec:discussion_pg1126} for
a closer look at the mini-BAL PG~1126$-$041 in this context).

\subsubsection{Thermal Wind and Blast Wave}
\label{sec:discussion_thermal}

For many years, ram-pressure acceleration of pre-existing clouds has
been considered a serious contender to explain BALs in quasars given
the need for a much hotter, rarefied medium to confine the clouds as
they are being accelerated \citep[e.g.][]{weymann1985}. However, it is
notoriouly difficult to accelerate dense gas clouds from rest up to
the typical (mini-)BAL velocities by a warm, fast thermal wind without
destroying them in the process through Rayleigh-Taylor fragmentation
and shear-driven Kelvin-Helmholtz instabilities
\citep[e.g.][]{cooper2009,scannapieco2015,schneider2015,schneider2017}. Radiative
cooling and magnetic fields may act to slow down cloud disruption
\citep{marcolini2005,cooper2009,banda-barragan2016,banda-barragan2018,banda-barragan2020,gronnow2018}. Radiative
cooling of the warm mixed gas can not only prevent disruption, but it
may cause the cloud to grow in mass
\citep[e.g.,][]{gronke2018,gronke2020,girichidis2021}, although there
are caveats \citep{schneider2020}. An alternative scenario is that the
BAL and mini-BAL clouds are created {\em in-situ} via thermal
instabilities and condensation from the hot gas with a cooling time
shorter than its dynamical time \citep{efstathiou2000,silich2003}.
This is the idea behind the ``blast wave'' simulations of
\citet{richings2018a,richings2018b}; see also
\citet{weymann1985,zubovas2012,zubovas2014,faucher-giguere2012,nims2015}.

In these simulations, a fast (presumably X-ray emitting) AGN wind with
outward radial velocity of 30,000 \kms\ is injected in the central 1
pc and collides violently with the host ISM. The resulting shocked
wind material reaches a very high temperature \citep[$\sim$10$^{10}$
  K;][]{nims2015} that does not efficienty cool, but instead
propagates outward as an adiabatic (energy-driven) hot bubble. This
expanding bubble sweeps up gas and drives an outer shock into the host
ISM raising its temperature to a few $\times$ 10$^7$ K
\citep{nims2015}. Radiative cooling of the shocked ISM eventually
becomes important and the outflowing material reforms as cool neutral
and molecular gas, but by that time, the outflowing material has
acquired a significant fraction of the initial kinetic energy of the
hot wind. These simulations predict that the cooling radius, i.e. the
radius at which the gas cools to 10$^4$ K, increases from 100 pc to 1
kpc for AGN with luminosities from 10$^{44}$ to 10$^{47}$
erg~s$^{-1}$, respectively \citep[Fig.\ 7 of][]{richings2018b}. This
cooling radius is also the expected location of the gas clouds
producing the N~V and O~VI mini-BALs, as the cooling gas rapidly
transitions from $\sim$ 10$^7$ K to $\sim$ 10$^4$ K. This large radius
is well outside the lower limit on the distance of the mini-BAL from
the quasars derived above so it is not inconsistent with our data.
However, one should note that the inner X-ray wind in quasars is
presumed to much smaller in reality than the value of 1 pc assumed in
the simulations so the cooling radius may have to be scaled down
accordingly. Moreover, these models do not address how the BELRs would
be restored after the passage of the blast wave. Finally, the detailed
analysis of the BAL in PG~1411$+$442 firmly rule out (at least in that
case) absorption at these large distances.

\subsubsection{Radiative Acceleration} 
\label{sec:discussion_radiation}

Although ram-pressure acceleration has been a serious contender,
overall the favored explanation for the large velocities of the BALs
and mini-BALs is that the gas absorbers have been accelerated by the
radiation pressure forces associated with the intense radiation field
that is emanating from the quasars
\citep[e.g.,][]{arav1994a,giustini2019}.
Strong support for this scenario comes from the observed trends for
the maximum velocity of the absorption to increase on average with
increasing optical, UV, or bolometric luminosity and the Eddington
ratio
\citep[e.g.,][]{perry1978,brandt2000,laor2002,ganguly2007,ganguly2008,gibson2009a,zhang2014}. Note,
however, that the overall correlations noted in these studies are
often quite modest and sometimes only visible when considering the
upper envelope of the velocity distribution and only when the sample
of AGN span 2-3 orders of magnitude in luminosity (sometimes combining
NALs, mini-BALs, and BALs together). While more recent studies
\citep[e.g.,][]{rankine2020} have confirmed and indeed strengthened
the existence of some of these correlations, all of the cited results
relate to the C~IV absorption, rather than the N~V and O~VI
features. The statistics on N~V and particularly O~VI absorbers are
much poorer.

{\em Far Ultraviolet Spectroscopic Explorer} ({\em FUSE}) observations
of Seyfert galaxies of relatively low luminosities (10$^{38}$ $-$
10$^{42}$ erg~s$^{-1}$) show either no or very weak trends of
increasing maximum velocities with increasing luminosities and no
trend at all with the Eddington ratio
\citep{kriss2004a,kriss2004b,dunn2008}. O~VI and N~V BALs in
high-redshift, high-luminosity quasars
\citep{baskin2013,hamann-herbst2019} have maximum velocities that
correlate with their C~IV counterparts, but the range in AGN
luminosity of their sample is too small to test the luminosity
dependence of the maximum velocities. More fundamentally, there is
also a trend between line widths and optical depth. The most extreme
example of this trend is P~V, which coexists with C~IV having the same
ionization requirements, but is always weaker and narrower than C~IV
\citep{hamann-herbst2019}. The reason is that P~V traces only the
highest column density regions with smaller covering fractions, while
C~IV can have significant absorption in more diffuse gas occupying a
larger volume. This evidence for optical depth-dependent covering
factors is a signature of inhomogeneous partial covering.

Overall, given the complex results from these previous studies, it is
perhaps not surprising to find no significant correlations in our
sample of QUEST quasars between (maximum) outflow velocities and the
AGN luminosities (Sec.\ \ref{sec:kinematics}). Theoretically, the
noise in the trends between the outflow kinematics and AGN luminosity
is {\em expected} in the radiative acceleration scenario given
projection effects that reduce the measured outflow velocities and
variance in both the (minimum) launching radius
\citep[e.g.][]{laor2002} and efficiency of radiative acceleration
associated with the complex micro-physics of the photon interaction
with the clouds - this complexity is hidden in the quantity
$\tilde{\tau}$ in equation \ref{eq:eom}. A similar trend of increasing
variance in the maximum velocity with increasing AGN luminosity is
observed in the other cooler gas phases of AGN-driven outflows
\citep[e.g.,][]{veilleux2020,fluetsch2020}.

Additional evidence that radiation pressure plays an important role in
accelerating the absorbers in quasars comes from the significant
dependence of the incidence rate, equivalent width, and weighted
outflow velocity dispersion of the blueshifted absorbers on the X-ray
properties of the quasars. This effect has been reported in numerous
studies of nearby and distant AGN, based largely on C~IV and Si~IV
\citep[e.g.,][]{laor2002,gibson2009a}, and is also clearly present in
our sample of quasars based on N~V and O~VI
(Sec.\ \ref{sec:incidence}, \ref{sec:ew}, and
\ref{sec:kinematics}). More specifically, we find that mini-BALs and
BALs are broader, stronger, and more common among X-ray faint quasars
with steep optical-to-X-ray slopes $\alpha_{\rm OX}$ $\la$ $-$1.7 and
hydrogen column densities $N_{\rm H}$ in excess of $\sim$ 10$^{22}$
cm$^{-2}$ (Sec.\ \ref{sec:results}). This result is expected in the
context of
radiative acceleration since the combined radiative force
\citep[``force multiplier'';][]{arav1994b} is greatly reduced when the
gas is over-ionized by the hard far-UV/X-rays, becoming too
transparent to be radiatively accelerated. This over-ionized
``failed-wind'' material may act as a radiative shield to produce the
spectral softening needed for efficient radiative acceleration of the
outflow material downstream
\citep{murray1995,proga2004,proga2007,sim2010}. However, the strong
near-UV absorption lines near systemic velocity expected in this
scenario are not observed \citep{hamann2013}. Alternatively, the
spectrum emerging from the accretion disk may be intrinsically
softer/fainter in the hard far-UV/X-rays than commonly assumed
\citep[e.g.][]{laor2014}. Weak-lined ``wind-dominated'' quasars, such
as Mrk~231, PHL 1811 and its analogs, which are intrinsically faint
and unabsorbed in the X-rays, may be naturally explained in this
fashion \citep{richards2011,wu2011,luo2015,veilleux2016}. While a
connection should exist between the X-ray warm absorbers and the UV
absorption-line outflows, a direct one-to-one kinematic correspondence
between the two classes of absorbers is often not seen because the gas
in the warm absorbers is too highly ionized to produce measurable
lines in the UV spectra \citep[][and references
  therein]{kaspi2000,kaspi2001,gabel2003,kraemer2001,krongold2003,arav2015,laha2021}. We
return to this point in Section \ref{sec:discussion_pg1126} when
discussing the mini-BAL in PG~1126$-$041 (the case of PG~1211$+$143 is
briefly discussed in Appendix \ref{appendix:detailed_results}).

Another observational characteristic of outflows that favors
radiative acceleration is the phenomenon of line-locking observed in
perhaps as many as $\sim$ 2/3 of all NAL and (mini-)BAL outflows
\citep[e.g.,][and references
  therein]{hamann2011,bowler2014,lu2018,mas-ribas2019}. This is
observed in outflows where multiple absorbers are present but are
separated by the exact same velocity separation as the doublet of C~IV
(499 \kms), N~V (962 \kms), or O~VI (1650 \kms). Previous studies have
shown that the probability of a line-locking signature accidentally
occurring over a relatively small redshift path is negligible
\citep[e.g.,][]{foltz1987,srianand2000,srianand2002,ganguly2003,benn2005}. Radiative
acceleration is a natural explanation for line-locking
\citep[e.g.,][]{mushotzky1972,scargle1973,braun1989}.

\begin{figure}[!htb]   
\epsscale{1.2}
\plotone{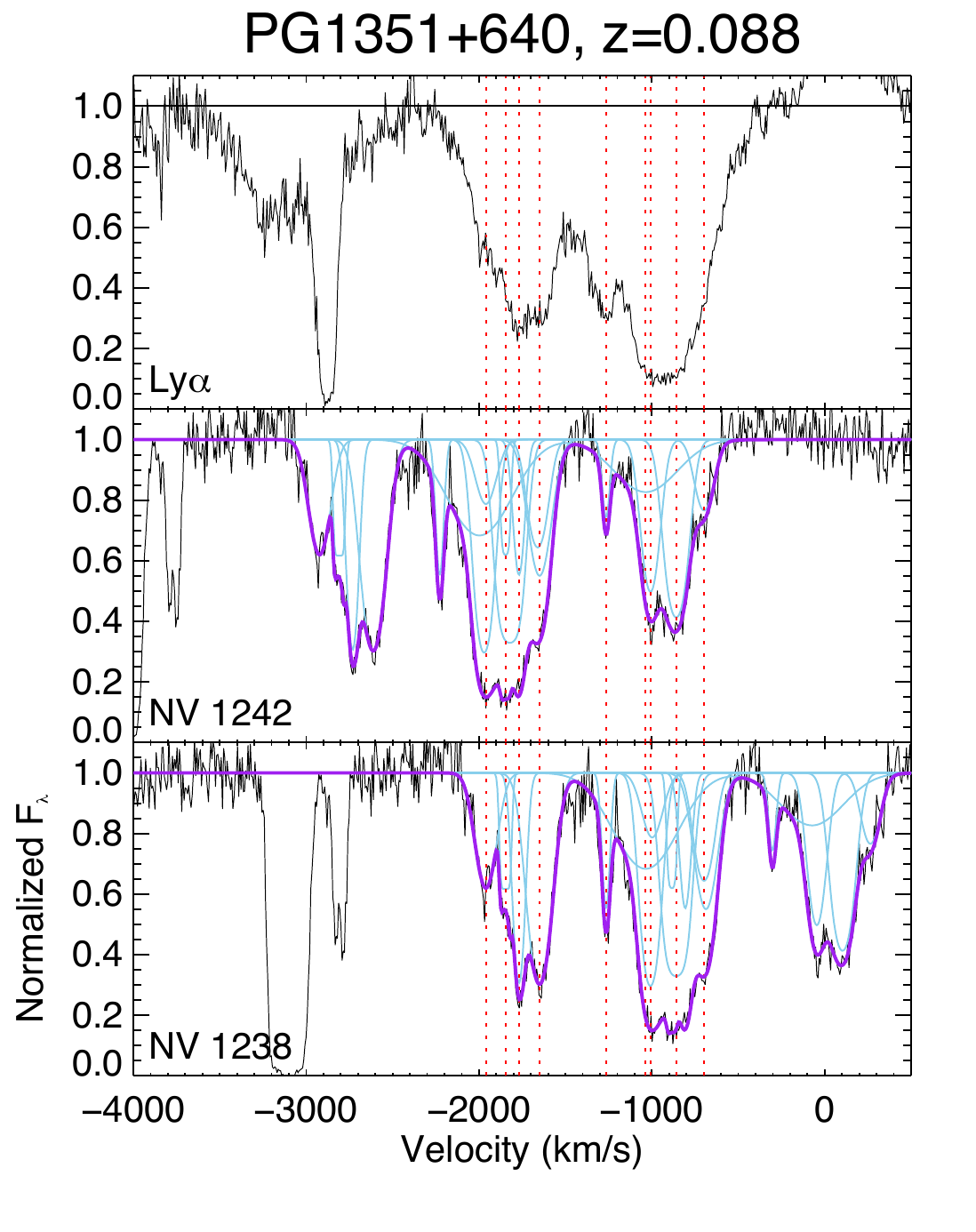}
\caption{Line-locking in PG~1351$+$640. The two deep Ly$\alpha$
  absorption throughs are separated by the velocity split of the N~V
  doublet lines, resulting in a N~V complex that looks like a
  ``triplet''.}
\label{fig:pg1351_vel}
\end{figure}

The best case for line-locking in our data is that of PG~1351$+$640,
where two deep absorption throughs are detected in \lya, extending
over $-$[2200, 1500] \kms\ and $-$[1400, 600] \kms, roughly
separated by the velocity split of the N~V doublet lines ($\Delta v$
$\approx$ 900 $-$ 1000 \kms\ $\approx$ $\Delta v_{\rm N V}$ ;
Fig.\ \ref{fig:pg1351_vel}). This results in a N~V mini-BAL that looks
like a ``triplet'' in this object. Another case of line-locking may be
present in PG~1126$-$041, where some of the deepest Ly$\alpha$
throughs are separated by $\Delta v$ $\approx$ 900 $-$ 1000
\kms\ $\approx$ $\Delta v_{\rm N V}$
(Fig.\ \ref{fig:example_pg1126_vel}).

Note, finally, that radiation pressure on dust grains has also been
invoked as an important contributor to the radiative acceleration
given that (mini-)BAL QSOs, particularly LoBALs and FeLoBALs, have
more reddened UV spectra than non-BAL QSOs \citep[e.g.,][and
  references therein]{allen2011,hamann-herbst2019}. This result has
been interpreted to mean that BAL and mini-BAL clouds have large
columns of ionized + neutral gas (log~$N_{\rm H}$(cm$^{-2}$) $\ga$ 23)
and enough dust to provide extinction equivalent to at least $A_V
\sim$ 1-2 mag.\ in some cases \citep[this is discussed in
  Sec.\ \ref{sec:discussion_pg1126} in the context of PG~1126$-$041,
  but see also the results on Mrk~231 and
  PG~1411$+$442;][respectively]{veilleux2016,hamann2019}. Under these
circumstances, the (mini-)BAL clouds may be subject to larger
radiative forces than dustless clouds since the dust cross section,
and thus $\tilde{\tau}$ in equation \ref{eq:eom}, in the UV optical is
$>$ 1-2 orders of magnitude than the Thompson scattering cross section
of electrons.
Unfortunately, we do not have a reliable reddening indicator of the
FUV continuum emission in our quasars, so we cannot directly compare
our data with those of BAL and non-BAL QSOs.
On the other hand, radiation that will be absorbed by the dust in the
broad absorption line regions (BALRs) will be re-emitted in the
infrared so our measurements of the infrared excess in our quasars may
serve as a surrogate for the amount of dust in the BALRs. The lack of
obvious correlation between BAL properties in our quasar sample and
the mid-, far-, and total (1 $-$ 1000 $\mu$m) infrared excesses (e.g.,
Figs.\ \ref{fig:weq_vs_quantities} and \ref{fig:vwtavg_vs_quantities})
indicates one of two things: (1) radiative acceleration on dust is not
important in the BALRs of these objects or (2) the various infrared
excesses are dominated by dust emission from outside of the BALR,
e.g. dust in the host galaxy itself.

\subsection{P~V Mini-BAL in PG~1126$-$041 and Other Quasars}
\label{sec:discussion_pg1126}

In this last section, we take a closer look at the mini-BAL system in
PG~1126$-$041. A mini-BAL extending from $-$1000 to $-$5000 \kms\ was
first reported in the N~V, C~IV, and S~IV absorption lines of this
object by \citet{wang1999a,wang1999b}, based on the analysis of old
low-resolution spectra obtained with {\em IUE} and the Goddard
High-Resolution Spectrograph (GHRS) on {\em HST}. Variable and much
faster ($\sim$16,500 \kms) X-ray absorption has also been detected in
this object \citep{wang1999a,wang1999b,giustini2011}. Interestingly,
this object is among the least luminous AGN (log~$L_{\rm BOL}/L_\odot$
= 11.52) in our sample, intermediate between quasars and typical
Seyfert 1 galaxies. Mini-BALs with outflow velocities of up to $\sim$
5000 \kms\ and widths (FWHM$_{\rm rms} \equiv 2.355~\sigma_{\rm
  rms}$) $>$ 1000 \kms\ are relatively rare in such low-luminosity
systems
\citep[e.g.,][]{kriss2004a,kriss2004b,dunn2008,crenshaw2012}. On the
other hand, PG~1126$-$041 is also the object in our sample with the
steepest X-ray to optical index ($\alpha_{\rm OX} = -2.13$, a virtual
tie with PG~1001$+$054, which also harbors a mini-BAL) and is among
those with the largest infrared excess
(Table \ref{tab:sample}), reinforcing the view expressed in Section
\ref{sec:discussion_radiation} that X-ray absorbed or intrinsically
weak quasars are more likely to host BALs and mini-BALs.

Apart from the line-locking signatures found in the N~V mini-BAL of
this object, which we argued in Sec.\ \ref{sec:discussion_driving}
favors radiative driving, the most remarkable aspect of this mini-BAL
is the detection of a narrow P~V $\lambda\lambda$1117, 1128 cloud at a
velocity of $-$2200 \kms\ (Fig.\ \ref{fig:example_pg1126_vel}). Large
ionized-gas column densities are needed to produce this feature given
the low abundance of phosphorus relative to hydrogen
\citep[log(P/H)$_\odot$ = $-$5.54;][]{lodders2003}. The multi-component
fit of each line in the P~V doublet requires two components with nearly
identifical median velocities ($-$2196 and $-$2203 \kms) but different
velocity dispersions (24 and 74 \kms), covering factors (0.37 and
0.10, respectively), and optical depths (0.7 and 2.6,
respectively). The total equivalent width of this doublet is 0.3 \AA.

Taken at face value, the results from the fits suggest that the P~V
lines are only moderately optically thick and therefore more reliable
indicators of the total ionized column densities of this cloud than
the highly saturated N~V and O~VI features. This is supported by the
$\sim$2:1 intensity ratio of the P~V lines. An optical depth of order
unity in P~V $\lambda$1128 implies
an ionized hydrogen column density log~$N_{\rm H}$(cm$^{-2}$)
$\approx$ 22.3, assuming a solar P/H abundance ratio and ionization
corrections based on detailed photoionization calculations for typical
BALs and mini-BALs \citep[ionization parameters log $U \ga$
  $-$0.5;][and references
  therein]{hamann1998,leighly2011,borguet2012,borguet2013,baskin2014,capellupo2017,moravec2017,hamann-herbst2019}.
This column density is remarkably consistent with the expectations
from radiation pressure confined cloud models \citep{baskin2014}.

This value of the total column density may be used to estimate the
minimum kinematic energy of this outflowing cloud using
\citep[eq.\ 2 from][]{hamann-herbst2019}
\begin{eqnarray}
E_{\rm kin} & = 1.5 \times 10^{52}
\left(\frac{Q}{0.15}\right)\left(\frac{N_{\rm H}}{2 \times
  10^{22}~{\rm cm}^{-2}}\right) \left(\frac{r}{1~{\rm pc}}\right)^2
\nonumber \\ & \times \left(\frac{v}{2200~{\rm
    km~s}^{-1}}\right)^2~{\rm ergs},
\label{eq:E_kin}
\end{eqnarray}
where $Q$ is an approximate global outflow covering factor based on
the incidence of mini-BALs in the SDSS quasars
\citep{trump2006,knigge2008,gibson2009a,allen2011} and $r$ = 1 pc is a
placeholder radial distance that we adopt for illustration purposes
\citep[it may underestimate the actual distance of the absorbers from
  the source; see Sec.\ \ref{sec:discussion_location_structure}
  and][and references therein]{arav2020}. Following
\citet{hamann2019}, we estimate the time-averaged kinetic energy
luminosity, $L_{\rm kin}$, by dividing $E_{\rm kin}$ by a
characteristic flow time, $r/v$ $\approx$ 450 yr. This yields $L_{\rm
  kin} \ga 1 \times 10^{42}$ erg s$^{-1}$.
In these units, $L_{\rm BOL}$ = 1.3 $\times$ 10$^{45}$ erg s$^{-1}$ so
$L_{\rm kin}/L_{\rm BOL}$ $\ga$ 0.001. Taken at face value, this ratio
is too small to significantly affect the evolution of the galaxy host
\citep[e.g., $L_{\rm kin}$ $\ga$ 0.005 $L_{\rm Edd}$ is needed
  according to][]{hopkins2010}, unless (1) $r$ is severely
underestimated or (2) the other clouds at lower and higher velocities
involved in this mini-BAL contribute significantly to $L_{\rm
  kin}$. This second possibility seems unlikely given the lack of P~V
detection in these clouds which suggests column densities log~$N_{\rm
  H}$(cm$^{-2}$) $\la$ 22.

Next, we use the total column density of the P~V cloud in PG~1126$-$041
to estimate the time-averaged momentum outflow rate of this cloud,
$\dot{p} = 2 L_{\rm kin} / v \approx$ 1 $\times$ 10$^{34}$ dynes,
and compare this value with the radiation pressure, $L_{\rm BOL}/c$ =
4 $\times$ 10$^{34}$ dynes. Given that $\dot{p}/(L_{\rm BOL}/c) \sim 0.2$,
radiation pressure can thus in principle accelerate this cloud,
although a contribution from a thermally driven wind as detailed in
Section \ref{sec:discussion_thermal} cannot be formally ruled out.

\begin{figure}[!htb]   
\epsscale{1.0}
\plotone{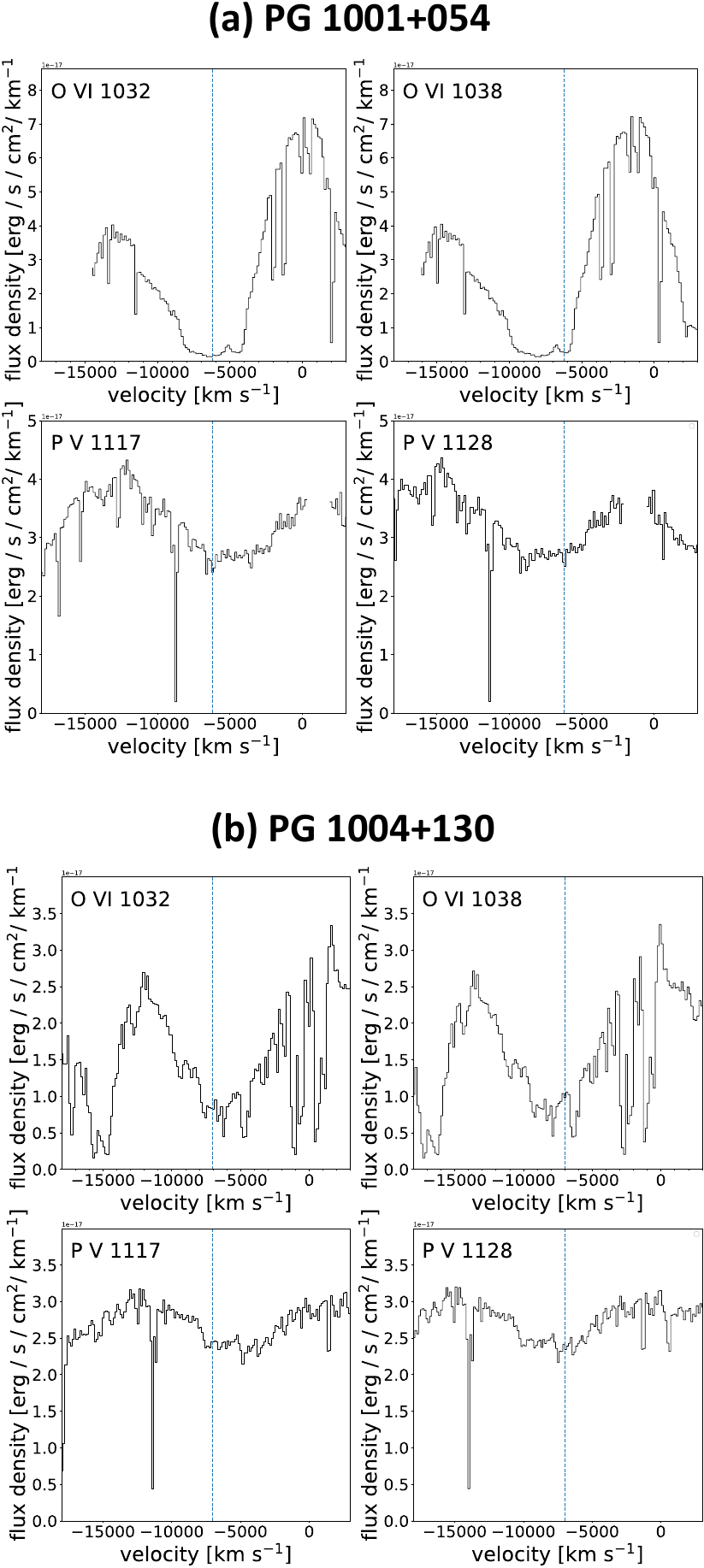}
\caption{Tentative detection of P~V in (a) PG~1001$+$054 and (b)
  PG~1004$+$130. The spectra have been heavily binned to emphasize the
  broad but shallow features. The deep and narrow absorption lines in
  the velocity range $-$[8,000, 14,000] \kms\ in the P~V 1117, 1128
  panels of both PG~1001$+$054 and PG~1004$+$130 are due to
  intervening MW ISM (Si II 1259, 1260 + Fe II 1260 and C~II 1334 +
  C~II$^*$ 1335, respectively).  }
\label{fig:pv_detection}
\end{figure}

Finally, we apply Equation \ref{eq:E_kin} and calculate $L_{\rm
  kin}/L_{\rm BOL}$ for the other mini-BALs in our sample with solid
and tentative P~V detections. For PG~1411$+$442, the only other
mini-BAL in the sample with a definite P~V detection, we get $L_{\rm
  kin}/L_{\rm BOL}$ $\ga$ 0.01 and $\dot{p}/(L_{\rm BOL}/c) \sim 1$,
for an outflow velocity of $-$1800 \kms\ (Table \ref{tab:fits}), a
total column density log~$N_{\rm H}$(cm$^{-2}$) $\ga$ 23.4, and a
BELR-like distance $\la$ 0.4 pc from the central light source derived
by \citet{hamann2019} using several absorption lines and detailed
photoionization simulations. This BAL may thus be sufficient to impact
the host galaxy evolution. P~V is also tentatively detected in
PG~1001$+$054 and PG~1004$+$130 at velocities of $\sim$ [$-$4000,
  $-$6,000] \kms\ (Fig.\ \ref{fig:pv_detection}). In both cases, the
equivalent widths of P~V 1117 and 1128 are very similar, implying
saturation and log~$N_{\rm H}$(cm$^{-2}$) $\ga$ 22.3. These numbers
yield outflows with kinetic-to-bolometric luminosity ratios that are
higher than that of PG~1126$-$041 but lower than that of
PG~1411$+$442, and thus marginally sufficient to impact the host
galaxy evolution. Overall, the mini-BALs in the QUEST quasars are less
powerful than the P~V BALs detected in $z_e \ga 1.6$ SDSS quasars
\citep{moravec2017} but perhaps more common (4/33 $\sim$ 10\%) than at
high luminosities/redshifts \citep[detection rate of only 3$-$6\%
  among the $z_e \ga 2.6$ BAL quasars in the SDSS-III BOSS quasar
  catalog][]{capellupo2017}.

\section{Summary}
\label{sec:summary}

As part I of a {\em HST} FUV spectroscopic study of the QUEST
(Quasar/ULIRG Evolutionary Study) sample of local quasars and ULIRGs,
we have conducted a uniform analysis of the COS spectra of 33 $z \la
0.3$ Palomar-Green quasars. The main conclusions from our analysis are
the following:

\begin{enumerate}
\item Highly ionized outflows traced by blueshifted N~V
  $\lambda\lambda$1238, 1243 and O~VI $\lambda\lambda$1032, 1038
  absorption lines with equivalent widths larger than $\sim$ 20
  m\AA\ are present in about 60\% of the QUEST quasars. This detection
  rate is similar to that of warm-ionized outflows traced by
  blueshifted C~IV $\lambda\lambda$1548, 1550 absorption lines in
  local Seyfert galaxies and more distant, higher luminosity quasars.
\item The N~V and O~VI features in the QUEST quasars span a broad
  range of properties, both in terms of equivalent widths (from 20
  m\AA\ to 25 \AA) and kinematics (outflow velocities from a few
  $\times$ 100 \kms\ up to $\sim$ 10,000 \kms).
\item The rate of incidence and equivalent widths of the highly
  ionized outflows are higher among X-ray weak sources with X-ray to
  optical spectral indices $\alpha_{\rm OX} \la -1.7$ and X-ray column
  densities log~$N_H$(cm$^{-2}$) $\ga$ 22. The weighted outflow
  velocity dispersions are highest in the X-ray weakest sources with
  X-ray to optical spectral indices $\alpha_{\rm OX} \la -2$. These
  results are qualitatively similar to AGN-driven warm ionized
  outflows traced by the C~IV $\lambda\lambda$1548, 1550 absorption
  lines. These results favor radiative acceleration of the
    absorbers, where the X-rays are either absorbed or intrinsically
    weak in the wind-dominated systems. Line-locking is detected in
    the Ly$\alpha$ absorption throughs of one or two objects,
    providing additional evidence that radiation pressure plays an
    important role in accelerating these absorbers.
  
\item There is no significant trend between the weighted average
  velocity of the highly ionized outflows and the properties of the
  quasars and host galaxies. This negative result is likely due 
    in part to the fact that the range of properties of the QUEST
  quasar sample is narrow in comparison to those of other studies.
\item Blueshifted P~V broad absorption lines are clearly detected in
  PG~1126$-$041 and PG~1411$+$442 \citep[previously reported
    in][]{hamann2019}, and also posssibly in PG~1001$+$054 and
  PG~1004$+$130.  Using the results from the analysis of
  \citet{hamann2019}, these features imply column densities of $\sim$
  10$^{22.3}$ cm$^{-2}$ or larger and time-averaged outflow kinetic
  power to bolometric luminosity ratios of $\ga$ 0.1\% if a
  conservatively small radial distance of 1 pc from the P~V absorbers
  is assumed.
\end{enumerate}

Paper II of this series (Liu et al.\ 2021, in prep.) will present the
results from our analysis of the COS spectra on the QUEST
ULIRGs. These results will be combined with those from the present
paper to provide a more complete picture of the gaseous environments
of quasars and ULIRGs as a function of host galaxy properties and age
across the merger sequence from ULIRGs to quasars.

\acknowledgements

The authors thank the anonymous referee for suggestions which improved
this paper. SV, WL, and TMT acknowledge partial support for this work
provided by NASA through grant numbers {\em HST} GO-1256901A and
GO-1256901B, GO-13460.001-A and GO-13460.001-B, and GO-15662.001-A and
GO-15662.001-B from the Space Telescope Science Institute, which is
operated by AURA, Inc., under NASA contract NAS 5-26555. Based on
observations made with the NASA/ESA Hubble Space Telescope, and
obtained from the Hubble Legacy Archive, which is a collaboration
between the Space Telescope Science Institute (STScI/NASA), the Space
Telescope European Coordinating Facility (ST-ECF/ESA) and the Canadian
Astronomy Data Centre (CADC/NRC/CSA).  The authors also made use of
NASA's Astrophysics Data System Abstract Service and the NASA/IPAC
Extragalactic Database (NED), which is operated by the Jet Propulsion
Laboratory, California Institute of Technology, under contract with
the National Aeronautics and Space Administration.

  \software{COSQUEST \citep{david_rupke_2021_5659382}, IFSFIT
  \citep{rupke2014,rupke2015,david_rupke_2021_5659520}, LINMIX\_ERR
  \citep{kelly2007}, pymccorrelation \citep{privon2020}, scipy
  \citep{scipy2020}, DRTOOLS \citep{david_rupke_2021_5659554}}
\clearpage

\appendix

\section{Detailed Results from the Spectral Analysis}
\label{appendix:detailed_results}

\begin{turnpage}
\begin{figure}[!htb]   
\plotone{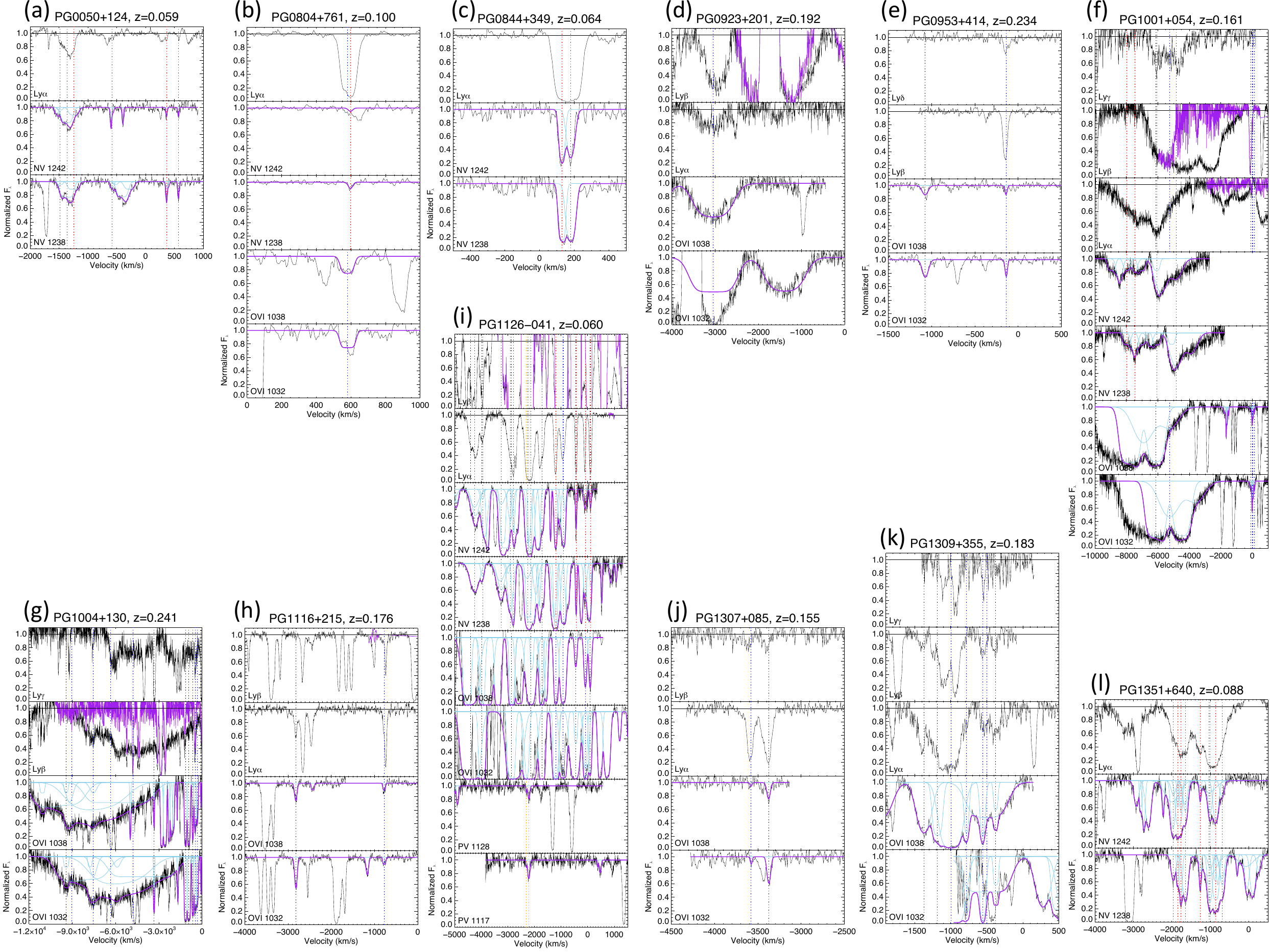}
\caption{[Part I] Interline comparison of the absorbing systems in
  each quasar with detected N~V or O~VI absorption lines, produced by
  plotting the normalized spectrum in velocity space relative to the
  quasar rest frame. The data are in black, the individual components
  used to fit the absorption profiles are shown in cyan, and the
  overall fit is shown in purple. The velocity centroids of the main
  absorbing systems are indicated by vertical red (\lya, \lyb, N~V,
  P~V) and blue (O~VI) dotted lines. }
\label{fig:qso_vel1}
\end{figure}
\end{turnpage}

Figures \ref{fig:qso_vel1}a$-$\ref{fig:qso_vel1}t present the fits to
the detected N~V, O~VI, and P~V absorption systems in our sample.  The
results from these fits are listed in Table \ref{tab:fits} in the
main body of the paper.  Here we summarize the results from our
spectral analysis for each object in the sample.

\vskip 0.1in

\indent{\em PG~0007$+$106.}---There are no associated N~V absorbers in
this system, although two deep blueshifted and redshifted \lya\
absorption features are present at $\vert v \vert < 400$ \kms.
\\
\indent{\em PG~0026$+$129.}---There are no associated N~V or O~VI
absorbers in this object.
\\
\indent{\em PG~0050$+$124 (I~Zw~1).}---N~V and \lya\ absorbers are detected at
$-$553, $-$1315, and $-$1467 \kms\ in this object.
Variable warm absorbers at $-$1870 and $-$2500 \kms\ have been
reported by \citet{silva2018} in {\em XMM-Newton} RGS spectra obtained
in 2015. \\
\indent{\em PG~0157$+$001 (Mrk~1014).}---There are no associated N~V or O~VI
absorbers in this system.
\\ \indent{\em PG~0804$+$761.}---This is a rare case for infall where
a strong redshifted absorption system at $+$600 \kms\ is observed in
\lya, and a corresponding weaker feature is also detected in N~V and
O~VI. The stronger features in the panel of Figure \ref{fig:qso_vel1}b
labeled O~VI 1037 are Fe~II features from the MW ISM.\\
\indent{\em PG~0838$+$770.}---A weak low-$\vert v \vert$ absorption
feature is seen in \lya\ but not N~V or O~VI.
\\
\indent{\em PG~0844$+$349.}---A strong slightly redshifted
double-component absorption feature, likely associated with tidal
debris, is present at $+$140 \kms\ and $+$190 \kms\ in both \lya\
and N~V, but the feature at $\sim$ 1288.9 \AA\ has no corresponding
N~V and is presumed to be \lya\ from intervening CGM.\\
\indent{\em PG~0923$+$201.}---A single broad absorber is observed at
$-$3300 \kms\ in \lya, \lyb, and O~VI 1032 and 1038 although the glare
of the geocoronal \lya\ airglow truncates the blue wing of the O~VI
1032 absorption line.  \\
\indent{\em PG~0953$+$414.}---Two faint O~VI features are detected at
$-$140 and $-$1074 \kms. The first feature is also
visible in \lya\ and \lyb, but the more blueshifted feature
is not. The strong feature near 1271 \AA\ cannot be fit with O~VI and
thus is likely \lya\ from intervening CGM.\\
\indent{\em PG~1001$+$054.}---The multi-epoch COS spectra shown in
Figure \ref{fig:mini-bal_variability} are co-added for this analysis
given the lack of variability in the absorption line profiles
(Sec.\ \ref{sec:discussion_location_structure}). This is a special case
because of the way the broad, high-velocity N~V absorption absorbs the
\lya\ profile and the O~VI absorption is deep, nearly dark, and highly
saturated. The three methods discussed in Section \ref{sec:analysis}
were attempted to deal with the \lya\ + N~V blend, and in the end
method \#3 was used for the final fit: (1) The use of
polynomials/splines to fit the blue and red sides of \lya\ is
problematic because it does not appear to yield a symmetric \lya\ line
and seems to miss N~V absorption that appears in \lya. (2) A
Lorentzian fit to \lya\ works reasonably well in that it yields a
symmetric line, but the line properties are highly unconstrained
(particularly in terms of height) and sensitive to the choice of which
continuum regions are fit. (3) After trying the QSO templates from
\citet{vandenberk2001},
\citet{stevans2014},
and \citet{harris2016}, we settled on the last one. We scaled the
template using a constant offset, a constant multiplier, and a scaling
according to $\lambda^p$ (arbitrary power $p$) to account for
differences between the spectral index and that of the template
\citep[following][]{harris2016}. We fit only the blue side of
\lya\ (and the far-red side) as well as some continuum regions in
between O~VI and \lya\/N~V that are fairly line-free. This
underpredicts the strength of highly-ionized lines, but is the best
compromise solution. If the lines are fit as well, the blue side of
\lya\ is not properly fit. The resulting fit to N~V seems to work
reasonably well, although the fit results are obviously illustrative
in terms of velocity space and certainly do not get the optical depth
correct. There seems to be a narrow \lya\ near $-$4000 \kms\ that is
barely detected in N~V, but the fit has difficulties capturing it. The
broad, highly saturated, nearly dark O~VI is fit over $-$[6000, 4000]
\kms\ but not $-$[8000, 6000] \kms\ to account for
\lyb\ contamination.  Broad blueshifted P~V 1117, 1128 at $\sim$
$-$[6000, 4000] \kms\ is also tentatively detected in this object but
no attempt is made to fit this faint feature
(Fig.\ \ref{fig:pv_detection}a).
\\
\indent{\em PG~1004$+$130.}---This is another special case
\citep[see][ for some previous analyses]{wills1999,brandt2000}. It is
difficult to fit the ``continuum'' shape and O~VI profile
simultaneously. The present fit is the best available compromise. It
is clear that \lyb\ / O~VI are interacting with each other so the O~VI
fit should be taken only as illustrative. There are higher-order Lyman
lines in the spectrum that are not considered. The Ly$\gamma$ region
is shown but is not considered in the fits to O~VI (the many narrow
features in the Ly$\gamma$ region are Galactic ISM lines).
Blueshifted P~V 1117 + 1128 at $-$[6000, 4000] \kms\ is tentatively
detected but no attempt is made to fit this faint feature
(Fig.\ \ref{fig:pv_detection}b).\\
\indent{\em PG~1116$+$215.}---There are several narrow absorption
features blueward of \lya, O~VI, and \lyb\ in this system, but only
two of them are paired in O~VI $\lambda$1032 and $\lambda$1038,
corresponding to $-$771 and $-$2825 \kms. Presumably, the orphan
features are intervening \lya\ absorbers. \citet{savage2014} also
report an O~VI system at 1175 and 1181 \AA, but it is not shown here
as it is very narrow and very high velocity, and is intervening CGM
\citep[there are nearby galaxies at the same redshift][]{tripp1998}.\\
\indent{\em PG~1126$-$041.}---Several separate COS/G130M spectra exist
for this object (Table \ref{tab:observations}).  Our analysis uses the
co-added spectrum. Broad blueshifted absorption features are clearly
detected in \lya, N~V, and O~VI (Fig.\ \ref{fig:example_pg1126_vel}).
A narrow P~V feature is observed at $-$2200 \kms, indicative of large
ionized column densities.  This P~V detection is discussed in more
detail in Section \ref{sec:discussion_pg1126}. 2014 and 2015
COS/G160M spectra centered around $\sim$ 1450 \AA\ show strong C~IV
$\lambda\lambda$1548, 1550 absorption throughs similar to those of N~V
and O~VI, but only weak and narrow blueshifted S~IV
$\lambda\lambda$1392, 1402 absorption features with $\vert v \vert$
$\la$ 500 \kms; the analysis of these features is beyond the scope of
the present study. \\
\indent{\em PG~1211$+$143.}---There are no associated N~V absorption
systems in this object. A broad absorption feature at the observed
wavelength of 1240 \AA\ has been interpreted by \citet{kriss2018} as a
highly blueshifted ($-$16,980 \kms) and broad (FWHM $\approx$ 1080 \kms)
\lya\ absorption from a fast wind. It may correspond to one of the
ultra-fast outflowing systems detected in the X-rays \citep[][and
  references therein]{danehkar2018}.  However, since it is not
detected in N~V
it is not included in the statistical analysis of Section
\ref{sec:results}. \\
\indent{\em PG~1226$+$023 (3C~273).}---The FUV spectrum of 3C~273 has been used
extensively to study the low-$z$ IGM
\citep[e.g.,][]{tripp2008,savage2014}.  There are no associated N~V or
O~VI absorption systems in this object. \\
\indent{\em PG~1229$+$204.}---There are no associated N~V absorption
systems in this object. The many unidentified features shortward of
\lya\ in the quasar rest-frame (e.g., 1289.5, 1282, 1223.2, 1220.4
\AA) may be \lya\ from intervening CGM.
\\
\indent{\em Mrk~231.}---The 2011 COS/G130M spectrum of this object was
presented in \citet{veilleux2013a},
while the 2014 COS/G140L and G230L spectra
were presented in \citet{veilleux2016}. The COS/G130M spectrum shows
\lya\ emission that is broad ($\ga$ 10,000 \kms) and highly blueshifted
(centroid at $\sim$ $-$3500 \kms). In contrast, blueshifted absorption
features are only present above $\sim$2200 \AA. These results have
been discussed in details in \citet{veilleux2016}, and this discussion
is not repeated here. This outflow is considered a non-detection in
our analysis in Section \ref{sec:results} since it has no N~V
absorption systems, but has all of the characteristics of a FeLoBAL at
visible and NUV wavelengths and is considered as such in our
discussion (Sec.\ \ref{sec:discussion}). \\
\indent{\em PG~1302$-$102.}---There are no associated O~VI absorbing
systems in this object, but several intervening \lya\ and
metal-line systems have been reported by \citet{cooksey2008}.\\
\indent{\em PG~1307$+$085.}---The weak O~VI absorber at $\sim$ $-$3400
\kms\ is also detected in \lya\ and \lyb. The other feature
at $\sim$ $-$3600 \kms\ is seen only in \lyb\ and \lya\ and
is presumed to be from intervening CGM. \\
\indent{\em PG~1309$+$355.}---A broad absorption feature is visible in
\lya, extending blueward to $\sim$ $-$1600 \kms. This feature is also
detected in both O~VI lines, but is truncated by the gap associated
with the strong geocoronal \lya\ emission. Strong absorption features
are detected coincident with P~V 1117 and 1128 near systemic velocity,
but given their equivalent widths they are likely of Galactic origin
(e.g. C~II 1334).\\
\indent{\em PG~1351$+$640.}---Two deep broad absorption throughs are
detected in N~V and \lya\ of this object, extending over [$-$2200,
  $-$1500] \kms\ and [$-$1400, $-$600] \kms. They are roughly
separated by the velocity separation of the N~V doublet lines,
resulting in a N~V ``triplet'', so this is a good case for line
locking (Sec.\ \ref{sec:discussion_radiation}). The template fit to
the \lya\ line emission works well in the very narrow region in which
it is used, but mostly because it gets the smooth profile in this
region better than a spline+polynomial fit. The central peak is
steeper than the template, and it is possible that the bluest
absorption region in \lya\ at $-$[3500, 2500] \kms\ is caused by a
weak peak rather than a true absorption, since it does not line up
with anything in N~V. The N~V template fit, again over this limited
range, is quite good and more trustworthy over a wider wavelength
range. It is actually quite comparable to the spline fit. It does not
get the steep central peak in \lya\ on the red side.
\\
\indent{\em PG~1411$+$442.}---This object has been the subject of a
detailed analysis by \citet{hamann2019}.  A deep broad absorption
through that extends over [$-$2800, $-$900] \kms\ is present in N~V
and \lya. Strong P~V absorption is also detected over $-$[2200, 1400]
\kms\ in this object. The absorption profiles presented here should be
taken only as illustrative. They are produced using a simple
spline/polynomical fit to the line emission + continuum. The quasar
template is much too broad to match the observed emission line
profiles of \lya\ and N~V. \\
\indent{\em PG~1435$-$067.}---There are no associated N~V or O~VI
absorption systems in this object, but a faint \lya\ absorber appears
to be present at $\sim$1369.3 \AA\ or $\sim$ $-$700 \kms. The other
bluer narrow absorption features are likely \lya\ absorption from the
intervening CGM.\\
  \indent{\em PG~1440$+$356.}---Two blueshifted N~V absorbers are
  observed at $-$2190 and $-$1610 \kms, and a fainter redshifted one
  at $+$460 \kms\ is also detected in both lines of the doublet. Note
  that the strongest of the blueshifted N~V systems does not have a
  good match in velocity space with \lya. The strong line near 1334
  \AA\ is produced in the Galactic ISM, while the strong line blueward
  of \lya\ is presumably produced by intervening CGM. The
  \lya\ line at $+$200 \kms\ may be systemic within the uncertainties
  on the redshift or a signature of inflow.\\
\indent{\em PG~1448$+$273.}---A broad (FWZI $\approx$ 700 \kms)
multi-component absorption feature is detected in both N~V and
\lya. The narrow \lya\ feature at $\sim$ $-$2700 \kms\ is
likely produced by intervening CGM, while the deep saturated feature
at 1280 \AA\ is C~I from Galactic ISM.  \\
\indent{\em PG~1501$+$106.}---Three strong absorption features are
seen blueward of \lya\ at $\sim$ 1253.0, 1255.8, and 1257.2 \AA,
corresponding to $\sim$ $-$1500, $-$800, and $-$500 \kms, respectively
(the absorption near 0 \kms\ is at least partly due to S~II from
Galactic ISM). Despite the chip gap at the position of the N~V
doublet, the present data allow us to rule out the presence of a N~V
$\lambda$1238 counterpart to the most highly blueshifted of these
\lya\ features since it would lie near $\sim$1276.5 \AA\ and is not
present in the data. \\
\indent{\em PG~1613$+$658.}---Two weak narrow absorption features are
detected at $-$3503 and $-$3764 \kms\ in N~V, O~VI, and \lya. The
other features around N~V are of Galactic ISM origin, while those in
O~VI and \lya\ are likely due to intervening CGM. \\
\indent{\em PG~1617$+$175.}---A broad ($\sim$1000 \kms) blueshifted
absorption feature centered around $\sim$ $-$3000 \kms\ is detected in
N~V, O~VI, and \lya. Three narrow absorbers at $-$3300, $-$1630, and
$-$1040 \kms\ are also detected in O~VI and \lya\ but are very weak or
absent in N~V.\\
\indent{\em PG~1626$+$554.}---The two distinct blueshifted \lya\
absorption features at 1378.8 \AA\ and 1374.7 \AA, corresponding to
$-$740 and $-$570 \kms\ in the quasar rest frame, are also detected in
\lyb\ but not in N~V. A faint depression at 1166.9 \AA\ may be the
$-$570 \kms\ counterpart of O~VI 1032 but it is not detected in the
fainter O~VI 1038. This feature was judged too uncertain to be
a detection in our analysis.\\
\indent{\em PG~2130$+$099.}---Two deep and narrow absorption features
at $\sim$ $-$1500 and 0 \kms\ are detected in N~V and \lya\ of
this object. \\
\indent{\em PG~2214$+$139.}---Our analysis is based on the co-added
spectrum of this object given the lack of variability in the
absorption lines between 2011 and 2012
(Fig.\ \ref{fig:mini-bal_variability}). This object shows complex N~V
  troughs that extend from $\sim$ $-$3400 to $-$400 \kms\ and are
  loosely matched to the complex absorption feature at \lya, except
  for the sharp \lya\ absorption feature near $-$100 \kms, which is
  not detected in N~V.  \\
\indent{\em PG~2233$+$134.}---This object shows a faint and narrow
absorber at $\sim$ $-$200 \kms\ in O~VI and \lyb. The \lyb\
absorber at $\sim$ $-$1300 \kms\ also seems to have a weak O~VI
counterpart but the fit is inconclusive and is therefore not included
in the statistics for this object. The other stronger lines in this
spectral region do not match Galactic ISM features so they are likely
produced by intervening CGM.\\
\indent{\em PG~2349$-$014.}---There are no N~V or O~V $\lambda$1032
absorbers in this object (O~VI $\lambda$1038 is lost in the glare of
geocoronal \lya).

\clearpage

\clearpage

\begin{turnpage}
\setcounter{figure}{12}
\begin{figure}[!htb]   
\plotone{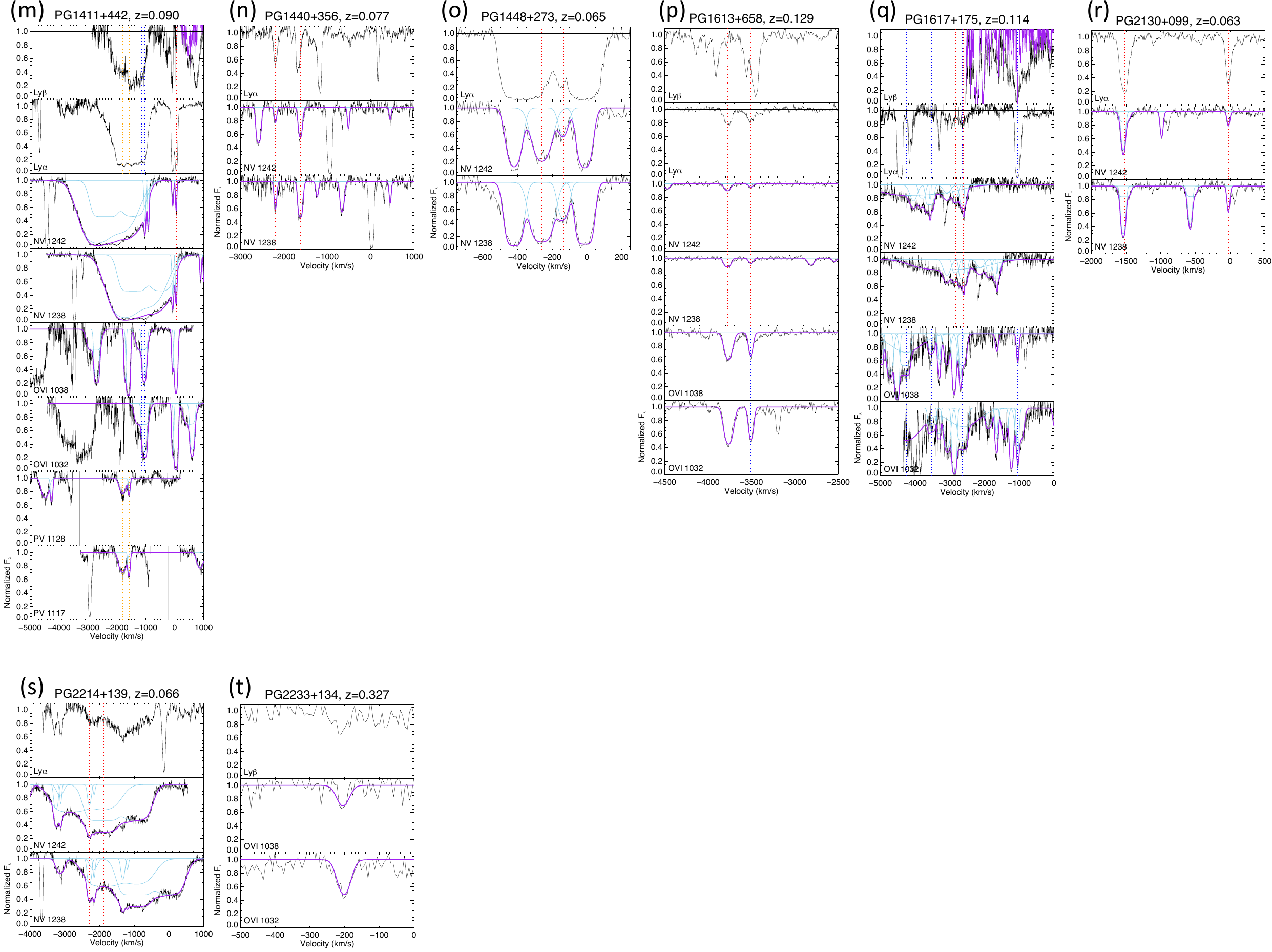}
\caption{[Part II] Same as previous figure.}
\label{fig:qso_vel2}
\end{figure}
\end{turnpage}
\clearpage

\bibliography{references.bib}

\end{document}